\documentclass[12pt]{article}
\usepackage{epsfig,amsfonts,amssymb}
\usepackage{hyperref}
\usepackage{cite}
\usepackage{cite}

\def\ZZZ{{\hbox{ Z\kern-1.6mm Z}}}
\def\RRR{{\hbox{ R\kern-2.4mm R}}}
\def\CCC{{\hbox{ C\kern-2.0mm C}}}
\def\zzz{{\hbox{z\kern-1mm z}}}

\newcommand{\qeq}{{\hbox{=\kern-2.3mm ? \kern.5mm }}}
\renewcommand{\qeq}{=}

\newcommand{\eps}{\epsilon}

\newcommand{\ve}{\varepsilon}

\newcommand{\II}{{\cal I}}

\newcommand{\FF}{{\cal F}}

\newcommand{\OO}{{\cal O}}

\newcommand{\EE}{{\cal E}}

\newcommand{\wt}{\widetilde}
\newcommand{\wh}{\widehat}

\newcommand{\RR}{{\cal R}}

\newcommand{\be}{\begin{equation}}
\newcommand{\ee}{\end{equation}}
\newcommand{\ben}{\begin{eqnarray}\displaystyle}
\newcommand{\een}{\end{eqnarray}}

\newcommand{\refb}[1]{(\ref{#1})}
\newcommand{\p}{\partial}
\newcommand{\sectiono}[1]{\section{#1}\setcounter{equation}{0}}

\def\one{{\hbox{ 1\kern-.8mm l}}}
\def\zero{{\hbox{ 0\kern-1.5mm 0}}}

\newcommand{\bea}[1]{\begin{eqnarray}\label{#1} }
\newcommand{\eea}{\end{eqnarray}}

\newcommand{\eqref}{\refb}

\usepackage{bm}
\usepackage[table]{xcolor}

\newcommand{\f}{\frac}

\newcommand{\non}{\nonumber}

\def\figgravitya{

\def\JPicScale{0.6}
\ifx\JPicScale\undefined\def\JPicScale{1}\fi
\unitlength \JPicScale mm
\begin{picture}(80,80)(0,0)
\linethickness{1mm}
\multiput(20,70)(0.14,-0.12){417}{\line(1,0){0.14}}
\linethickness{1mm}
\multiput(20,20)(0.14,0.12){417}{\line(1,0){0.14}}
\linethickness{0.4mm}
\put(32,60){\line(1,0){35}}
\linethickness{0.4mm}
\put(50,60){\line(0,1){20}}
\put(20,75){\makebox(0,0)[cc]{$p_a$}}

\put(80,75){\makebox(0,0)[cc]{$p_b$}}

\put(50,25){\makebox(0,0)[cc]{$\cdots$}}

\put(55,80){\makebox(0,0)[cc]{$k$}}

\put(40,65){\makebox(0,0)[cc]{$\ell$}}

\put(60,65){\makebox(0,0)[cc]{$k-\ell$}}

\put(30,50){\makebox(0,0)[cc]{$p_a+\ell$}}

\put(72,50){\makebox(0,0)[cc]{$p_b+k-\ell$}}

\put(45,60){\makebox(0,0)[cc]{$g$}}

\put(60,60){\makebox(0,0)[cc]{$g$}}

\put(50,70){\makebox(0,0)[cc]{$g$}}

\end{picture}

}

\def\figqedcGR{

\def\JPicScale{0.5}
\ifx\JPicScale\undefined\def\JPicScale{1}\fi
\unitlength \JPicScale mm
\begin{picture}(110,85)(0,0)
\linethickness{1mm}
\multiput(30,80)(0.12,-0.12){333}{\line(1,0){0.12}}
\linethickness{1mm}
\multiput(70,40)(0.12,0.12){333}{\line(1,0){0.12}}
\linethickness{1mm}
\multiput(40,10)(0.12,0.12){250}{\line(1,0){0.12}}
\linethickness{1mm}
\multiput(70,40)(0.12,-0.12){250}{\line(1,0){0.12}}
\linethickness{0.4mm}
\put(40,70){\line(1,0){60}}
\linethickness{0.4mm}
\put(25,55){\line(1,0){30}}
\put(35,85){\makebox(0,0)[cc]{$p_a$}}

\put(25,60){\makebox(0,0)[cc]{$k$}}

\put(62,65){\makebox(0,0)[cc]{$p_a+q $}}

\put(73,75){\makebox(0,0)[cc]{$\leftarrow q $}}

\put(40,45){\makebox(0,0)[cc]{$p_a+k+q $}}

\put(110,85){\makebox(0,0)[cc]{$p_b$}}

\put(94,50){\makebox(0,0)[cc]{$p_b-q $}}

\put(72,20){\makebox(0,0)[cc]{$\cdots$}}

\put(72,0){\makebox(0,0)[cc]{(c)}}

\put(90,70){\makebox(0,0)[cc]{g}}

\put(40,55){\makebox(0,0)[cc]{g}}

\end{picture}

}

\def\figonea{

\def\JPicScale{0.5}
\ifx\JPicScale\undefined\def\JPicScale{1}\fi
\unitlength \JPicScale mm
\begin{picture}(135,90)(0,0)
\linethickness{0.3mm}
\multiput(45,90)(0.12,-0.18){167}{\line(0,-1){0.18}}
\linethickness{0.3mm}
\multiput(100,55)(0.12,0.12){250}{\line(1,0){0.12}}
\linethickness{0.3mm}
\multiput(100,25)(0.18,-0.12){167}{\line(1,0){0.18}}
\linethickness{0.3mm}
\multiput(30,5)(0.18,0.12){167}{\line(1,0){0.18}}
\put(110,55){\makebox(0,0)[cc]{$r_1$}}

\put(55,60){\makebox(0,0)[cc]{$r_2$}}

\put(110,26){\makebox(0,0)[cc]{$r_1'$}}

\put(51,26){\makebox(0,0)[cc]{$r_2'$}}

\put(80,40){\makebox(0,0)[cc]{$\RR$}}

\linethickness{0.3mm}
\put(105.56,40.03){\line(0,1){0.5}}
\multiput(105.55,41.03)(0.01,-0.5){1}{\line(0,-1){0.5}}
\multiput(105.53,41.53)(0.02,-0.5){1}{\line(0,-1){0.5}}
\multiput(105.5,42.04)(0.03,-0.5){1}{\line(0,-1){0.5}}
\multiput(105.46,42.54)(0.04,-0.5){1}{\line(0,-1){0.5}}
\multiput(105.41,43.04)(0.05,-0.5){1}{\line(0,-1){0.5}}
\multiput(105.35,43.54)(0.06,-0.5){1}{\line(0,-1){0.5}}
\multiput(105.28,44.03)(0.07,-0.5){1}{\line(0,-1){0.5}}
\multiput(105.2,44.53)(0.08,-0.5){1}{\line(0,-1){0.5}}
\multiput(105.11,45.02)(0.09,-0.49){1}{\line(0,-1){0.49}}
\multiput(105.01,45.52)(0.1,-0.49){1}{\line(0,-1){0.49}}
\multiput(104.9,46.01)(0.11,-0.49){1}{\line(0,-1){0.49}}
\multiput(104.78,46.5)(0.12,-0.49){1}{\line(0,-1){0.49}}
\multiput(104.65,46.98)(0.13,-0.49){1}{\line(0,-1){0.49}}
\multiput(104.51,47.47)(0.14,-0.48){1}{\line(0,-1){0.48}}
\multiput(104.37,47.95)(0.15,-0.48){1}{\line(0,-1){0.48}}
\multiput(104.21,48.42)(0.16,-0.48){1}{\line(0,-1){0.48}}
\multiput(104.04,48.9)(0.17,-0.47){1}{\line(0,-1){0.47}}
\multiput(103.87,49.37)(0.18,-0.47){1}{\line(0,-1){0.47}}
\multiput(103.68,49.84)(0.09,-0.23){2}{\line(0,-1){0.23}}
\multiput(103.49,50.3)(0.1,-0.23){2}{\line(0,-1){0.23}}
\multiput(103.28,50.76)(0.1,-0.23){2}{\line(0,-1){0.23}}
\multiput(103.07,51.21)(0.11,-0.23){2}{\line(0,-1){0.23}}
\multiput(102.85,51.66)(0.11,-0.23){2}{\line(0,-1){0.23}}
\multiput(102.62,52.11)(0.12,-0.22){2}{\line(0,-1){0.22}}
\multiput(102.38,52.55)(0.12,-0.22){2}{\line(0,-1){0.22}}
\multiput(102.13,52.99)(0.12,-0.22){2}{\line(0,-1){0.22}}
\multiput(101.87,53.42)(0.13,-0.22){2}{\line(0,-1){0.22}}
\multiput(101.61,53.85)(0.13,-0.21){2}{\line(0,-1){0.21}}
\multiput(101.33,54.27)(0.14,-0.21){2}{\line(0,-1){0.21}}
\multiput(101.05,54.69)(0.14,-0.21){2}{\line(0,-1){0.21}}
\multiput(100.76,55.1)(0.15,-0.21){2}{\line(0,-1){0.21}}
\multiput(100.46,55.5)(0.15,-0.2){2}{\line(0,-1){0.2}}
\multiput(100.15,55.9)(0.1,-0.13){3}{\line(0,-1){0.13}}
\multiput(99.84,56.29)(0.1,-0.13){3}{\line(0,-1){0.13}}
\multiput(99.52,56.68)(0.11,-0.13){3}{\line(0,-1){0.13}}
\multiput(99.19,57.06)(0.11,-0.13){3}{\line(0,-1){0.13}}
\multiput(98.85,57.43)(0.11,-0.12){3}{\line(0,-1){0.12}}
\multiput(98.5,57.79)(0.11,-0.12){3}{\line(0,-1){0.12}}
\multiput(98.15,58.15)(0.12,-0.12){3}{\line(0,-1){0.12}}
\multiput(97.79,58.5)(0.12,-0.12){3}{\line(1,0){0.12}}
\multiput(97.43,58.85)(0.12,-0.11){3}{\line(1,0){0.12}}
\multiput(97.06,59.19)(0.12,-0.11){3}{\line(1,0){0.12}}
\multiput(96.68,59.52)(0.13,-0.11){3}{\line(1,0){0.13}}
\multiput(96.29,59.84)(0.13,-0.11){3}{\line(1,0){0.13}}
\multiput(95.9,60.15)(0.13,-0.1){3}{\line(1,0){0.13}}
\multiput(95.5,60.46)(0.13,-0.1){3}{\line(1,0){0.13}}
\multiput(95.1,60.76)(0.2,-0.15){2}{\line(1,0){0.2}}
\multiput(94.69,61.05)(0.21,-0.15){2}{\line(1,0){0.21}}
\multiput(94.27,61.33)(0.21,-0.14){2}{\line(1,0){0.21}}
\multiput(93.85,61.61)(0.21,-0.14){2}{\line(1,0){0.21}}
\multiput(93.42,61.87)(0.21,-0.13){2}{\line(1,0){0.21}}
\multiput(92.99,62.13)(0.22,-0.13){2}{\line(1,0){0.22}}
\multiput(92.55,62.38)(0.22,-0.12){2}{\line(1,0){0.22}}
\multiput(92.11,62.62)(0.22,-0.12){2}{\line(1,0){0.22}}
\multiput(91.66,62.85)(0.22,-0.12){2}{\line(1,0){0.22}}
\multiput(91.21,63.07)(0.23,-0.11){2}{\line(1,0){0.23}}
\multiput(90.76,63.28)(0.23,-0.11){2}{\line(1,0){0.23}}
\multiput(90.3,63.49)(0.23,-0.1){2}{\line(1,0){0.23}}
\multiput(89.84,63.68)(0.23,-0.1){2}{\line(1,0){0.23}}
\multiput(89.37,63.87)(0.23,-0.09){2}{\line(1,0){0.23}}
\multiput(88.9,64.04)(0.47,-0.18){1}{\line(1,0){0.47}}
\multiput(88.42,64.21)(0.47,-0.17){1}{\line(1,0){0.47}}
\multiput(87.95,64.37)(0.48,-0.16){1}{\line(1,0){0.48}}
\multiput(87.47,64.51)(0.48,-0.15){1}{\line(1,0){0.48}}
\multiput(86.98,64.65)(0.48,-0.14){1}{\line(1,0){0.48}}
\multiput(86.5,64.78)(0.49,-0.13){1}{\line(1,0){0.49}}
\multiput(86.01,64.9)(0.49,-0.12){1}{\line(1,0){0.49}}
\multiput(85.52,65.01)(0.49,-0.11){1}{\line(1,0){0.49}}
\multiput(85.02,65.11)(0.49,-0.1){1}{\line(1,0){0.49}}
\multiput(84.53,65.2)(0.49,-0.09){1}{\line(1,0){0.49}}
\multiput(84.03,65.28)(0.5,-0.08){1}{\line(1,0){0.5}}
\multiput(83.54,65.35)(0.5,-0.07){1}{\line(1,0){0.5}}
\multiput(83.04,65.41)(0.5,-0.06){1}{\line(1,0){0.5}}
\multiput(82.54,65.46)(0.5,-0.05){1}{\line(1,0){0.5}}
\multiput(82.04,65.5)(0.5,-0.04){1}{\line(1,0){0.5}}
\multiput(81.53,65.53)(0.5,-0.03){1}{\line(1,0){0.5}}
\multiput(81.03,65.55)(0.5,-0.02){1}{\line(1,0){0.5}}
\multiput(80.53,65.56)(0.5,-0.01){1}{\line(1,0){0.5}}
\put(80.03,65.56){\line(1,0){0.5}}
\multiput(79.52,65.55)(0.5,0.01){1}{\line(1,0){0.5}}
\multiput(79.02,65.53)(0.5,0.02){1}{\line(1,0){0.5}}
\multiput(78.52,65.5)(0.5,0.03){1}{\line(1,0){0.5}}
\multiput(78.02,65.46)(0.5,0.04){1}{\line(1,0){0.5}}
\multiput(77.52,65.41)(0.5,0.05){1}{\line(1,0){0.5}}
\multiput(77.02,65.35)(0.5,0.06){1}{\line(1,0){0.5}}
\multiput(76.52,65.28)(0.5,0.07){1}{\line(1,0){0.5}}
\multiput(76.03,65.2)(0.5,0.08){1}{\line(1,0){0.5}}
\multiput(75.53,65.11)(0.49,0.09){1}{\line(1,0){0.49}}
\multiput(75.04,65.01)(0.49,0.1){1}{\line(1,0){0.49}}
\multiput(74.55,64.9)(0.49,0.11){1}{\line(1,0){0.49}}
\multiput(74.06,64.78)(0.49,0.12){1}{\line(1,0){0.49}}
\multiput(73.57,64.65)(0.49,0.13){1}{\line(1,0){0.49}}
\multiput(73.09,64.51)(0.48,0.14){1}{\line(1,0){0.48}}
\multiput(72.61,64.37)(0.48,0.15){1}{\line(1,0){0.48}}
\multiput(72.13,64.21)(0.48,0.16){1}{\line(1,0){0.48}}
\multiput(71.66,64.04)(0.47,0.17){1}{\line(1,0){0.47}}
\multiput(71.19,63.87)(0.47,0.18){1}{\line(1,0){0.47}}
\multiput(70.72,63.68)(0.23,0.09){2}{\line(1,0){0.23}}
\multiput(70.26,63.49)(0.23,0.1){2}{\line(1,0){0.23}}
\multiput(69.8,63.28)(0.23,0.1){2}{\line(1,0){0.23}}
\multiput(69.34,63.07)(0.23,0.11){2}{\line(1,0){0.23}}
\multiput(68.89,62.85)(0.23,0.11){2}{\line(1,0){0.23}}
\multiput(68.44,62.62)(0.22,0.12){2}{\line(1,0){0.22}}
\multiput(68,62.38)(0.22,0.12){2}{\line(1,0){0.22}}
\multiput(67.57,62.13)(0.22,0.12){2}{\line(1,0){0.22}}
\multiput(67.13,61.87)(0.22,0.13){2}{\line(1,0){0.22}}
\multiput(66.71,61.61)(0.21,0.13){2}{\line(1,0){0.21}}
\multiput(66.29,61.33)(0.21,0.14){2}{\line(1,0){0.21}}
\multiput(65.87,61.05)(0.21,0.14){2}{\line(1,0){0.21}}
\multiput(65.46,60.76)(0.21,0.15){2}{\line(1,0){0.21}}
\multiput(65.06,60.46)(0.2,0.15){2}{\line(1,0){0.2}}
\multiput(64.66,60.15)(0.13,0.1){3}{\line(1,0){0.13}}
\multiput(64.27,59.84)(0.13,0.1){3}{\line(1,0){0.13}}
\multiput(63.88,59.52)(0.13,0.11){3}{\line(1,0){0.13}}
\multiput(63.5,59.19)(0.13,0.11){3}{\line(1,0){0.13}}
\multiput(63.13,58.85)(0.12,0.11){3}{\line(1,0){0.12}}
\multiput(62.76,58.5)(0.12,0.11){3}{\line(1,0){0.12}}
\multiput(62.4,58.15)(0.12,0.12){3}{\line(1,0){0.12}}
\multiput(62.05,57.79)(0.12,0.12){3}{\line(0,1){0.12}}
\multiput(61.71,57.43)(0.11,0.12){3}{\line(0,1){0.12}}
\multiput(61.37,57.06)(0.11,0.12){3}{\line(0,1){0.12}}
\multiput(61.04,56.68)(0.11,0.13){3}{\line(0,1){0.13}}
\multiput(60.72,56.29)(0.11,0.13){3}{\line(0,1){0.13}}
\multiput(60.4,55.9)(0.1,0.13){3}{\line(0,1){0.13}}
\multiput(60.1,55.5)(0.1,0.13){3}{\line(0,1){0.13}}
\multiput(59.8,55.1)(0.15,0.2){2}{\line(0,1){0.2}}
\multiput(59.51,54.69)(0.15,0.21){2}{\line(0,1){0.21}}
\multiput(59.22,54.27)(0.14,0.21){2}{\line(0,1){0.21}}
\multiput(58.95,53.85)(0.14,0.21){2}{\line(0,1){0.21}}
\multiput(58.68,53.42)(0.13,0.21){2}{\line(0,1){0.21}}
\multiput(58.43,52.99)(0.13,0.22){2}{\line(0,1){0.22}}
\multiput(58.18,52.55)(0.12,0.22){2}{\line(0,1){0.22}}
\multiput(57.94,52.11)(0.12,0.22){2}{\line(0,1){0.22}}
\multiput(57.71,51.66)(0.12,0.22){2}{\line(0,1){0.22}}
\multiput(57.49,51.21)(0.11,0.23){2}{\line(0,1){0.23}}
\multiput(57.27,50.76)(0.11,0.23){2}{\line(0,1){0.23}}
\multiput(57.07,50.3)(0.1,0.23){2}{\line(0,1){0.23}}
\multiput(56.87,49.84)(0.1,0.23){2}{\line(0,1){0.23}}
\multiput(56.69,49.37)(0.09,0.23){2}{\line(0,1){0.23}}
\multiput(56.51,48.9)(0.18,0.47){1}{\line(0,1){0.47}}
\multiput(56.35,48.42)(0.17,0.47){1}{\line(0,1){0.47}}
\multiput(56.19,47.95)(0.16,0.48){1}{\line(0,1){0.48}}
\multiput(56.04,47.47)(0.15,0.48){1}{\line(0,1){0.48}}
\multiput(55.9,46.98)(0.14,0.48){1}{\line(0,1){0.48}}
\multiput(55.78,46.5)(0.13,0.49){1}{\line(0,1){0.49}}
\multiput(55.66,46.01)(0.12,0.49){1}{\line(0,1){0.49}}
\multiput(55.55,45.52)(0.11,0.49){1}{\line(0,1){0.49}}
\multiput(55.45,45.02)(0.1,0.49){1}{\line(0,1){0.49}}
\multiput(55.36,44.53)(0.09,0.49){1}{\line(0,1){0.49}}
\multiput(55.28,44.03)(0.08,0.5){1}{\line(0,1){0.5}}
\multiput(55.21,43.54)(0.07,0.5){1}{\line(0,1){0.5}}
\multiput(55.15,43.04)(0.06,0.5){1}{\line(0,1){0.5}}
\multiput(55.1,42.54)(0.05,0.5){1}{\line(0,1){0.5}}
\multiput(55.06,42.04)(0.04,0.5){1}{\line(0,1){0.5}}
\multiput(55.03,41.53)(0.03,0.5){1}{\line(0,1){0.5}}
\multiput(55.01,41.03)(0.02,0.5){1}{\line(0,1){0.5}}
\multiput(55,40.53)(0.01,0.5){1}{\line(0,1){0.5}}
\put(55,40.03){\line(0,1){0.5}}
\multiput(55,40.03)(0.01,-0.5){1}{\line(0,-1){0.5}}
\multiput(55.01,39.52)(0.02,-0.5){1}{\line(0,-1){0.5}}
\multiput(55.03,39.02)(0.03,-0.5){1}{\line(0,-1){0.5}}
\multiput(55.06,38.52)(0.04,-0.5){1}{\line(0,-1){0.5}}
\multiput(55.1,38.02)(0.05,-0.5){1}{\line(0,-1){0.5}}
\multiput(55.15,37.52)(0.06,-0.5){1}{\line(0,-1){0.5}}
\multiput(55.21,37.02)(0.07,-0.5){1}{\line(0,-1){0.5}}
\multiput(55.28,36.52)(0.08,-0.5){1}{\line(0,-1){0.5}}
\multiput(55.36,36.03)(0.09,-0.49){1}{\line(0,-1){0.49}}
\multiput(55.45,35.53)(0.1,-0.49){1}{\line(0,-1){0.49}}
\multiput(55.55,35.04)(0.11,-0.49){1}{\line(0,-1){0.49}}
\multiput(55.66,34.55)(0.12,-0.49){1}{\line(0,-1){0.49}}
\multiput(55.78,34.06)(0.13,-0.49){1}{\line(0,-1){0.49}}
\multiput(55.9,33.57)(0.14,-0.48){1}{\line(0,-1){0.48}}
\multiput(56.04,33.09)(0.15,-0.48){1}{\line(0,-1){0.48}}
\multiput(56.19,32.61)(0.16,-0.48){1}{\line(0,-1){0.48}}
\multiput(56.35,32.13)(0.17,-0.47){1}{\line(0,-1){0.47}}
\multiput(56.51,31.66)(0.18,-0.47){1}{\line(0,-1){0.47}}
\multiput(56.69,31.19)(0.09,-0.23){2}{\line(0,-1){0.23}}
\multiput(56.87,30.72)(0.1,-0.23){2}{\line(0,-1){0.23}}
\multiput(57.07,30.26)(0.1,-0.23){2}{\line(0,-1){0.23}}
\multiput(57.27,29.8)(0.11,-0.23){2}{\line(0,-1){0.23}}
\multiput(57.49,29.34)(0.11,-0.23){2}{\line(0,-1){0.23}}
\multiput(57.71,28.89)(0.12,-0.22){2}{\line(0,-1){0.22}}
\multiput(57.94,28.44)(0.12,-0.22){2}{\line(0,-1){0.22}}
\multiput(58.18,28)(0.12,-0.22){2}{\line(0,-1){0.22}}
\multiput(58.43,27.57)(0.13,-0.22){2}{\line(0,-1){0.22}}
\multiput(58.68,27.13)(0.13,-0.21){2}{\line(0,-1){0.21}}
\multiput(58.95,26.71)(0.14,-0.21){2}{\line(0,-1){0.21}}
\multiput(59.22,26.29)(0.14,-0.21){2}{\line(0,-1){0.21}}
\multiput(59.51,25.87)(0.15,-0.21){2}{\line(0,-1){0.21}}
\multiput(59.8,25.46)(0.15,-0.2){2}{\line(0,-1){0.2}}
\multiput(60.1,25.06)(0.1,-0.13){3}{\line(0,-1){0.13}}
\multiput(60.4,24.66)(0.1,-0.13){3}{\line(0,-1){0.13}}
\multiput(60.72,24.27)(0.11,-0.13){3}{\line(0,-1){0.13}}
\multiput(61.04,23.88)(0.11,-0.13){3}{\line(0,-1){0.13}}
\multiput(61.37,23.5)(0.11,-0.12){3}{\line(0,-1){0.12}}
\multiput(61.71,23.13)(0.11,-0.12){3}{\line(0,-1){0.12}}
\multiput(62.05,22.76)(0.12,-0.12){3}{\line(0,-1){0.12}}
\multiput(62.4,22.4)(0.12,-0.12){3}{\line(1,0){0.12}}
\multiput(62.76,22.05)(0.12,-0.11){3}{\line(1,0){0.12}}
\multiput(63.13,21.71)(0.12,-0.11){3}{\line(1,0){0.12}}
\multiput(63.5,21.37)(0.13,-0.11){3}{\line(1,0){0.13}}
\multiput(63.88,21.04)(0.13,-0.11){3}{\line(1,0){0.13}}
\multiput(64.27,20.72)(0.13,-0.1){3}{\line(1,0){0.13}}
\multiput(64.66,20.4)(0.13,-0.1){3}{\line(1,0){0.13}}
\multiput(65.06,20.1)(0.2,-0.15){2}{\line(1,0){0.2}}
\multiput(65.46,19.8)(0.21,-0.15){2}{\line(1,0){0.21}}
\multiput(65.87,19.51)(0.21,-0.14){2}{\line(1,0){0.21}}
\multiput(66.29,19.22)(0.21,-0.14){2}{\line(1,0){0.21}}
\multiput(66.71,18.95)(0.21,-0.13){2}{\line(1,0){0.21}}
\multiput(67.13,18.68)(0.22,-0.13){2}{\line(1,0){0.22}}
\multiput(67.57,18.43)(0.22,-0.12){2}{\line(1,0){0.22}}
\multiput(68,18.18)(0.22,-0.12){2}{\line(1,0){0.22}}
\multiput(68.44,17.94)(0.22,-0.12){2}{\line(1,0){0.22}}
\multiput(68.89,17.71)(0.23,-0.11){2}{\line(1,0){0.23}}
\multiput(69.34,17.49)(0.23,-0.11){2}{\line(1,0){0.23}}
\multiput(69.8,17.27)(0.23,-0.1){2}{\line(1,0){0.23}}
\multiput(70.26,17.07)(0.23,-0.1){2}{\line(1,0){0.23}}
\multiput(70.72,16.87)(0.23,-0.09){2}{\line(1,0){0.23}}
\multiput(71.19,16.69)(0.47,-0.18){1}{\line(1,0){0.47}}
\multiput(71.66,16.51)(0.47,-0.17){1}{\line(1,0){0.47}}
\multiput(72.13,16.35)(0.48,-0.16){1}{\line(1,0){0.48}}
\multiput(72.61,16.19)(0.48,-0.15){1}{\line(1,0){0.48}}
\multiput(73.09,16.04)(0.48,-0.14){1}{\line(1,0){0.48}}
\multiput(73.57,15.9)(0.49,-0.13){1}{\line(1,0){0.49}}
\multiput(74.06,15.78)(0.49,-0.12){1}{\line(1,0){0.49}}
\multiput(74.55,15.66)(0.49,-0.11){1}{\line(1,0){0.49}}
\multiput(75.04,15.55)(0.49,-0.1){1}{\line(1,0){0.49}}
\multiput(75.53,15.45)(0.49,-0.09){1}{\line(1,0){0.49}}
\multiput(76.03,15.36)(0.5,-0.08){1}{\line(1,0){0.5}}
\multiput(76.52,15.28)(0.5,-0.07){1}{\line(1,0){0.5}}
\multiput(77.02,15.21)(0.5,-0.06){1}{\line(1,0){0.5}}
\multiput(77.52,15.15)(0.5,-0.05){1}{\line(1,0){0.5}}
\multiput(78.02,15.1)(0.5,-0.04){1}{\line(1,0){0.5}}
\multiput(78.52,15.06)(0.5,-0.03){1}{\line(1,0){0.5}}
\multiput(79.02,15.03)(0.5,-0.02){1}{\line(1,0){0.5}}
\multiput(79.52,15.01)(0.5,-0.01){1}{\line(1,0){0.5}}
\put(80.03,15){\line(1,0){0.5}}
\multiput(80.53,15)(0.5,0.01){1}{\line(1,0){0.5}}
\multiput(81.03,15.01)(0.5,0.02){1}{\line(1,0){0.5}}
\multiput(81.53,15.03)(0.5,0.03){1}{\line(1,0){0.5}}
\multiput(82.04,15.06)(0.5,0.04){1}{\line(1,0){0.5}}
\multiput(82.54,15.1)(0.5,0.05){1}{\line(1,0){0.5}}
\multiput(83.04,15.15)(0.5,0.06){1}{\line(1,0){0.5}}
\multiput(83.54,15.21)(0.5,0.07){1}{\line(1,0){0.5}}
\multiput(84.03,15.28)(0.5,0.08){1}{\line(1,0){0.5}}
\multiput(84.53,15.36)(0.49,0.09){1}{\line(1,0){0.49}}
\multiput(85.02,15.45)(0.49,0.1){1}{\line(1,0){0.49}}
\multiput(85.52,15.55)(0.49,0.11){1}{\line(1,0){0.49}}
\multiput(86.01,15.66)(0.49,0.12){1}{\line(1,0){0.49}}
\multiput(86.5,15.78)(0.49,0.13){1}{\line(1,0){0.49}}
\multiput(86.98,15.9)(0.48,0.14){1}{\line(1,0){0.48}}
\multiput(87.47,16.04)(0.48,0.15){1}{\line(1,0){0.48}}
\multiput(87.95,16.19)(0.48,0.16){1}{\line(1,0){0.48}}
\multiput(88.42,16.35)(0.47,0.17){1}{\line(1,0){0.47}}
\multiput(88.9,16.51)(0.47,0.18){1}{\line(1,0){0.47}}
\multiput(89.37,16.69)(0.23,0.09){2}{\line(1,0){0.23}}
\multiput(89.84,16.87)(0.23,0.1){2}{\line(1,0){0.23}}
\multiput(90.3,17.07)(0.23,0.1){2}{\line(1,0){0.23}}
\multiput(90.76,17.27)(0.23,0.11){2}{\line(1,0){0.23}}
\multiput(91.21,17.49)(0.23,0.11){2}{\line(1,0){0.23}}
\multiput(91.66,17.71)(0.22,0.12){2}{\line(1,0){0.22}}
\multiput(92.11,17.94)(0.22,0.12){2}{\line(1,0){0.22}}
\multiput(92.55,18.18)(0.22,0.12){2}{\line(1,0){0.22}}
\multiput(92.99,18.43)(0.22,0.13){2}{\line(1,0){0.22}}
\multiput(93.42,18.68)(0.21,0.13){2}{\line(1,0){0.21}}
\multiput(93.85,18.95)(0.21,0.14){2}{\line(1,0){0.21}}
\multiput(94.27,19.22)(0.21,0.14){2}{\line(1,0){0.21}}
\multiput(94.69,19.51)(0.21,0.15){2}{\line(1,0){0.21}}
\multiput(95.1,19.8)(0.2,0.15){2}{\line(1,0){0.2}}
\multiput(95.5,20.1)(0.13,0.1){3}{\line(1,0){0.13}}
\multiput(95.9,20.4)(0.13,0.1){3}{\line(1,0){0.13}}
\multiput(96.29,20.72)(0.13,0.11){3}{\line(1,0){0.13}}
\multiput(96.68,21.04)(0.13,0.11){3}{\line(1,0){0.13}}
\multiput(97.06,21.37)(0.12,0.11){3}{\line(1,0){0.12}}
\multiput(97.43,21.71)(0.12,0.11){3}{\line(1,0){0.12}}
\multiput(97.79,22.05)(0.12,0.12){3}{\line(1,0){0.12}}
\multiput(98.15,22.4)(0.12,0.12){3}{\line(0,1){0.12}}
\multiput(98.5,22.76)(0.11,0.12){3}{\line(0,1){0.12}}
\multiput(98.85,23.13)(0.11,0.12){3}{\line(0,1){0.12}}
\multiput(99.19,23.5)(0.11,0.13){3}{\line(0,1){0.13}}
\multiput(99.52,23.88)(0.11,0.13){3}{\line(0,1){0.13}}
\multiput(99.84,24.27)(0.1,0.13){3}{\line(0,1){0.13}}
\multiput(100.15,24.66)(0.1,0.13){3}{\line(0,1){0.13}}
\multiput(100.46,25.06)(0.15,0.2){2}{\line(0,1){0.2}}
\multiput(100.76,25.46)(0.15,0.21){2}{\line(0,1){0.21}}
\multiput(101.05,25.87)(0.14,0.21){2}{\line(0,1){0.21}}
\multiput(101.33,26.29)(0.14,0.21){2}{\line(0,1){0.21}}
\multiput(101.61,26.71)(0.13,0.21){2}{\line(0,1){0.21}}
\multiput(101.87,27.13)(0.13,0.22){2}{\line(0,1){0.22}}
\multiput(102.13,27.57)(0.12,0.22){2}{\line(0,1){0.22}}
\multiput(102.38,28)(0.12,0.22){2}{\line(0,1){0.22}}
\multiput(102.62,28.44)(0.12,0.22){2}{\line(0,1){0.22}}
\multiput(102.85,28.89)(0.11,0.23){2}{\line(0,1){0.23}}
\multiput(103.07,29.34)(0.11,0.23){2}{\line(0,1){0.23}}
\multiput(103.28,29.8)(0.1,0.23){2}{\line(0,1){0.23}}
\multiput(103.49,30.26)(0.1,0.23){2}{\line(0,1){0.23}}
\multiput(103.68,30.72)(0.09,0.23){2}{\line(0,1){0.23}}
\multiput(103.87,31.19)(0.18,0.47){1}{\line(0,1){0.47}}
\multiput(104.04,31.66)(0.17,0.47){1}{\line(0,1){0.47}}
\multiput(104.21,32.13)(0.16,0.48){1}{\line(0,1){0.48}}
\multiput(104.37,32.61)(0.15,0.48){1}{\line(0,1){0.48}}
\multiput(104.51,33.09)(0.14,0.48){1}{\line(0,1){0.48}}
\multiput(104.65,33.57)(0.13,0.49){1}{\line(0,1){0.49}}
\multiput(104.78,34.06)(0.12,0.49){1}{\line(0,1){0.49}}
\multiput(104.9,34.55)(0.11,0.49){1}{\line(0,1){0.49}}
\multiput(105.01,35.04)(0.1,0.49){1}{\line(0,1){0.49}}
\multiput(105.11,35.53)(0.09,0.49){1}{\line(0,1){0.49}}
\multiput(105.2,36.03)(0.08,0.5){1}{\line(0,1){0.5}}
\multiput(105.28,36.52)(0.07,0.5){1}{\line(0,1){0.5}}
\multiput(105.35,37.02)(0.06,0.5){1}{\line(0,1){0.5}}
\multiput(105.41,37.52)(0.05,0.5){1}{\line(0,1){0.5}}
\multiput(105.46,38.02)(0.04,0.5){1}{\line(0,1){0.5}}
\multiput(105.5,38.52)(0.03,0.5){1}{\line(0,1){0.5}}
\multiput(105.53,39.02)(0.02,0.5){1}{\line(0,1){0.5}}
\multiput(105.55,39.52)(0.01,0.5){1}{\line(0,1){0.5}}

\end{picture}

}

\begin{document}

\baselineskip 24pt

\begin{center}

{\Large \bf Proof of the Classical Soft Graviton Theorem in D=4}


\end{center}

\vskip .6cm
\medskip

\vspace*{4.0ex}

\baselineskip=18pt

\centerline{\large \rm Arnab Priya Saha, Biswajit Sahoo and Ashoke Sen}

\vspace*{4.0ex}

\centerline{\large \it Harish-Chandra Research Institute, HBNI}
\centerline{\large \it  Chhatnag Road, Jhusi,
Allahabad 211019, India}

\vspace*{1.0ex}
\centerline{\small E-mail:  arnabpriyasaha@hri.res.in, biswajitsahoo@hri.res.in, sen@hri.res.in}

\vspace*{5.0ex}

\centerline{\bf Abstract} \bigskip

Classical subleading 
soft graviton theorem in four space-time dimensions determines the gravitational wave-form
at late and early retarded time, generated during a scattering or explosion,
in terms of the
four momenta of the ingoing and outgoing objects. 
This result was `derived' earlier by taking the classical limit of the quantum soft graviton theorem, and making
some assumptions about how to deal with the infrared divergences of the soft factor.
In this paper we give a direct proof of this result by analyzing the classical equations of motion of gravity
coupled to matter. We also extend the result to the electromagnetic wave-form generated during scattering of
charged particles, and present a new 
conjecture on subsubleading corrections to the
gravitational wave-form at early and late retarded time.

\vfill \eject

\baselineskip=18pt

\tableofcontents

\sectiono{Introduction and summary}

In a quantum theory of gravity, soft graviton theorem gives an amplitude with a 
set of finite energy external particles
and
one or
more low energy external gravitons, in terms of the amplitude without 
the low energy gravitons\cite{weinberg2,1103.2981,1401.7026,1404.4091,1405.3533,1406.6987,
1408.2228,1706.00759,1503.04816,1504.05558,1607.02700,1707.06803,1809.01675,1811.01804}. 
However
when we take the classical limit, there is a different manifestation of the same theorem --
it determines
the low frequency component of the 
gravitational wave-form produced during a scattering process
in terms of the momenta and spin of the incoming and outgoing objects, without any
reference to the interactions responsible for  the 
scattering\cite{1801.07719}. 
Although initially this
result was derived by taking the classical limit of quantum amplitudes, this has now been
proved directly in the classical theory\cite{1906.08288} in five or more space-time dimensions.

In four space-time dimensions there are additional subtleties. In quantum theory these 
are related to infrared divergences of the S-matrix. In the classical limit these
manifest themselves 
in the logarithmic corrections to the asymptotic trajectories of the objects due
to the long range force operating between these objects. Due to these logarithmic
corrections, the orbital angular momenta of external objects, that enter the expression
for the gravitational wave-form, become ill defined. Refs.\cite{1804.09193,1808.03288} proposed 
a specific way of regulating these logarithmic divergences by suggesting that we
use the wave-length
of the soft graviton as the infrared cut-off. This introduced terms proportional to
$\ln\omega$ in the soft graviton theorem where $\omega$ is the angular
frequency of the soft graviton. After Fourier transformation, these terms control
the gravitational wave-form produced during a scattering at late and early retarded
time\cite{1806.01872}.\footnote{A possible
explanation of these logarithmic 
terms using asymptotic symmetries has been discussed
recently\cite{1903.09133,appear}.}

Our goal in this paper will be to prove the classical soft graviton theorem in four 
space-time dimensions directly in the classical theory. In particular we 
prove the following result. Let us consider a scattering process in which a set of $m$
objects carrying four 
momenta $p_1',\cdots, p_m'$ come together, scatter via some (unknown) interactions and disperse as $n$ objects carrying momenta $p_1,\cdots, p_n$. 
The special case $m=1$ will describe an explosion in which a single bound system
fragments into many objects, including radiation.
We shall
choose the origin of the space-time coordinate system so that the scattering takes place
within a finite neighborhood of the origin. Let us also suppose that we have a gravitational
wave detector
placed at a faraway point $\vec x$, and define
\be
R = |\vec x|, \qquad \hat n = {\vec x\over R}, \qquad n=(1,  \hat n)\, .
\ee
We shall consider the limit of large $R$ and analyze only the terms of order $1/R$
in the gravitational wave-form. We define the retarded time at the detector:
\be 
u \equiv t - R + 2\, G\, \ln R \sum_{b=1}^n p_b.n\, .
\ee
Here $t-R$ is the usual retarded time and the $2\, G\, \ln R \sum_{b=1}^n p_b.n$
takes into account the effect of the long range gravitational force on the gravitational
wave as it travels from the scattering center to the detector. $G$ denotes the
Newton's constant. We have used units in which the velocity of light $c$ has been set equal
to 1, -- this is the unit we shall use throughout the paper.
We also define the deviation of the metric $g_{\mu\nu}$ from flat metric via:
\be 
h_{\mu\nu} \equiv (g_{\mu\nu}-\eta_{\mu\nu})/2, \qquad e_{\mu\nu} \equiv h_{\mu\nu} 
- {1\over 2}\, \eta_{\mu\nu} \, \eta^{\rho\sigma} \,  h_{\rho\sigma}\, .
\ee
Let us first assume that the objects do not carry charge so that gravity is the only
long range force acting on the objects at late and early time, although during the scattering
they may undergo complicated interactions. Then at late and early retarded time,
our result for
the gravitational wave-form at the detector is given by:
\ben\label{e3.66int}
e^{\mu\nu}(t, R, \hat n) &=& {2\, G\over R} \, \left[-\sum_{a=1}^{n}  
p_a^\mu \, p_a^\nu\, 
{1\over n.p_a} + \sum_{a=1}^{m}  
p_a^{\prime\mu} \, p_a^{\prime\nu}\, 
{1\over n.p_a'} \right]\non\\
&& \hskip -1in -\, {4\, G^2\over R\, u} \left[
\sum_{a=1}^n \sum_{b=1\atop b\ne a}^n 
{ p_a.p_b\over 
\{(p_a.p_b)^2 
-p_a^2 p_b^2\}^{3/2}} \, 
\left\{{3\over 2} p_a^2 p_b^2 - (p_a.p_b)^2\right\} \, {n_\rho p_a^\mu \over n.p_a}\,
 (p_b^\rho p_a^\nu - p_b^\nu p_a^\rho) \right.
\non\\ && \left. \hskip -.5in - 
 \sum_{b=1}^n\, p_b.n  \left\{\sum_{a=1}^{n} \, {1\over p_a.n} \, p_a^\mu p_a^\nu
 - \sum_{a=1}^{m} \, {1\over p_a'.n} \, p_a^{\prime\mu} p_a^{\prime \nu}\right\}
 \right]\ + \ \OO(u^{-2})\, , \quad \hbox{as $u\to \infty$}\non\\
 e^{\mu\nu}(t, R, \hat n) &=&  {4\, G^2\over R\, u} \Bigg[ \sum_{a=1}^m \sum_{b=1\atop b\ne a}^m 
{ p_a'.p_b'\over 
\{(p_a'.p_b')^2 
-p_a^{\prime 2} p_b^{\prime 2}\}^{3/2}} \, 
\left\{{3\over 2} p_a^{\prime 2} p_b^{\prime 2} - (p_a'.p_b')^2\right\} \non\\
&& \hskip .8in \times\ {n_\rho p_a^{\prime \mu} \over n.p'_a}\,
 (p_b^{\prime\rho} p_a^{\prime\nu} - p_b^{\prime\nu} p_a^{\prime\rho})
 \Bigg]\ + \ \OO(u^{-2})\, , \quad 
 \hbox{as $u\to -\infty$}\, ,
\een
where $\OO(u^{-2})$ includes terms of order $u^{-2}\ln|u|$.
The term on the right hand side of the first line represents a
constant jump in $h_{\mu\nu}$ during the passage of the gravitational wave, and
is known as the memory 
effect\cite{mem1,mem2,mem3,mem4,christodoulou,thorne,bondi,1003.3486,1401.5831,1312.6871}. 
This is related to the leading soft 
theorem\cite{1411.5745}.
The terms of order $1/u$ are related to logarithmic corrections to the 
subleading soft theorem. These have been verified in various examples via explicit 
calculations\cite{peters,1812.08137,1901.10986}. The sum over $a$ in \refb{e3.66int} also includes
the contribution from finite frequency radiation emitted during the scattering.

As already discussed in 
\cite{1806.01872,1808.03288}, in case of decay ($m=1$), if at most  
one of the final objects is massive and the rest are
massless, including radiation, then the 
terms proportional to $1/u$ in the expression for $e^{\mu\nu}$ cancel. This will be the case for binary
black hole merger where the initial state is a single bound system, and 
the final state consists of a single massive black hole and gravitational radiation.
Therefore absence of $1/u$ tails in such decays can be taken as a test of general theory of relativity. 

If the objects participating in the scattering process are charged, with the incoming 
objects carrying charges $q'_1,\cdots, q'_m$ and outgoing objects carrying 
charges $q_1,\cdots , q_n$, then there are further corrections to \refb{e3.66int} due
to long range electromagnetic forces between the incoming and the outgoing objects.
These corrections have been given in 
\refb{e4.28em}.

A similar result can be given for the profile of the electromagnetic vector potential
$a_\mu$ at the detector at late and early retarded time. The results are given in 
\refb{eemabs1}, \refb{eemabs2}.

Although these results are derived in this paper for the first time, they have been 
conjectured earlier from soft graviton theorem following the chain of arguments given
at the beginning of this section. Emboldened by the success of these arguments,
we describe in \S\ref{s2} a new conjecture for terms of order $u^{-2} \ln|u|$ at late
and early retarded time. These have been given in \refb{eaddcon3}, \refb{eaddcon4}
and \refb{eaddcon5}. 

\sectiono{Some useful results} \label{suse}

In this section we shall review some simple mathematical results that will be useful for
our analysis.

\subsection{Different Fourier transforms}

We shall deal with functions of four variables $x\equiv 
(t,\vec x)\equiv(x^0,x^1,x^2, x^3)$ describing
the space-time coordinates. Given any such function $F(x)$, we shall introduce
the following different kinds of Fourier transforms:
\be \label{edefft}
\wh F(k)\equiv \int d^4 x\, e^{-i k. x} \, F(t,\vec x), \quad
\bar F(t,\vec k)\equiv \int d^3 x\, e^{-i\vec k.\vec x} \, F(t,\vec x), \quad
\wt F(\omega,\vec x)\equiv \int dt\, e^{i\omega t} \, F(t,\vec x)
\, .
\ee
The inverse relations are
\be\label{edefinvft}
F(t,\vec x)= \int {d^4 k\over (2\pi)^4}\, e^{i k. x} \, \wh F(k), \quad
F(t,\vec x)= \int {d^3 k\over (2\pi)^3} \, e^{i\vec k.\vec x} \, \bar F(t,\vec k), \quad
F(t,\vec x) =  \int {d\omega\over 2\pi}\, e^{-i\omega t} \, \wt F(\omega,\vec x)
\, .
\ee
Note that we are using the convention $k.x\equiv \eta_{\mu\nu}k^\mu x^\nu
= -k^0 x^0+\vec k.\vec x$.

\subsection{Radiative field at large distance} \label{s2.2}

Let us consider a differential equation of the form:
\be
\square F(x) = - j(x), \qquad \square\equiv \eta^{\alpha\beta}\, \p_\alpha\, \p_\beta\, ,
\ee
where $j(x)$ is some given function. 
The retarded solution to this equation is given by
\be\label{e2.4}
F(x) = - \int d^4 y \, G_r(x,y)\, j(y)\, ,
\ee
where $G_r(x,y)$ is the retarded Green's function:
\be 
G_r(x,y) = \int {d^4\ell\over (2\pi)^4} \, e^{i\ell.(x-y)} \, 
{1\over (\ell^0+i\eps)^2 -\vec\ell^2}\, .
\ee
Using \refb{edefft} we get
\be
\wt F(\omega, \vec x) = - \int d^4 y \,  j(y)\, 
\int{d^3\ell\over (2\pi)^3} \, e^{i\omega y^0 +i\vec\ell.(\vec x-\vec y)}
\, {1\over (\omega+i\eps)^2 -\vec\ell^2}\, .
\ee
For large $|\vec x|$, we can evaluate this integral using a saddle point approximation
as follows\cite{1801.07719}. Defining $\vec\ell_\parallel$ and $\vec\ell_\perp$ as 
components of $\vec \ell$ along $\vec x-\vec y$ and transverse to $\vec x-\vec y$
respectively, we get
\be
\wt F(\omega, \vec x) = - \int d^4 y \, j(y) \, 
\int{d^2\ell_\perp\over (2\pi)^2} \, {d\ell_\parallel\over 2\pi}
\, e^{i\omega y^0 +i\ell_\parallel \, |\vec x-\vec y|}
\, {1\over (\omega+i\eps)^2 - \ell_\parallel^2- \vec\ell_\perp^2}\, .
\ee
First consider the case $\omega>0$.
We now close the $\ell_\parallel$ integration contour in the upper half plane, picking
up residue at the pole at $\sqrt{(\omega+i\eps)^2 -\vec \ell_\perp^2}$. This gives
\be
\wt F(\omega, \vec x) = i\, \int d^4 y \, j(y)\,
\int{d^2\ell_\perp\over (2\pi)^2} \,
 e^{i\omega y^0 +i \, |\vec x-\vec y|\, \sqrt{(\omega+i\eps)^2 - \vec\ell_\perp^2}}
\, {1\over 2\sqrt{(\omega+i\eps)^2 - \vec\ell_\perp^2}}\, .
\ee
For large $|\vec x-\vec y|$ the exponent is a rapidly varying function of $\vec\ell_\perp$
and therefore we can carry out the integration over $\vec\ell_\perp$ using saddle point
approximation. The saddle point is located at $\vec \ell_\perp=0$. Expanding
the exponent to order $\vec \ell_\perp^2$ and carrying out gaussian integration
over $\vec\ell_\perp$ we get:
\be\label{efinF}
\wt F(\omega, \vec x) = i\, \int d^4 y \, j(y)\,
 e^{i\omega y^0 +i \, (\omega+i\eps) \, |\vec x-\vec y|}\, 
 {\omega+i\eps\over 2\, \pi\, i\, |\vec x-\vec y|}
\, {1\over 2(\omega+i\eps)} \simeq {1\over 4\pi R} \, e^{i\omega R}
\int d^4 y \, e^{-ik.y} \, j(y)\, ,
\ee
where we have made the approximation $|\vec x|>>|\vec y|$, and,
\be \label{edefkn}
k\equiv \omega(1,\hat n), \quad \hat n \equiv \vec x/|\vec x|, \quad R\equiv |\vec x|\, .
\ee
A similar analysis can be carried out for $\omega<0$, leading to the same final
expression. Using \refb{edefft}, eq.\refb{efinF} may be written as
\be \label{efinFin}
\wt F(\omega, \vec x) \simeq  {1\over 4\pi R} \, e^{i\omega R}
\, \wh j(k)\, .
\ee

This is a known formula (see {\it e.g.} \cite{1611.03493}), but the derivation given above also 
gives its limitations. 
In arriving at the right hand side of \refb{efinF}
we used the approximation $|\vec x|>> |\vec y|$. Therefore in the integration over $\vec y$ 
there is
a natural infrared cut-off given by $|\vec x|=R$.  
If the $y$ integral is convergent then there is
no need of such a cut-off, but in case the $y$ integral diverges from the large $y$ region, we need 
to explicitly impose the cut-off. We can
implement the cut-off by putting a cut-off on $y^0$, since typically the source $j(y)$ 
will have support inside
the light-cone $|\vec y|\le |y^0|$ for large $y$. For example, for positive $y^0$
we can implement the infrared cut-off by adding to
$k^0$ an imaginary part $i \, \Lambda \, R^{-1}$ for some fixed
number $\Lambda$. In that case for $y^0<< R/\Lambda$ this additional
factor has no effect on \refb{efinF}, but for $y^0>>  R/\Lambda$ there is an exponential
suppression factor that cuts off the integration over $y$. For negative $y^0$ the corresponding
modification of $k$ corresponds to adding an imaginary part $-i \, \Lambda \, R^{-1}$ to $k^0$.

\subsection{Late and early time behaviour from Fourier transformation} \label{s2.3}

In our analysis we shall encounter functions
$\wt F(\omega,\vec x)$ that
are non-analytic as $\omega\to 0$, -- having singularities either of the form $1/\omega$ or 
of the form $\ln \omega$. On general grounds we expect these singular small 
$\omega$ behaviour to be related
to the behaviour of $F(t,\vec x)$ as $t\to\pm\infty$. We shall now determine the
precise correspondence between the small $\omega$ behaviour of 
$\wt F(\omega,\vec x)$ and large $|t|$ behaviour of $F(t,\vec x)$. Since the analysis will be
carried out at fixed $\vec x$, we shall not display the $\vec x$ dependence of various quantities 
in subsequent discussions.

First we shall consider
singularities of the form $1/\omega$ for small $\omega$. For this consider a function of the form:
\be
\wt F(\omega) = C\, e^{i\omega \phi}\,
{1\over \omega} \, f(\omega)\, .
\ee
Here $C$ and $\phi$ are constants that could depend on $\vec x$. $f(\omega)$
is a function of $\omega$ that is smooth at $\omega=0$ with $f(0)=1$ 
and falls off sufficiently fast
as $\omega\to \infty$ so as to make the Fourier integral  over $\omega$ well defined.
Our final result will not depend on $f(\omega)$, but for definiteness we shall choose
\be
f(\omega) = {1\over \omega^2+1}\, .
\ee
This gives
\be \label{e2.14}
F(t) =\int {d\omega\over 2\pi} e^{-i\omega t} \wt F(\omega)
= C \, \int{d\omega\over 2\pi} \, e^{-i\omega u} {1\over \omega} \, f(\omega), \qquad
u\equiv t - \phi\, .
\ee
In order to define the integral around $\omega=0$,
we need to choose an appropriate $i\eps$ prescription. However since $1/(\omega+i\eps)$ and $1/(\omega-i\eps)$
differ by a term proportional to $\delta(\omega)$, whose Fourier transform is a $u$ independent constant, the difference
will not be of interest to us. For definiteness , we shall work with $1/(\omega+i\eps)$. 
Then we have 
\be\label{e1.9pre}
F(t) =  C \,{1\over 2\pi} \int d\omega \,
e^{-i\omega u} {1\over \omega+ i\eps} f(\omega) =  -i \, C\, H(u) + \OO(e^{-u})\, ,
\ee
where $H$ is the Heaviside step function.
This result is obtained by closing the contour in the lower (upper) half plane 
for positive (negative) $u$, and
picking up the residues at the poles. The order $e^{-u}$ contribution comes from 
the residues at the poles
of $f(\omega)$. The step function $H(u)$ gives a jump in $e_{\mu\nu}$ 
between $u\to -\infty$ and
$u\to \infty$, leading to the memory effect\cite{mem1,mem2,mem3,mem4}.

Let us now turn to the Fourier transform of the logarithmic terms. 
We consider functions of the form:
\be
\wt F(\omega) = C \, e^{i\omega \phi}\,
\ln\omega \, f(\omega)\, .
\ee
Again we need to consider the different
$i\eps$ prescriptions, and this time the difference between the two choices is not trivial. 
Therefore we consider\footnote{If $F(t)$ is real, we must have from \refb{edefft}
$\wh F(\omega) = \wh F(-\omega)^*$.
Now since $\ln(-\omega+i\eps)^* = \ln(-\omega-i\eps) = \ln(\omega+i\eps) -i\pi$, we see that
$\ln(\omega+i\eps)$ is not a good candidate for $\wt F(\omega)$. This can be rectified by
averaging over $\ln(\omega+i\eps)$ and $\ln(-\omega-i\eps)$. However since the two differ by a constant,
whose Fourier transform, being proportional to $\delta(u)$, does not affect the behavior at large $|u|$, we
shall ignore this complication. A similar remark holds for $\ln(\omega-i\eps)$. \label{fo11}
}
\be\label{e2.17}
F_\pm(t) = C\, \int{d\omega\over 2\pi} \, e^{-i\omega t} \, e^{i\omega \phi}\,
\ln(\omega\pm i\eps) \, f(\omega)
=  C \,\int{d\omega\over 2\pi} \, e^{-i\omega u} \, 
\ln(\omega\pm i\eps) \, f(\omega)\, .
\ee
For $u>0$ we can close the contour in the lower half plane. In this case $F_-$ gets contribution only from 
the poles of $f(\omega)$ and therefore is suppressed by factors of $e^{-u}$. Similarly for $u<0$, $F_+$ is
suppressed by powers of $e^{-u}$. Furthermore, using $\ln(\omega+i\eps)=\ln(\omega-i\eps) + 2\,\pi\, i\,
H(-\omega)$, we have
\be
F_+ - F_- = i \, C \,
\int_{-\infty}^0 d\omega \, e^{-i\omega u} f(\omega) \simeq - 
{C\over u}, \quad \hbox{for \, $u\to \pm\infty$}\, .
\ee
Using these results we get
\ben \label{e1.13ag}
&& F_+\equiv C\, \int{d\omega\over 2\pi} \, e^{-i\omega u} \, 
\ln(\omega + i\eps) \, f(\omega)\to\cases{ -\, 
\displaystyle{C\over u} \quad \hbox{for} \quad u\to \infty, \cr \cr 0  \quad \hbox{for} \quad u\to -\infty,
} \nonumber \\ \nonumber\\
&& F_-\equiv C\,\int{d\omega\over 2\pi} \, e^{-i\omega u} \, 
\ln(\omega - i\eps)\, f(\omega) \to \cases {0 \quad \hbox{for}   \quad u\to \infty, \cr \cr 
\displaystyle{C\over u}  \quad \hbox{for}   \quad
u\to -\infty.
}
\een

Next we shall consider the integrals:
\be
G_\pm \equiv C\, \int{d\omega\over 2\pi} \, e^{-i\omega u} \, \omega\, 
\{\ln(\omega\pm i\eps)\}^2 \, f(\omega)\, .
\ee
As before, $G_+$ vanishes for large negative $u$ and $G_-$ vanishes for large positive $u$ up to exponentially
suppressed corrections. Furthermore we have
\ben
G_+-G_- &=& 4\, \pi \, i \, C\, \int_{-\infty}^0
{d\omega\over 2\pi} \, e^{-i\omega u} \, \omega\, \left\{ \ln(\omega-i\eps)+i\pi\right\}\,f(\omega)\non\\
&=& - 4\, \pi \, C\, {d\over du} \int_{-\infty}^0
{d\omega\over 2\pi} \, e^{-i\omega u} \, \left\{ \ln(\omega-i\eps)+i\pi\right\}\,f(\omega)\, .
\een
Changing integration variable to $v=\omega\, u$ we can express this as
\ben
G_+-G_- &=& -2 \, C\, {d\over du}\left[u^{-1}\, \int_{-\infty\times {\rm sign}\, u}^0
dv \, e^{-iv} \, \left\{ \ln(v-i\eps)-\ln u+i\pi\right\}\, f(v/u)\right]\non\\
&=& -2 \, C\, {d\over du}\left[-i\, u^{-1}\, \ln u + \OO(u^{-1})\right] = -2\, i\, C 
\, u^{-2} \, \ln |u| + \OO(u^{-2})  \, .
\een
This gives
\ben \label{eGlimit}
&& G_+\equiv C\, \int{d\omega\over 2\pi} \, e^{-i\omega u} \, \omega
\{\ln(\omega + i\eps)\}^2 \, f(\omega)\to\cases{ -\, 
2\, i\, C\, u^{-2} \, \ln|u| \quad \hbox{for} \quad u\to \infty, \cr \cr 0  \quad \hbox{for} \quad u\to -\infty,
} \nonumber \\ 
&& G_-\equiv C\, \int{d\omega\over 2\pi} \, e^{-i\omega u} \, \omega
\{\ln(\omega - i\eps)\}^2 \, f(\omega)\to\cases{ 0 \quad \hbox{for} \quad u\to \infty, \cr \cr 
2\, i\, C\, u^{-2} \, \ln|u|  \quad \hbox{for} \quad u\to -\infty\, ,
} 
\een
up to corrections of order $u^{-2}$.

Finally we consider the integral:
\be
H \equiv C\, \int{d\omega\over 2\pi} \, e^{-i\omega u} \, \omega\, 
\ln(\omega + i\eps)\, \ln(\omega- i\eps) \, f(\omega)\, .
\ee
For evaluating this we use the result:
\ben
G_+ + G_- - 2\, H &=& C\, \int{d\omega\over 2\pi} \, e^{-i\omega u} \, \omega\, 
\{\ln(\omega+ i\eps) -\ln(\omega-i\eps)\}^2 \, f(\omega)\non\\
&=& -\, 2\, \pi \, C \, \int_{-\infty}^0 {d\omega} \, e^{-i\omega u} \, \omega\, 
 \, f(\omega) = \OO(u^{-2})\, .
\een
Using \refb{eGlimit} we now get:
\be \label{eHlimit}
H \equiv C\, \int{d\omega\over 2\pi} \, e^{-i\omega u} \, \omega\, 
\ln(\omega + i\eps)\, \ln(\omega- i\eps) \, f(\omega)
\to\cases{ -\, 
i\, C\, u^{-2} \, \ln|u| \quad \hbox{for} \quad u\to \infty\, , \cr \cr i\, C\, 
u^{-2} \, \ln|u|   \quad \hbox{for} \quad u\to -\infty\, .
} 
\ee

\sectiono{Proof of classical soft graviton theorem} \label{sgrav}

We consider a scattering event in asymptotically flat space-time in which $m$ objects
carrying masses $\{m'_a\}$, four velocities $\{v'_a\}$ and 
four momenta $\{p'_a=m'_a \, v'_a\}$ for $1\le a\le m$ come close, undergo complicated interactions,
and disperse as 
$n$ objects carrying masses $\{m_a\}$, four velocities $\{v_a\}$  and
four momenta $\{p_a\}$ for $1\le a\le n$.  We do not assume that the interactions
are weak, and they could involve  exchange of
energy and other quantum numbers, fusion and splitting. Our goal will be to compute
the gravitational wave-form emitted during this scattering event at early and
late retarded time. As discussed in \S\ref{s2.3}, this is related to the behaviour of the Fourier
transform of the wave-form in the
low frequency limit.

Since we shall be interested in the long wavelength gravitational waves emitted by the system, we can
represent the leading contribution to the energy momentum tensor of the incoming and outgoing objects 
by the energy momentum tensor of point
particles, and include the effect of internal structure of the
objects by adding subleading contributions involving higher derivative 
terms\cite{Tulczyjew,Dixon,0511061,0511133,0604099,0804.0260,1709.06016,1712.09250,1812.06895}. 
In fact, to the order
at which we shall be working, it will be sufficient to keep just the leading term. For this reason, we shall
henceforth refer to the incoming and outgoing objects as particles.

The strategy we shall follow will be to iteratively solve the coupled equations of
motion of matter and gravity using Feynman diagram like techniques. This method
has been widely used in recent years\cite{0409156,0912.4254,1203.2962,1211.6095,1907.02869}, 
most notably in \cite{1611.03493,1711.09493,1705.09263,1806.07388}. 
However  the
main difference between our approach and the earlier ones is in setting up the boundary
conditions. In the usual approach we set the initial condition and evolve the
system using the equations of motion, computing both the trajectories and the
emitted radiation during this process. In our approach we take the initial and
final momenta as given, but allow the interactions during the scattering to be
arbitrary. Therefore while solving the equations we need to evolve the initial
particle trajectories forward in time and the final particle
trajectories backward in time, and compute the
net gravitational wave emitted during the scattering.

For simplicity, in this section we shall
consider the situation where the particles are uncharged so that there are
no long range 
electromagnetic interactions between the asymptotic particles. The
effect of such interactions will be incorporated in \S\ref{sgen.3}.

\subsection{General set-up} \label{egrav.0}

\begin{figure}
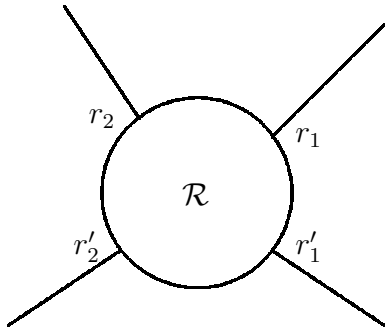


\begin{center}

\figonea

\end{center}

\caption{A scattering process in which the particles interact
strongly inside the region $\RR$ via some unspecified forces, but
outside the region $\RR$ the only force operative between the particles 
is the long range
gravitational force.  \label{fig1}}

\end{figure}

We choose the origin
of the space-time coordinate system to be somewhere within the region where the scattering 
takes place and denote by $\RR$ a large but finite region of space-time so that
the non-trivial part of the scattering occurs within the region $\RR$. In particular
we shall choose
$\RR$ to be sufficiently large so that outside the region $\RR$ 
the only interaction that exists
between the particles is the long range gravitational interaction.
This has been shown in
Fig.~\ref{fig1}.
We shall denote by $L$ the linear size of $\RR$ and analyze gravitational radiation at
retarded time $u$ for $|u|>>L$.

We define:
\be
h_{\mu\nu}= {1\over 2} (g_{\mu\nu}-\eta_{\mu\nu}), \quad e_{\mu\nu} =
h_{\mu\nu} -{1\over 2} \eta_{\mu\nu} \, \eta^{\rho\sigma}h_{\rho\sigma} \,\, 
\Leftrightarrow \,\, h_{\mu\nu} =e_{\mu\nu} -{1\over 2} \eta_{\mu\nu} \, \eta^{\rho\sigma}e_{\rho\sigma}\, .
\ee
We denote by $X_a(\sigma)$ for $1\le a\le n$ the outgoing particle trajectories parametrized by the
proper time\footnote{More precisely, $\sigma$ is a parameter labelling the trajectory, that is set equal to the
proper time after deriving the equations of motion.}
$\sigma$  in the range $0\le \sigma< \infty$, with $\sigma=0$ labelling the point where
the trajectory exits the region $\RR$.
Similarly 
$X_a'(\sigma)$ for $1\le a\le m$ will denote the incoming particle trajectories parametrized by the
proper time $\sigma$  in the range $-\infty< \sigma\le 0$, with $\sigma=0$ labelling the point where
the trajectory enters the region $\RR$.
We now consider the Einstein's action coupled to these particles:
\ben \label{eaction}
S &=& {1\over 16\pi G} \, \int d^4 x \, \sqrt{-\det g} \, R - \sum_{a=1}^n \int_0^\infty d\sigma \, m_a \left\{
-g_{\mu\nu}(X(\sigma))\, {d X_a^\mu\over d\sigma} \, {d X_a^\nu\over d\sigma}\right\}^{1/2}\nonumber \\ &&
- \sum_{a=1}^m \int_{-\infty}^0 d\sigma \, m_a' \left\{
-g_{\mu\nu}(X'(\sigma))\, {d X_a^{\prime\mu}\over d\sigma} \, {d X_a^{\prime\nu}\over d\sigma}\right\}^{1/2} \, .
\een
Note that we have included in the action the contribution only from part of the particle trajectories 
that lie outside the region
$\RR$. We shall argue later that this action is sufficient for determining 
the gravitational wave-form at late and early time. We now 
derive the equations of motion for 
$e_{\mu\nu}$ by extremizing the action \refb{eaction} with respect to $e_{\mu\nu}$. This takes the form:
\be\label{efulleom}
\sqrt{-\det g} \left(R^{\mu\nu} - {1\over 2} g^{\rho\sigma} R_{\rho\sigma} \, g^{\mu\nu}\right) = 8\, \pi\, G\, 
T^{X\mu\nu}\, ,
\ee
where,
\be\label{etmatter}
T^{X\mu\nu}\equiv \sum_{a=1}^n
m_a\, \int_0^\infty d\sigma\, \delta^{(4)}(x - X_a(\sigma)) \, {dX_a^\mu\over d\sigma} \, {dX_a^\nu\over d\sigma}
+ \sum_{a=1}^m
m_a'\, \int_{-\infty}^0 d\sigma\, \delta^{(4)}(x - X_a'(\sigma)) \, {dX_a^{\prime\mu}\over d\sigma} \, {dX_a^{\prime\nu}\over d\sigma}\,  .
\ee 
Note the factor of $\sqrt{-\det g}$ 
and the raised indices on the
left hand side of \refb{efulleom} -- this makes the right hand side independent of the metric.
After imposing
the
de Donder gauge:
\be
\eta^{\mu\nu} \p_\mu h_{\nu\lambda} - {1\over 2}
\p_\lambda \left(\eta^{\rho\sigma}h_{\rho\sigma}\right)=0 \quad \Leftrightarrow \quad \eta^{\mu\nu}\, \p_\mu\, 
e_{\nu\lambda}=0\, ,
\ee
and expanding the left hand side of \refb{efulleom} in power series in $h_{\mu\nu}$, we can express the
equations of motion of the metric as:
\be \label{egrbox}
\eta^{\alpha\mu}\, \eta^{\beta\nu}\, 
\eta^{\rho\sigma} \p_\rho\p_\sigma e_{\alpha\beta} = - 
8\, \pi \, G\, T^{\mu\nu}(x), \qquad 
T^{\mu\nu}\equiv T^{X\mu\nu} + T^{h\mu\nu}\, ,
\ee
where $T^{h\mu\nu}$ denotes the gravitational
stress tensor, defined as what we obtain by taking all $e_{\alpha\beta}$ dependent
terms on the left hand side of \refb{efulleom}, except the 
terms linear in $e_{\alpha\beta}$, 
to the right hand side and dividing it by $8\,\pi\, G$.  In 
all subsequent equations, the indices will be raised and lowered by $\eta_{\mu\nu}$.

Our goal is to compute $e_{\mu\nu}(t, \vec x)$ at a point far away from the scattering
center. We shall label $\vec x$ as $R\, \hat n$ where $\hat n$ is a unit vector and 
$R\equiv |\vec x|$. It follows from \refb{e2.4} and 
\refb{efinFin} that the retarded solution to \refb{egrbox} is given by\cite{1801.07719}\footnote{\refb{e55gr} 
can also be 
written as
$$
e_{\mu\nu} (t, R , \hat n) = {2\, G\over R}\, \bar T_{\mu\nu}(t-R, \vec k) +\OO(R^{-2})\, .
$$}
\be\label{e55gr}
\tilde e_{\mu\nu}(\omega, R, \hat n) = {2\, G\over R}\, e^{i\omega R}\, 
\wh T_{\mu\nu}(k)+\OO(R^{-2})\, ,
\ee
where
\be\label{edefThat}
\wh T_{\mu\nu}(k)\equiv \int d^4 x \, e^{-ik.x}\, T_{\mu\nu}(x)\, ,
\ee
is the Fourier transform of $T_{\mu\nu}(x)$ in all the variables and
$k =\omega(1, \hat n)$ as defined in \refb{edefkn}.
Therefore we need to compute $\wh T_{\mu\nu}(k)$. 
Furthermore, it follows from the analysis of \S\ref{s2.3} that to extract the late and early 
time behaviour
of $e_{\mu\nu}(t,\vec x)$ we need to examine the non-analytic part of 
$\tilde e_{\mu\nu}(\omega, R, \hat n)$ as a function of $\omega$ -- in particular terms of
order $1/\omega$ and $\ln\omega$.
For this, we can restrict
the integration over $x$ in \refb{edefThat} to outside the region $\RR$, since integration over a finite
region of space-time will give an infrared finite contribution and 
cannot generate a singularity as $\omega\to 0$. This justifies the omission of the contribution to the action
\refb{eaction} from particle trajectories inside the region $\RR$.

We shall compute
$\wh T_{\mu\nu}$  by solving the following
equations iteratively:
\ben \label{eitergrpre}
&& T^{\mu\nu}(x) = T^{X\mu\nu}(x) 
+ T^{h\mu\nu}(x), \nonumber \\
&& T^{X\mu\nu}(x) \equiv
\sum_{a=1}^n m_a\, \int_0^\infty d\sigma\, \delta^{(4)}(x - X_a(\sigma)) \, {dX_a^\mu\over d\sigma} \, {dX_a^\nu\over d\sigma} \non\\ && \hskip .5in
+ \ \sum_{a=1}^m
m_a'\, \int_{-\infty}^0 d\sigma\, \delta^{(4)}(x - X_a'(\sigma)) \, {dX_a^{\prime\mu}\over d\sigma} \, {dX_a^{\prime\nu}\over d\sigma}\, ,
 \nonumber \\
&& \square \, e_{\mu\nu} = -8\, \pi\, G\, T_{\mu\nu}\equiv - 8\, \pi\, G\, \eta_{\mu\alpha}\,
\eta_{\nu\beta} \, T^{\alpha\beta}\,  , \cr \nonumber \\ &&
{d^2 X_a^\mu\over d\sigma^2} = - \Gamma^\mu_{\nu\rho}(X(\sigma)) \,  {dX_a^\nu\over d\sigma} \, {dX_a^\rho\over d\sigma} , \qquad  
{d^2 X_a'^\mu\over d\sigma^2} = - \Gamma^\mu_{\nu\rho}(X'(\sigma)) \,  {dX_a^{\prime \nu}\over d\sigma} \, 
{dX_a^{\prime\rho}\over d\sigma}\, , 
\een
with boundary conditions:
\be \label{eboundpre}
X_a^\mu(\sigma=0)=r_a^\mu, \quad \lim_{\sigma\to\infty} {dX_a^\mu \over d\sigma}=v_a^\mu ={1\over m_a}\, p_a^\mu,
\quad
X_a^{\prime\mu}(\sigma=0)=r_a^{\prime \mu}, \quad \lim_{\sigma\to-\infty} {dX_a^{\prime\mu} \over d\sigma}=v_a^
{\prime \mu} ={1\over m_a'}\, p_a^{\prime\mu}\, .
\ee
Here $\Gamma^\mu_{\nu\rho}$ denotes the
Christoffel symbol constructed from the metric $\eta_{\mu\nu}+2\,h_{\mu\nu}$. $r_a$ denotes the point where
the trajectory of the
$a$-th outgoing particle intersects the boundary of $\RR$ and $r_a'$ denotes the
point where
the trajectory of the
$a$-th incoming particle intersects the boundary of $\RR$.
$T^h$ is the stress tensor of gravity, as defined below \refb{egrbox}. 
$h_{\mu\nu}$ and hence $e_{\mu\nu}$ is required to satisfy retarded boundary condition.
The starting
solution for the iteration is taken to be
\be\label{e511grpre}
e_{\mu\nu}=0, \quad X_a^\mu(\sigma) = r_a^\mu + v_a^\mu\, \sigma = r_a^\mu + {1\over m_a}\, p_a^\mu\, \sigma ,
\quad X_a^{\prime\mu}(\sigma) = r_a^{\prime\mu} + v_a^{\prime\mu}\, \sigma = r_a^{\prime\mu} + 
{1\over m_a'}\, p_a^{\prime\mu}\, \sigma\, .
\ee

We can give a uniform treatment of the incoming and the outgoing particles by
defining:
\ben \label{edefprime}
&& X^\mu_{a+n}(\sigma) = X_a^{\prime\mu}(-\sigma), \quad m_{a+n}=m_a', \quad
v_{a+n}^\mu = -v_a^{\prime \mu}, \quad r_{a+n}^\mu = r_a^{\prime\mu},\quad
p_{a+n}^\mu = -p_a^{\prime \mu}, \non\\ && \hskip 3in   \hbox{for 
$1\le a\le m$}\, .
\een
In this case we can express \refb{eitergrpre} and \refb{eboundpre} as:
\ben \label{eitergr}
&& T^{\mu\nu}(x) = T^{X\mu\nu} (x)
+ T^{h\mu\nu}(x), \nonumber \\
&& T^{X\mu\nu}(x) \equiv
\sum_{a=1}^{m+n} m_a\, \int_0^\infty d\sigma\, \delta^{(4)}(x - X_a(\sigma)) \, {dX_a^\mu\over d\sigma} \, {dX_a^\nu\over d\sigma}
 \nonumber \\
&& \square e_{\mu\nu} = -8\, \pi\, G\, T_{\mu\nu},  \qquad
{d^2 X_a^\mu\over d\sigma^2} = - \Gamma^\mu_{\nu\rho}(X(\sigma)) \,  {dX_a^\nu\over d\sigma} \, {dX_a^\rho\over d\sigma}\, , \quad \hbox{for $1\le a\le m+n$}\, , \nonumber \\
\een
and
\be \label{ebound}
X_a^\mu(\sigma=0)=r_a^\mu, \quad \lim_{\sigma\to\infty} {dX_a^\mu \over d\sigma}=v_a^\mu ={1\over m_a}\, p_a^\mu,
\quad \hbox{for $1\le a\le m+n$}\, .
\ee
Also the starting solution \refb{e511grpre} for iteration may be written as
\be\label{e511gr}
e_{\mu\nu}=0, \quad X_a^\mu(\sigma) = r_a^\mu + v_a^\mu\, \sigma = r_a^\mu + {1\over m_a}\, p_a^\mu\, \sigma ,
\quad \hbox{for $1\le a\le m+n$}\, .
\ee
From now on we shall follow this convention, with the understanding that the sum over $a$ always runs from 
1 to $(m+n)$ unless stated otherwise.

\subsection{Leading order contribution}

At the leading order in the expansion in powers of $G$, $T^{h\mu\nu}$ vanishes, and we have:
\ben\label{e4.3gr}
\wh T^{\mu\nu}(k) \ = \ \wh T^{X\mu\nu}(k) &=& \int d^4 x \, e^{-ik.x} \,
\sum_{a=1}^{m+n} m_a\, \int_0^\infty d\sigma \, \delta^{(4)}(x - X_a(\sigma)) \, {dX_a^\mu\over d\sigma}
\, {dX_a^\nu\over d\sigma}  \nonumber \\ &=& 
\sum_{a=1}^{m+n} 
m_a\, \int_0^\infty d\sigma \, e^{-ik.X(\sigma)} \, {dX_a^\mu\over d\sigma}
\, {dX_a^\nu\over d\sigma} 
\, , 
\een
where, as mentioned earlier, we have restricted the region of integration over $x$ to
outside the region $\RR$. Using the leading order solution \refb{e511gr} we get
\ben\label{eleadgr}
\wh T^{\mu\nu}(k) &=& \sum_{a=1}^{m+n} m_a\, \int_0^\infty d\sigma \, e^{-ik.(v_a \, \sigma+ r_a)} \, v_a^\mu v_a^\nu
=  \sum_{a=1}^{m+n} m_a\, 
{1\over i(k.v_a-i\eps)} \, e^{-ik.r_a} \, v_a^\mu v_a^\nu
\nonumber \\ 
&=& \sum_{a=1}^{m+n}  
p_a^\mu \, p_a^\nu\, e^{-ik.r_a} \, 
{1\over i(k.p_a-i\eps)}   \,.
\een
The $i\eps$ prescription is obtained by noting that addition of a small negative imaginary part to $k.v_a$
makes the $\sigma$ integrals convergent. Therefore the poles must be
in the upper half $k.v_a$ plane.

Since we are looking for terms that are singular at $\omega\to 0$, i.e.\ $k^\mu\to 0$, we can replace the
$e^{-ik.r_a}$  factors by 1. This gives the leading soft factor associated with the memory effect.

\subsection{First order correction to the gravitational field} \label{s3.3}

We now turn to the next order contribution. We first solve for $e_{\mu\nu}$ satisfying
the third equation in \refb{eitergr} as
\ben\label{e514gr}
&& \wh e_{\mu\nu}(k) = -8\, \pi\, G\, G_r(k) \, \wh T_{\mu\nu}(k)
= -8\, \pi\, G\, \sum_{a=1}^{m+n}  
p_{a\mu} \, p_{a\nu}\, e^{-ik.r_a} \, G_r(k) \, 
{1\over i(k.p_a-i\eps)}, \nonumber \\ &&
G_r(k) \equiv {1\over (k^0+i\eps)^2 -\vec k^2}\, .
\een
One comment is in order here. The expression \refb{e4.3gr} for 
$\wh T^{\mu\nu}(k)$, which we are using in \refb{e514gr}, ignores the contribution from
the region of integration $\RR$. This was justified earlier since we were
computing the singular part of $\wh T_{\mu\nu}$.  However, now we need the contribution to
$\wh e_{\mu\nu}$ from the full $\wh T_{\mu\nu}$ since our goal will be to use this to
compute $T^h_{\mu\nu}$, and also to compute the corrections to the particle trajectories, which, in
turn, give corrections to $T^{X}_{\mu\nu}$. 
Once we compute these, we use \refb{e55gr} to compute $\wt e_{\mu\nu}$. At this stage, we can again
restrict the integration region to outside $\RR$ while taking the Fourier transform
to compute the corrected $\wh T_{\mu\nu}$. To address this issue, we first analyze the possible
correction $\delta T^{X}_{\mu\nu}$ to $T^{X}_{\mu\nu}$ due to gravitational fields generated from inside $\RR$.
Since
in four space-time dimensions the retarded Green's function has support on the future light-cone,
the field sourced by energy momentum tensor inside $\RR$ will have support 
on the future light-cone
emerging from points inside $\RR$. These intersect  the time-like trajectories of the outgoing (or
incoming) particles 
emerging from $\RR$ only within a distance of order $L$ -- the size of $\RR$. 
Therefore $\delta T^X_{\mu\nu}$ is affected only in this region. Since
integration over this region will not produce a singular contribution to $\wh T^X_{\mu\nu}(k)$ 
in the $\omega\to 0$ limit, this effect
may be ignored.
However the gravitational field produced from the sources inside $\RR$ could 
give significant contribution to $T^h$,
since we are not assuming the interactions inside $\RR$ to be weak. 
We take this into account by
regarding
the contribution to $\wh e_{\mu\nu}(k) =-8\,\pi\, G\,G_r(k)
\wh T_{\mu\nu}(k)$ from inside the region $\RR$
as a flux of finite wavelength gravitational waves produced by $T_{\mu\nu}(x)$ inside $\RR$, and 
include this in the sum over $a$. Therefore the outgoing momenta $\{p_a\}$ not
only will include finite mass particles, but also the finite wave-length
`massless gravitons' emitted during the scattering process.

Using \refb{e514gr} we can calculate, at the next order,
\ben\label{eabmugr}
e^{(b)}_{\mu\nu}(x) &=& -8\, \pi\, G\, 
\int {d^4 \ell\over (2\pi)^4} \, e^{i\ell . (x-r_b)} G_r(\ell)  \, p_{b\mu} 
\, p_{b\nu}\, 
{1\over i(\ell.p_b-i\eps)}, \nonumber \\
h^{(b)}_{\mu\nu}(x) &=& -8\, \pi\, G\, 
\int {d^4 \ell\over (2\pi)^4} \, e^{i\ell . (x-r_b)} G_r(\ell)  \, \left\{
p_{b\mu} 
\, p_{b\nu} -{1\over 2} \, p_b^2 \, \eta_{\mu\nu}\right\} \, 
{1\over i(\ell.p_b-i\eps)}\, ,
\een
where $e^{(b)}_{\mu\nu}$ is the gravitational field due to the $b$-th particle.
This gives
\ben\label{eGam}
\Gamma^{(b)\mu}_{\nu\rho}(x) &=&  \eta^{\mu\alpha}
\left\{\p_\nu \, h^{(b)}_{\alpha\rho}+\p_\rho \, h^{(b)}_{\alpha\nu} 
- \p_\alpha \, h^{(b)}_{\nu\rho}\right\}\nonumber \\
&=& 
- 8\, \pi\, G\,  \int{d^4 \ell\over (2\pi)^4} e^{i\ell.(x-r_b)}\, G_r(\ell)  \, {1\over (\ell.p_b-i\eps)}
\bigg[\left\{\ell_\nu p_{b}^\mu p_{b\rho}+\ell_\rho p_{b}^\mu p_{b\nu}
-\ell^\mu p_{b\nu}p_{b\rho}\right\} \nonumber \\ &&
-{1\over 2}\, p_b^2\, \left\{ \ell_\nu \delta^\mu_\rho +  \ell_\rho \delta^\mu_\nu - \ell^\mu \, \eta_{\nu\rho}
\right\}
\bigg]\, .
\een
These results will be used for two purposes.
We shall substitute \refb{eGam} into the last equation in \refb{eitergr} to compute the
correction to the outgoing particle trajectories and hence to $T^X_{\mu\nu}$. We
shall also use \refb{eabmugr} to compute the leading contribution to $T^h_{\mu\nu}$.

Note that $e^{(b)}_{\mu\nu}(x)$ given in \refb{eabmugr} satisfies:
\be \label{ededon}
\p_\mu e^{\mu\nu}= \sum_{b=1}^{m+n} \p_\mu\,  e^{(b)\mu\nu}(x) = -8\, \pi\, G\, \sum_{b=1}^{m+n}\,
\int {d^4 \ell\over (2\pi)^4} \, e^{i\ell . (x-r_b)} G_r(\ell)  \, 
\, p_b^\nu\, .
\ee
As long as we restrict the integration range of $\ell$ to values for which $\ell.(r_c-r_a)$ is small for every
pair $a,c$, we can take $e^{-i\ell. r_b}$ to be approximately independent of $b$, and
the right hand side of \refb{ededon} vanishes due to momentum conservation law
$\sum_{b=1}^{m+n}
p_b^\mu=0$. Therefore $e^{\mu\nu}$ at this order satisfies the
de Donder gauge condition:
\be
\p^\mu e_{\mu\nu}=0\,.
\ee
At the next order there is apparent violation of this condition
due to the $\ell.r_b$ factors coming from
the expansion of the exponential factor. This can be compensated by some boundary terms on
$\p\RR$
coming from integration inside the region $\RR$\cite{1906.08288}, but since these terms will not contribute to
the singular terms that are of interest to us, we shall ignore them.

In the next two subsections we shall compute the correction to $\wh T^X$ and $\wh T^h$ using these results.
It is also possible to argue that in order to calculate the logarithmic terms of interest, we can stop at this order.
The natural dimensionless 
expansion parameter is $G M \omega$ where $M$ denotes the typical energy of the incoming / outgoing
particles. Since the leading term \refb{eleadgr} is of order $1/\omega$,
the subleading corrections that we shall compute will be of order $\omega^0$ multiplied by powers of
$\ln\omega$. Higher order terms will involve higher powers of $\omega$ and will not be needed for our
analysis.

\subsection{Subleading contribution to the matter stress tensor} \label{sgrav.3}

We begin by computing correction to the particle trajectory \refb{e511gr}.
Let $Y_a^\mu$ denote the correction:
\be\label{ethisgr}
X_a^\mu(\sigma) = v_a^\mu\, \sigma +r_a^\mu + Y_a^\mu(\sigma)\, .
\ee
Then $Y_a^\mu$ satisfies the differential equation and boundary conditions:
\be\label{eYeqgr}
{d^2 Y_a^\mu\over d\sigma^2} =-\Gamma^\mu_{\nu\rho}(v_a\, \sigma+r_a)\,  v_a^\nu \,
v_a^\rho, \quad Y_a^\mu\to 0\, \, \hbox{as $\sigma\to  0$} , \quad  \, {d Y_a^\mu\over d\sigma}
\to 0 \, \,
\hbox{as $\sigma\to\infty$}\, ,
\ee
where\footnote{The self-force effects\cite{0306052} will not be important at this order.}
\be \label{e4.11gr}
\Gamma^\mu_{\nu\rho}=\sum_{b=1\atop b\ne a}^{m+n} \Gamma^{(b)\mu}_{\nu\rho} \, ,
\ee
captures the effect of the gravitational field produced by all particles other than $a$. Some of these terms
must vanish, {\it e.g.} the gravitational field produced by an outgoing particle should not affect an incoming
particle. This however will follow automatically from the equations that we shall derive, and need not be 
imposed externally.
Integrating \refb{eYeqgr} we get
\be\label{eYY1}
{dY_a^\mu(\sigma)\over d\sigma} = \int_\sigma^\infty\, d\sigma' \, \Gamma^\mu_{\nu\rho}(v_a\, \sigma'+r_{a}) \, 
v_a^\nu\, 
v_a^\rho\, ,
\ee
and
\be \label{eYY2}
Y_a^\mu(\sigma) =\int_0^\sigma\, d\sigma' \, \int_{\sigma'}^\infty\, d\sigma'' \, \Gamma^\mu_{\nu\rho}(v_a\, \sigma''+r_{a}) \, v_a^\nu\, v_a^\rho \, .
\ee
Substituting \refb{ethisgr}
into \refb{e4.3gr} we get $\wh T^{X\mu\nu}$ to subleading order:
\ben\label{e4.10grpre}
\wh T^{X\mu\nu}(k) &=&  
\sum_{a=1}^{m+n} m_a\, \int_0^\infty d\sigma \, e^{-ik.(v_a\, \sigma+r_a)} \, \left\{1- ik. Y_a(\sigma)\right\}
\left\{v_a^\mu+{dY_a^\mu\over d\sigma}\right\}
\left\{v_a^\nu+{dY_a^\nu\over d\sigma}\right\}\nonumber \\ 
&& \hskip -.8in =
 \sum_{a=1}^{m+n} m_a\, \int_0^\infty d\sigma \, e^{-ik.(v_a\, \sigma+r_a)} \, \left[v_a^\mu v_a^\nu - ik.Y_a(\sigma) \, v_a^\mu
v_a^\nu
+  {dY_a^\mu\over d\sigma}\, v_a^\nu + v_a^\mu \, {dY_a^\nu\over d\sigma}\right]\, .
\een
Using \refb{eYY1}, \refb{eYY2}, we can express this as,
\ben\label{e4.10gr}
\wh T^{X\mu\nu}(k) \hskip-.1in &=& \hskip-.1in
 \sum_{a=1}^{m+n} m_a\, \int_0^\infty d\sigma \, e^{-ik.(v_a\, \sigma+r_a)} \, \Bigg[v_a^\mu v_a^\nu- ik_\rho \,  
\int_0^\sigma\, d\sigma' \, \int_{\sigma'}^\infty\, d\sigma'' \, \Gamma^\rho_{\alpha\beta}(v_a\, \sigma''+r_a) v_a^\alpha
v_a^\beta \, v_a^\nu\, v_a^\mu
\nonumber \\ &&
+ \int_\sigma^\infty\, d\sigma' \, \Gamma^\mu_{\alpha\beta}(v_a\, \sigma'+r_a)\, v_a^\alpha v_a^\beta \, v_a^\nu
+ \int_\sigma^\infty\, d\sigma' \, \Gamma^\nu_{\alpha\beta}(v_a\, \sigma'+r_a)\, v_a^\alpha v_a^\beta \, v_a^\mu \Bigg]\, .
\een
Substituting \refb{eGam} and \refb{e4.11gr} into \refb{e4.10gr}, 
 and
dropping the leading term given in \refb{eleadgr}, 
we get the first order correction to $\wh T^X$: 
\ben\label{ejmpregr}
\Delta\wh T^{X\mu\nu}(k) &=& -8\, \pi\, G\, \, \sum_{a=1}^{m+n} \sum_{b\ne a}  {m_a} \, \int{d^4 \ell\over (2\pi)^4} \,
{1\over \ell.p_b-i\eps} \, G_r(\ell) \, 
\Bigg[\int_0^\infty d\sigma \int_0^\sigma\, d\sigma' \, \int_{\sigma'}^\infty\, d\sigma''  \nonumber \\ && 
\hskip -.2in e^{-i \, k.v_a\sigma}\,
e^{i\ell.v_a\sigma''}
\bigg\{  -i \,   v_a. p_b\, (2k.p_b \, v_a.\ell -k.\ell \, v_a.p_b)
+ {i\over 2}  \, p_b^2 (2k.v_a \, v_a.\ell -k.\ell \, v_a^2)
\bigg\}\, v_a^\nu\, v_a^\mu 
\nonumber \\ && \hskip 1in
 + \int_0^\infty d\sigma  \, \int_{\sigma}^\infty\, d\sigma' \, e^{-i \, k.v_a\sigma}\,
e^{i\ell.v_a\sigma'}
\nonumber \\ && \hskip .3in \bigg\{ 2 \, \ell.v_a \, v_a.p_b \, \Big(v_a^\nu p_b^\mu + v_a^\mu p_b^\nu\Big)- (v_a.p_b)^2 
 \Big(\ell^\mu v_a^\nu +\ell^\nu v_a^\mu\Big) \nonumber \\ && \hskip .5in 
 - 2\,\ell.v_a \, p_b^2 \, v_a^\mu \, v_a^\nu + {1\over 2} \, v_a^2p_b^2 
 \Big(\ell^\mu v_a^\nu +\ell^\nu v_a^\mu\Big) 
 \bigg\}
 \Bigg] e^{-i k.r_a-i\ell. (r_b-r_a)}\, .
\een
After carrying out the integrations over $\sigma,\sigma',\sigma''$, and using $p_a^\mu=m_a v_a^\mu$, we get
\ben\label{ejmgr}
\Delta\wh T^{X\mu\nu}(k) &=& -8\, \pi\, G\, \, \sum_{a=1}^{m+n} \sum_{b\ne a}  \int{d^4 \ell\over (2\pi)^4} \,
{1\over \ell.p_b-i\eps} \, G_r(\ell) \, e^{-i k.r_a-i\ell. (r_b-r_a)} \nonumber \\ &&  \hskip -1in
\Bigg[ \Big(2\, p_a. p_b \, k.p_b \, p_a.\ell -k.\ell \, (p_a.p_b)^2  - p_b^2 \, p_a.k \, p_a.\ell +{1\over 2} k.\ell\, p_a^2
\, p_b^2 \Big)
\, p_a^\nu\, p_a^\mu\, 
{1\over \ell.p_a}\, {1\over k.p_a} \, {1\over (\ell-k).p_a}
\nonumber \\ && \hskip -1in
 - \bigg\{ 2 \, p_a. p_b\, \ell.p_a \, \Big(p_a^\nu p_b^\mu + p_a^\mu p_b^\nu\Big)- (p_a.p_b)^2
 \Big(\ell^\mu p_a^\nu +\ell^\nu p_a^\mu\Big) -2\, p_b^2 \, \ell.p_a \, p_a^\mu \, p_a^\nu \nonumber \\ && \hskip 1in 
 + {1\over 2} \, p_a^2 \, p_b^2 \, 
 \Big(\ell^\mu p_a^\nu +\ell^\nu p_a^\mu\Big) 
 \bigg\} \times \, 
{1\over \ell.p_a}\, {1\over (\ell-k).p_a}
 \Bigg]\, .
\een
For $|r_a^\mu- r_b^\mu|\sim L$, the ultraviolet divergence in the integration over $\ell$ is cut-off at $L^{-1}$
due to the oscillatory
phase factor $e^{-i\ell. (r_b-r_a)}$.

In order to evaluate the integral, we need to determine the $i\eps$ prescription for the poles in 
\refb{ejmgr}.
The $i\eps$ prescription for the $1/\ell.p_b$ term has already been determined before.
Similarly, since the $1/\ell.p_a$ factor comes from an integral in
\refb{ejmpregr} of
the form $\int_{\sigma'}^\infty d\sigma''\, e^{i\ell.v_a\sigma''}$ or $\int_\sigma^\infty d\sigma'\, e^{i\ell.v_a\sigma'}$, the $i\eps$ prescription will be to replace 
$1/\ell.p_a$ by $1/(\ell.p_a+i\eps)$. The $1/k.p_a$ factor comes from 
an integral of
the form $\int_0^\infty d\sigma\, e^{-ik.v_a\sigma}$, and 
the $i\eps$ prescription will be to replace 
$1/k.p_a$ by $1/(k.p_a-i\eps)$. Finally, the $1/(\ell-k).p_a$ factor in \refb{ejmgr} arises from an integral of
the form $\int_0^\infty d\sigma\, e^{i(\ell-k).v_a\sigma}$, and the correct $i\eps$ prescription for this term is
$1/((\ell-k).p_a+i\eps)$. Therefore, \refb{ejmgr} should be written as
\ben\label{ejmfingr}
\Delta\wh T^{X\mu\nu}(k) &=& -8\, \pi\, G\, \, \sum_{a=1}^{m+n} 
\sum_{b\ne a}  \, \int{d^4 \ell\over (2\pi)^4} \,
{1\over \ell.p_b-i\eps} \, G_r(\ell) \, e^{-i k.r_a-i\ell. (r_b-r_a)} \nonumber \\ &&  \hskip -1in
\Bigg[ \Big(2\, p_a. p_b \, k.p_b \, p_a.\ell -k.\ell \, (p_a.p_b)^2  - p_b^2 \, p_a.k \, p_a.\ell +{1\over 2} k.\ell\, p_a^2
\, p_b^2 \Big)
\, p_a^\nu\, p_a^\mu\nonumber \\ && \hskip 1in \times \, 
{1\over \ell.p_a+i\eps}\, {1\over k.p_a-i\eps} \, {1\over (\ell-k).p_a+i\eps}
\nonumber \\ && \hskip -1.1in
 - \bigg\{ 2 \, p_a. p_b\, \ell.p_a \, \Big(p_a^\nu p_b^\mu + p_a^\mu p_b^\nu\Big)- (p_a.p_b)^2
 \Big(\ell^\mu p_a^\nu +\ell^\nu p_a^\mu\Big) -2\, p_b^2 \, \ell.p_a \, p_a^\mu \, p_a^\nu + {1\over 2} \, p_a^2 \, p_b^2 \, 
 \Big(\ell^\mu p_a^\nu +\ell^\nu p_a^\mu\Big) 
 \bigg\}\nonumber \\ && \hskip 1in \times \, 
{1\over \ell.p_a+i\eps}\, {1\over (\ell-k).p_a+i\eps}
 \Bigg]\, .
\een

Since we are interested in the singular term proportional to $\ln\omega$, we can simplify the analysis of the integral
as follows. Since the expression is Lorentz covariant, we could evaluate it in a special frame in which $p_a$ and $p_b$
have only third component of spatial momenta. Let us denote by $\ell_\perp=(\ell^1,\ell^2)$ the transverse component of 
$\ell$. Now  since
$p_a.\ell$ and $p_b.\ell$ are both linear in $\ell^0$ and $\ell^3$, we can use $p_a.\ell$ and $p_b.\ell$ as independent variables
instead of $\ell^0$ and $\ell^3$. Then, if we ignore the poles of $G_r(\ell)$, we see that 
we have one pole in the $p_b.\ell$ plane and two poles on the same side of the real axis in the
$p_a.\ell$ plane. Therefore 
we can deform the $p_a.\ell$ and $p_b.\ell$ integration contours 
away from the poles. 
However due to the presence of the $G_r(\ell)$ factor there are also poles at
\be\label{e3.34}
(\ell^0+i\eps +\ell^3) (\ell^0+i\eps -\ell^3) = \ell_\perp^2\, .
\ee
Therefore, for small but fixed $\ell_\perp$, if we  deform the $(\ell^0 +\ell^3)$ contour to a distance 
of order $|\ell_\perp|$ away 
from the origin, 
a pole will
approach the origin within a distance of order $|\,\ell_\perp|$
in the complex $(\ell^0-\ell^3)$ plane. The 
integration contour could then be pinched between this pole  and one of the poles of the 
$(\ell.p_a+i\eps)^{-1} \{(\ell-k).p_a+i\eps\}^{-1}(\ell.p_b-i\eps)^{-1}$ factor.
However it is clear that in the complex
$\ell^0$ and complex $\ell^3$ plane, the integration contour can be deformed 
so that the contour maintains a minimum distance of order $|\ell_\perp|$ from all the poles, which themselves
are situated within a distance of order $|\ell_\perp|$ of the origin.
This shows that while estimating the integrand to examine possible sources of 
singularity of the integral, we can take all the components
of $\ell$ to be of order $\ell_\perp$ and need not worry about the regions where one or more components are
smaller than the others. Since for $\ell^\mu\sim \ell_\perp$ the 
integration measure gives a factor of $|\ell_\perp|^4$, we see that in order to get a logarithmic correction, the integrand
must be of order $|\ell_\perp|^{-4}$.

We now note that in both terms the integrand of \refb{ejmfingr} grow 
as $|\ell_\perp|^{-3}$ for $|\ell^\mu|\sim |\ell_\perp|
<<\omega$ and
therefore there are no logarithmic corrections from this region. For 
$|r_b^\mu-r_a^\mu|^{-1}\sim L^{-1}>>|\ell^\mu|>>\omega$ 
we can replace $(k-\ell).p_a$ by $-\ell.p_a$, and drop the $e^{-i\ell. (r_b-r_a)}$ factor. In this case
the integrand is of order $|\ell_\perp|^{-4}$ and the integral could have logarithmic contributions. To compute this,
we note that in this region of integration
the integral may be approximated as
\ben\label{ejmapproxgr}
\Delta\wh T^{X\mu\nu}(k) &\simeq& -8\, \pi\, G\, \, \sum_a \sum_{b\ne a} \int{d^4 \ell\over (2\pi)^4} \,
{1\over \ell.p_b-i\eps} \, G_r(\ell) \, e^{-i k.r_a} \nonumber \\ &&  \hskip -1in
\Bigg[ \Big(2\, p_a. p_b \, k.p_b \, p_a.\ell -k.\ell \, (p_a.p_b)^2  - p_b^2 \, p_a.k \, p_a.\ell +{1\over 2} k.\ell\, p_a^2
\, p_b^2 \Big)
\, p_a^\nu\, p_a^\mu\, 
{1\over (\ell.p_a+i\eps)^2}\, {1\over k.p_a-i\eps} 
\nonumber \\ && \hskip -1in
 - \bigg\{ 2 \, p_a. p_b\, \ell.p_a \, \Big(p_a^\nu p_b^\mu + p_a^\mu p_b^\nu\Big)- (p_a.p_b)^2
 \Big(\ell^\mu p_a^\nu +\ell^\nu p_a^\mu\Big) -2\, p_b^2 \, \ell.p_a \, p_a^\mu \, p_a^\nu \nonumber \\ && \hskip 1in 
 + {1\over 2} \, p_a^2 \, p_b^2 \, 
 \Big(\ell^\mu p_a^\nu +\ell^\nu p_a^\mu\Big) 
 \bigg\} \times \, 
{1\over (\ell.p_a+i\eps)^2}
 \Bigg]\, .
\een
It will be understood that in this integral the integration over $\vec \ell_\perp$ is restricted to the region
$L^{-1}>> |\vec \ell_\perp|>>\omega$. Since for fixed $\vec \ell_\perp$, the integration over 
$\ell^0$ and $\ell^3$ are finite, we do not need to impose separate cut-off on the $\ell^0$ and $\ell^3$
integrals. 
All the terms in \refb{ejmapproxgr}  can be expressed in terms of the basic integral
\be \label{e3.29}
\int{d^4 \ell\over (2\pi)^4} \, {1\over \ell.p_b-i\eps} \, G_r(\ell) \, {1\over 
(\ell.p_a+i\eps)^2}\, \ell_\alpha = -{\p\over \p p_a^\alpha} \, J_{ab}\, ,
\ee
where
\be\label{e3.29J}
J_{ab}= 
\int{d^4 \ell\over (2\pi)^4} \, {1\over \ell.p_b-i\eps} \, G_r(\ell) \, {1\over 
\ell.p_a+i\eps}\, .
\ee
It has been shown in appendix \ref{sintegral} that $J_{ab}$ vanishes when $a$ represents an incoming particle 
and $b$ represents an
outgoing particle or vice versa. On the other hand when $a$ and $b$ are both ingoing particles or both outgoing
particles, we have, from \refb{ejabfinfin},
\be \label{e3.29JJ}
J_{ab}={1\over 4\pi} \, \ln\{L(\omega+i\eps\eta_a)\}\,  {1\over \sqrt{(p_a.p_b)^2 
-p_a^2 p_b^2}}\, ,
\ee
where $\eta_a$ is a number that takes value $1$ for outgoing particles ($1\le a\le n$)
and $-1$ for incoming particles 
($n+1\le a\le m+n$).\footnote{This is opposite to the convention used in
\cite{1808.03288}.}
Using \refb{e3.29JJ} we can express \refb{e3.29} as
\ben \label{e3.37}
&& \int{d^4 \ell\over (2\pi)^4} \, {1\over \ell.p_b-i\eps} \, G_r(\ell) \, {1\over 
(\ell.p_a+i\eps)^2}\, \ell_\alpha \\ &=&
-{1\over 4\pi} \, \ln\{L(\omega+i\eps\eta_a)\}\, {\p\over \p p_a^\alpha}\, {1\over \sqrt{(p_a.p_b)^2 
-p_a^2 p_b^2}}=-{1\over 4\pi} \, \ln\{L(\omega+i\eps\eta_a)\}\,  {p_b^2\,
p_{a\alpha} 
- p_a.p_b \, p_{b\alpha}\over 
\{(p_a.p_b)^2 
-p_a^2 p_b^2\}^{3/2}
}\, .\non
\een
We now use this to evaluate the right hand side of \refb{ejmapproxgr}.
We can also replace $e^{ik.r_a}$ by 1 since the difference is higher order in the small $\omega$ limit.
This gives
\ben\label{efingratot}
\Delta\wh T^{X\mu\nu}(k) &=& 2\, G \, \sum_{a=1}^{m+n} {\sum_{b\ne a\atop \eta_a\eta_b=1}} \,
{\ln\{L(\omega+i\eps\eta_a)\}\over 
\{(p_a.p_b)^2 
-p_a^2 p_b^2\}^{3/2}}\Bigg[{k.p_b\over k.p_a}\, p_a^\mu p_a^\nu \, p_a.p_b
\left\{{3\over 2} p_a^2 p_b^2 - (p_a.p_b)^2\right\}  \nonumber \\ &&
\hskip .1in
+ {1\over 2} p_a^\mu p_a^\nu \, p_a^2 \, (p_b^2)^2 - \{ p_a^\mu p_b^\nu + p_a^\nu p_b^\mu\} \,  p_a.p_b\,
\left\{{3\over 2} p_a^2 p_b^2 - (p_a.p_b)^2\right\} 
\Bigg]\, .
\een
The constraint $\eta_a\eta_b=1$ 
means that the sum over $b$ runs over incoming particles if $a$ represents an incoming particle
and runs over outgoing particles if $a$ represents an outgoing particle.

\subsection{Subleading contribution from the gravitational stress tensor} \label{sgrsubgr}

Let us now turn to the computation of $T^{h\mu\nu}$ defined via \refb{egrbox}.
A detailed calculation shows that to quadratic order in $h_{\mu\nu}$, it has the form:
\ben \label{equadth}
8\pi\, G\, T^{h\mu\nu}
&=&\ -2\Big[\f{1}{2}\p^{\mu}h_{\alpha\beta}\p^{\nu}h^{\alpha\beta}+h^{\alpha\beta}\p^{\mu}\p^{\nu}h_{\alpha\beta}-h^{\alpha\beta}\p^{\nu}\p_{\beta}h_{\alpha}^{~\mu}-h^{\alpha\beta}\p^{\mu}\p_{\beta}h_{\alpha}^{~\nu}
+h^{\alpha\beta}\p_{\alpha}\p_{\beta}h^{\mu\nu}\non\\
&&\hskip -.5in +\,
\p^{\beta}h^{\nu\alpha}\p_{\beta}h_{\alpha}^{~\mu}-\p^{\beta}h^{\alpha\nu}\p_{\alpha}h_{\beta}^{~\mu}\Big]
+  h^{\mu\nu}\p_{\rho}\p^{\rho}h - 2\, h^{\mu}_{~\rho} \p^\sigma\p_\sigma h^{\nu\rho}
- 2\, h^{\nu}_{~\rho} \p^\sigma\p_\sigma h^{\mu\rho}  
\non\\
&& \hskip -.5in +\, \eta^{\mu\nu}\Big[ \f{3}{2}\p^{\rho}h_{\alpha\beta}\p_{\rho}h^{\alpha\beta}+2h^{\alpha\beta}\p^{\rho}\p_{\rho}h_{\alpha\beta}-\p^{\beta}h^{\alpha\rho}\p_{\alpha}h_{\beta\rho}\Big]
+h\Big[ \p^{\rho}\p_{\rho}h^{\mu\nu}-\f{1}{2}\p^{\rho}\p_{\rho}h\, \eta^{\mu\nu}\Big]\, , \non\\
\een
where we have used de Donder gauge condition to simplify the expression. To the order that we shall be working,
this is allowed due to 
the observation made below \refb{ededon}. This expression differs from some of the more standard expressions
given {\it e.g.} in \cite{weinbergbook}, since we have defined $8\, \pi\, G\, T^{h\mu\nu}$ 
as the
collection of the quadratic terms in the expansion of
$-\sqrt{-\det g}\, (R^{\mu\nu} - g^{\mu\nu} R/2)$. As already mentioned,
all indices in \refb{equadth} are raised and lowered using the flat metric $\eta$.

We shall manipulate \refb{equadth} 
by expressing $h_{\alpha\beta}$ in the momentum space as given in
\refb{eabmugr}.  This gives a general expression of the form:
\ben \label{etmninitial}
\widehat{T}^{h\mu\nu}(k)\ &=&\ -8\,\pi\, G\, 
\sum_{a,b}e^{-ik.r_{a}}\int \f{d^{4}\ell}{(2\pi)^{4}}\ e^{i\ell.(r_{a}-r_{b})}\ G_{r}(k-\ell)G_{r}(\ell)\ \f{1}{p_{b}.\ell-i\epsilon}\ \f{1}{p_{a}.(k-\ell)-i\epsilon}\non\\
&&\times \Big{\lbrace} p_{b\alpha}p_{b\beta}-\f{1}{2}p_{b}^{2}\eta_{\alpha\beta}\Big{\rbrace}\ \mathcal{F}^{\mu\nu, \alpha\beta , \rho\sigma}(k,\ell)\ \Big{\lbrace}p_{a\rho}p_{a\sigma}-\f{1}{2}p_{a}^{2}\eta_{\rho\sigma}\Big{\rbrace}\, ,\een
where,
\ben \label{edefffmn}
&&\mathcal{F}^{\mu\nu , \alpha\beta , \rho\sigma }(k,\ell)\non\\
 &=&\ 2\Big[\f{1}{2}\ell^{\mu}(k-\ell)^{\nu}\eta^{\rho\alpha}\eta^{\sigma\beta}+(k-\ell)^{\mu}(k-\ell)^{\nu}\eta^{\rho\alpha}\eta^{\sigma\beta}-(k-\ell)^{\nu}(k-\ell)^{\beta}\eta^{\rho\alpha}\eta^{\sigma\mu}
\non\\
&& 
 -(k-\ell)^{\mu}(k-\ell)^{\beta}\eta^{\rho\alpha}\eta^{\sigma\nu}+(k-\ell)^{\alpha}(k-\ell)^{\beta}\eta^{\rho\mu}\eta^{\sigma\nu}+(k-\ell).\ell\, \eta^{\beta\nu}\eta^{\alpha\rho}\eta^{\sigma\mu}
 \non\\
&&-\ell^{\rho}(k-\ell)^{\alpha}\eta^{\beta\nu}\eta^{\sigma\mu}-\f{1}{2}(k-\ell)^{2}\eta^{\alpha\mu}\eta^{\beta\nu}\eta^{\rho\sigma}\ +\eta^{\alpha\mu}\eta^{\beta\rho}\eta^{\nu\sigma}(k-\ell)^{2}\ +\ \eta^{\alpha\nu}\eta^{\beta\rho}\eta^{\mu\sigma}(k-\ell)^{2}
\Big]\non\\
&&\ -\eta^{\mu\nu}\Big[\f{3}{2}(k-\ell).\ell\, 
\eta^{\rho\alpha}\eta^{\sigma\beta}+2(k-\ell)^{2}\eta^{\rho\alpha}\eta^{\sigma\beta}-\ell^{\sigma}(k-\ell)^{\alpha}\eta^{\rho\beta}\Big]\non\\
&&\ -\eta^{\alpha\beta}(k-\ell)^{2}\eta^{\rho\mu}\eta^{\sigma\nu}+\f{1}{2}\eta^{\alpha\beta}(k-\ell)^{2}\eta^{\rho\sigma}\eta^{\mu\nu}\, .
\een
In the $\ell^\mu\to 0$ limit the integrand diverges as $|\ell^\mu|^{-4}$ and therefore the integral has logarithmic infrared
divergence.
As discussed below \refb{efinFin}, 
the lower cut-off on the $\ell^\mu$ integral in this case is provided by $R^{-1}$
where $R$ is the distance to the detector (measured in flat metric). 
Formally, this can be achieved by adding to $k^0=\omega$ a
small imaginary part proportional to $R^{-1}$. 

Now in \refb{etmninitial} 
the $G_r(\ell) \, G_r(k-\ell)$ factor takes the form:
\be\label{egrdecomp}
G_r(\ell) \, G_r(k-\ell) = {1\over (\ell^0+i\eps)^2 - \vec \ell^2}\, {1\over (k^0-\ell^0+i\eps)^2 - (\vec k-\vec \ell)^2}\, .
\ee
As a result the poles of the two denominators in the $\ell^0$ plane are on the opposite sides of the integration
contour lying along the real axis. We shall express this as:
\be \label{e345gr}
G_r(\ell)^* G_r(k-\ell) - 2\, i\, \pi\, \delta(\ell^2) \left\{H(\ell^0) - H(-\ell^0)\right\} G_r(k-\ell)\, ,
\ee
where $H$ is the Heaviside step function.
In this case 
in the first term the poles in both factors are in the upper half $\ell^0$ plane. This allows us to deform
the $\ell^0$ contour away from these poles till we hit the zeros of the other denominators. 
In particular, following the argument given in the paragraph containing \refb{e3.34}, one can argue that for
$|\ell_\perp|>\omega$, we can deform the contours such that it maintains a distance of order $\ell_\perp$ from 
all the poles. 
We shall show in
appendix \ref{sreal} that the contribution from the terms proportional to $\delta(\ell^2)$
in \refb{e345gr} represents the 
contribution to $\wh T^{\mu\nu}$ from the gravitational radiation (real gravitons) 
emitted during the scattering.
Since this contribution has already been
included by including the radiation contribution in the sum over $a$, we shall not discuss
them any further in this section.

We shall now analyze possible logarithmic contribution to \refb{etmninitial} with $G_r(\ell)$ replaced
by $G_r(\ell)^*$. These can arise from three regions:
$R^{-1}<<|k^\mu-\ell^\mu|<<\omega$, $R^{-1}<<|\ell^\mu|<<\omega$ and 
$L^{-1}>>|\ell^\mu|>>\omega$. Since each term
in \refb{edefffmn} has at least one power of $(k-\ell)$, one finds by simple power counting that there is no
logarithmic contribution from the region $R^{-1}<<|k^\mu-\ell^\mu|<<\omega$. For $R^{-1}<<|\ell^\mu|<<\omega$ the
integrand has four powers of $\ell$ in the denominator and could give logarithmic contribution. In this region
we can replace the integrand by its leading term in the $\ell\to 0$ limit. 
In particular $\FF^{\mu\nu}_{\alpha\beta;\rho\sigma}(k,\ell)$ may be approximated as
\be \label{efapprox}
\FF^{\mu\nu}_{\alpha\beta;\rho\sigma}(k,\ell)\simeq 2\, k^\mu \, k^\nu \,  
\eta_{\alpha\rho}\, \eta_{\beta\sigma}
- 2\, k^\nu \, k_\beta\, \eta_{\rho\alpha}\, \delta^\mu_\sigma
- 2\, k^\mu \, k_\beta\, \eta_{\rho\alpha}\,
\delta^\nu_\sigma + 2 k_\alpha \, k_\beta \, \delta^\mu_\rho \delta^\nu_\sigma\, ,
\ee
where we have used $k^\rho k_\rho=0$. A further simplification is possible by noting that eventually we shall
use the $\wh T^{h\mu\nu}(k)$ 
computed from \refb{etmninitial} to calculate its contribution to 
subleading correction to asymptotic  $\wt e_{\mu\nu}$ via \refb{e55gr}. Since $\wt e_{\mu\nu}$ is determined
only up to a gauge transformation
\be
\wt e_{\mu\nu}\to \wt e_{\mu\nu} + k_\mu \xi_\nu + k_\nu \xi_\mu - k.\xi \, \eta_{\mu\nu}\, ,
\ee
for any vector $\xi$, addition of a similar term to $\wh T^{h}_{\mu\nu}$ and hence to 
$\FF^{\mu\nu}_{\alpha\beta;\rho\sigma}(k,\ell)$ 
will not have any effect of $\wt e_{\mu\nu}$. Using this we
can simplify \refb{efapprox} to:
\be\label{efapproxa}
\FF^{\mu\nu}_{\alpha\beta;\rho\sigma}(k,\ell)\simeq - 2\, k_\sigma \, k_\beta\, \eta_{\rho\alpha}\, \eta^{\mu\nu}
+ 2\, k_\alpha \, k_\beta \, \delta^\mu_\rho \delta^\nu_\sigma\, .
\ee
We can also make the approximations:
\ben\label{e3.51}
&& {1\over p_a.(k-\ell)-i\eps}\simeq {1\over p_a.k-i\eps}, \quad  G_r(\ell)^*={1\over (\ell^0-i\eps)^2 
- \vec\ell^2}, \non\\ && G_r(k-\ell) = {1\over (k^0-\ell^0+i\eps)^2 - (\vec k-\vec\ell)^2} \simeq
{1\over 2(k.\ell + i\eps\omega)}\simeq {1\over 2(k.\ell + i\eps)}\, .
\een
Substituting these into \refb{etmninitial}, 
we get the logarithmic contribution from the $R^{-1}<<|\ell^\mu|<<\omega$ region, 
denoted by
$\wh T^{(1)\mu\nu}(k)$:
\ben\label{et1mn4}
\wh T^{(1)\mu\nu}(k) &=& 8\pi G\,
\sum_{a,b=1}^{m+n} \, {1\over p_a.k-i\eps} \, \int {d^4\ell\over (2\pi)^4} \, {1\over k.\ell + i\eps}\, 
{1\over p_b.\ell-i\eps}\, {1\over (\ell^0-i\eps)^2 
- \vec\ell^2}\non\\ &&
\left\{ p_a.p_b \, k.p_a\, k.p_b\, \eta^{\mu\nu} - {1\over 2} \, p_b^2 \,  (k.p_a)^2\, \eta^{\mu\nu}
- (k.p_b)^2 \, p_a^\mu \, p_a^\nu\right\}\, .
\een
This integral, called $K_b'$ in \refb{eKpab}, has been
evaluated in \refb{eKpabfin}, and gives:
\be \label{eKpabfinpre}
 \int {d^4\ell\over (2\pi)^4} \, {1\over k.\ell + i\eps}\, 
{1\over p_b.\ell-i\eps}\, {1\over (\ell^0-i\eps)^2 
- \vec\ell^2}= {1\over 4\pi}\, \delta_{\eta_b, 1}\, \ln\{(\omega+i\eps)R\} \, {1\over k.p_b}\, .
\ee
Using this, we get:
\ben
\wh T^{(1)\mu\nu}(k) &=& 2\, G\, \ln\{(\omega+i\eps) R\}\, 
\sum_{a=1}^{m+n} \, \sum_{b=1}^n\, {1\over p_a.k-i\eps} \, {1\over p_b.k-i\eps}\, \non\\&&
\left\{ p_a.p_b \, k.p_a\, k.p_b\, \eta^{\mu\nu} - {1\over 2} \, p_b^2 \,  (k.p_a)^2\, \eta^{\mu\nu}
- (k.p_b)^2 \, p_a^\mu \, p_a^\nu\right\}\, .
\een
The terms proportional to $p_a.k\, p_b.k$ and $(k.p_a)^2$ 
inside the curly bracket cancel the denominator factor of $k.p_a$,
and the result vanishes by momentum conservation after summing over $a$. Therefore we have
\be \label{et1mnfingr}
\wh T^{(1)\mu\nu}(k) = -2\, G\, \ln\{(\omega+i\eps) R\}\, 
\sum_{a=1}^{m+n} \, \sum_{b=1}^n\, {p_b.k\over p_a.k-i\eps} \, p_a^\mu p_a^\nu\, .
\ee

Next we turn to the contribution from the region $L^{-1}>>|\ell^\mu|>>\omega$.
Simple power counting shows that the integrand goes as $|\ell|^{-4}$ in this
region. Therefore, in order to extract the logarithmic term, we need to keep only the
leading term in the integrand for large $\ell^\mu$. In particular, 
in the expression for $\mathcal{F}^{\mu\nu, \alpha\beta , \rho\sigma}(k,\ell)$,
we need to keep only quadratic terms in $\ell$.
Therefore we need to evaluate the integral:
\be \label{edefiab}
\II_{ab}^{\alpha\beta} \equiv \int \f{d^{4}\ell}{(2\pi)^{4}}\ e^{i\ell.(r_{a}-r_{b})}\ 
G_{r}(k-\ell)G_{r}(\ell)^*\ \f{1}{p_{b}.\ell-i\epsilon}\ \f{1}{p_{a}.(\ell-k)+i\epsilon} 
\ell^\alpha \, \ell^\beta\, .
\ee
Using $L^{-1}>>|\ell^\mu|>>\omega$, we can further 
approximate \refb{edefiab} by
\ben \label{edefiabap}
\II_{ab}^{\alpha\beta} &\simeq& \int \f{d^{4}\ell}{(2\pi)^{4}}\ 
{1 \over \{(\ell^0-i\eps)^2 -\vec \ell^2\}^2} \ \f{1}{p_{b}.\ell-i\epsilon}\ \f{1}{p_{a}.\ell+i\epsilon} \,
\ell^\alpha \ell^\beta \non\\ &=&  {1\over 2}\, 
\int \f{d^{4}\ell}{(2\pi)^{4}}\ {\p\over \p\ell_\alpha}\left\{
{1 \over (\ell^0-i\eps)^2 -\vec \ell^2}\right\} \ \f{1}{p_{b}.\ell-i\epsilon}\ \f{1}{p_{a}.\ell+i\epsilon} \,
\ell^\beta \non\\ &=&
-{1\over 2}\, 
\int \f{d^{4}\ell}{(2\pi)^{4}}\ 
{1 \over (\ell^0-i\eps)^2 -\vec \ell^2} \ \f{1}{p_{b}.\ell-i\epsilon}\ \f{1}{p_{a}.\ell+i\epsilon} \,
\left\{ \eta^{\alpha\beta} - {p_a^\alpha\ell^\beta\over p_{a}.\ell+i\epsilon}-
{p_b^\alpha\ell^\beta\over p_{b}.\ell-i\epsilon}\right\} \non\\ &=& 
-{1\over 2}
\left\{ \eta^{\alpha\beta} + p_a^\alpha {\p\over p_{a\beta}} + p_b^\alpha {\p\over \p p_{b\beta}}\right\}
\int \f{d^{4}\ell}{(2\pi)^{4}}\ 
{1 \over (\ell^0-i\eps)^2 -\vec \ell^2} \ \f{1}{p_{a}.\ell+i\epsilon}\ \f{1}{p_{b}.\ell-i\epsilon}
\, .
\een
In the third step we have carried out an integration by parts.\footnote{This can be justified as follows.
First, following arguments similar to the one given below \refb{ejmfingr}, we can consider the integration region to be
$\omega<< |\ell_\perp|<<L^{-1}$, without any restriction on $\ell^0$ and $\ell^3$. Integration by parts will then
give boundary contributions from $|\ell_\perp|=\omega$ and $|\ell_\perp|=L^{-1}$. These involve angular integration 
and do not generate any logarithmic terms. 
}
We can now evaluate the integral using the
result for $J_{ab}$ given in \refb{e3.29Japp}. The result vanishes when $a$ represents an ingoing particle and
$b$ represents an outgoing particle or vice versa. When both particles are ingoing or both particles are outgoing,
the result is given in \refb{ejabfinfin}. This gives
\ben\label{eiabfin}
I_{ab}^{\alpha\beta} &\simeq& -{1\over 8\pi} \,  \ln\{L\, (\omega+i\eps\eta_a)\}\,  
\left\{ \eta^{\alpha\beta} + p_a^\alpha {\p\over p_{a\beta}} + p_b^\alpha {\p\over \p p_{b\beta}}\right\}
{1\over \sqrt{(p_a.p_b)^2 
-p_a^2 p_b^2}}\non\\ &=&
-{1\over 8\pi} \,  \ln\{L\, (\omega+i\eps\eta_a)\}\,{1\over \left\{(p_a.p_b)^2 
-p_a^2 p_b^2\right\}^{3/2}} \non\\ &&
\times\,  \left[\eta^{\alpha\beta} \{ (p_a.p_b)^2 - p_a^2 p_b^2\} + p_a^2 \, p_b^\alpha \, p_b^\beta + p_b^2 \,
p_a^\alpha \, p_a^\beta
- p_a. p_b \, (p_a^\alpha \, p_b^\beta + p_a^\beta \, p_b^\alpha)
\right]\, . 
\een

Note that the result diverges for $a=b$. This can be traced to the fact that if we replace the $p_{a}.(\ell-k)+i\epsilon$ factor
in the denominator of \refb{edefiab} by $p_a.\ell+i\eps$ from the beginning, then for $b=a$  the contour is pinched
by the poles from both sides with separation of order $\eps$, and we shall get a divergence in the
$\eps\to 0$ limit. This shows that for $a=b$ we have
to be more careful in evaluating the integral. We proceed by working with \refb{edefiab} without making any approximation
at the beginning. If we work in the rest frame of $p_a$, then we can evaluate the $\ell^0$ integral by closing the contour
in the lower half plane, picking the residue at $\ell^0=0$  for outgoing $p_a=p_b$ and at $\ell^0=k^0$ for incoming
$p_a=p_b$. Let us for definiteness consider the case where the particle is outgoing,
so that we pick up the residue
from the pole at $\ell^0=0$. This reduces the integral to
\be
I_{aa}^{\alpha\beta} = -2\, \pi\, i \, {1\over p_a^0}\, {1\over p_a.k -i\eps}\, \int \f{d^{3}\ell}{(2\pi)^{4}}\ 
{1\over \vec\ell^2} \, {1\over \vec\ell^2 - 2\vec k.\vec\ell -i\eps} \, \ell^\alpha\, \ell^\beta\, .
\ee
Since this is potentially linearly divergent from the region of large $|\vec \ell|$, we expand the integrand in
power series expansion in inverse powers of $\ell$, keeping up to the first subleading term:
\be
I_{aa}^{\alpha\beta} \simeq -2\, \pi\, i \, {1\over p_a^0}\, {1\over p_a.k -i\eps}\, \int \f{d^{3}\ell}{(2\pi)^{4}}\ 
{1\over \vec\ell^2} \, \left[{1\over \vec\ell^2} +{ 2\vec k.\vec\ell \over (\vec\ell^2)^2}\right] \, \ell^\alpha \,
\ell^\beta\, .
\ee
The leading linearly divergent term, where we pick the $\ell^\alpha\ell^\beta$ 
term from the numerator, represents the usual
infinite self energy of a classical point particle, and is regulated by the intrinsic size of the particle. 
In any case, this does not lead to any logarithmic terms. The potentially logarithmically divergent subleading
contribution actually vanishes by $\vec\ell\to -\vec\ell$ symmetry since it has to be evaluated at $\ell^0=0$. Therefore
we conclude that $I_{aa}^{\alpha\beta}$ does not have any logarithmic correction. A similar analysis can be carried
out for the incoming particles, leading to the same conclusion.

Substituting \refb{eiabfin} into \refb{etmninitial} for $a\ne b$,  we get
the logarithmic contribution to $\wh T^{h\mu\nu}(k)$ from the region 
$\omega<<|\ell^\mu|<<L^{-1}$, which we shall denote by $\wh T^{(2)\mu\nu}(k)$:
\ben\label{efingrhh}
\wh T^{(2)\mu\nu}(k)
&=& G \, \sum_{a=1}^{m+n}  \ln \{L(\omega+i\eps\eta_a)\}\,\sum_{b=1\atop b\ne a, \eta_a\eta_b=1}^{m+n} 
{1\over 
\{(p_a.p_b)^2 
-p_a^2 p_b^2\}^{3/2}}\Bigg[- p_b^\mu p_b^\nu \, (p_a^2)^2 \, (p_b^2) \non\\ &&
\hskip 1in + \{ p_a^\mu p_b^\nu + p_a^\nu p_b^\mu\} \, 
p_a.p_b\, \left\{{3\over 2} p_a^2 p_b^2 - (p_a.p_b)^2\right\} 
\Bigg]\, .
\een

\subsection{Gravitational wave-form at early and late time }

Adding \refb{eleadgr}, \refb{efingratot}, \refb{et1mnfingr} and 
\refb{efingrhh} we get the net logarithmic 
contribution to $\wh T^{\mu\nu}(k)$ to the subleading order in the
small $\omega$ expansion:\footnote{The quantum computation of \cite{1808.03288} 
gave rise to additional terms in the soft factor, but
they do not seem to play any role in the classical gravitational wave-form found here.}
\ben\label{efingrabs}
&& \wh T^{\mu\nu}(k) \non\\
&=& \sum_{a=1}^{n}  
p_a^\mu \, p_a^\nu\, 
{1\over i(k.p_a-i\eps)} - \sum_{a=1}^{m}  
p_a^{\prime\mu} \, p_a^{\prime\nu}\, 
{1\over i(k.p_a'+i\eps)}  \non\\ &+&
2\, G \, \ln\{L(\omega+i\eps)\}\,\sum_{a=1}^n \sum_{b=1\atop b\ne a}^n 
{ p_a.p_b\over 
\{(p_a.p_b)^2 
-p_a^2 p_b^2\}^{3/2}} \, 
\left\{{3\over 2} p_a^2 p_b^2 - (p_a.p_b)^2\right\} \, {k_\rho p_a^\mu \over k.p_a}\,
 (p_b^\rho p_a^\nu - p_b^\nu p_a^\rho) 
\nonumber \\ &&\hskip -.3in +\
2\, G \, \ln\{L(\omega-i\eps)\}\,\sum_{a=1}^m \sum_{b=1\atop b\ne a}^m 
{ p_a'.p_b'\over 
\{(p_a'.p_b')^2 
-p_a^{\prime 2} p_b^{\prime 2}\}^{3/2}} \, 
\left\{{3\over 2} p_a^{\prime 2} p_b^{\prime 2} - (p_a'.p_b')^2\right\} \, {k_\rho p_a^{\prime \mu} \over k.p'_a}\,
 (p_b^{\prime\rho} p_a^{\prime\nu} - p_b^{\prime\nu} p_a^{\prime\rho})\non\\  
 &-& 2\, G\, \ln\{(\omega+i\eps) R\}\, 
 \sum_{b=1}^n\, p_b.k  \left[\sum_{a=1}^{n} \, {1\over p_a.k-i\eps} \, p_a^\mu p_a^\nu
 - \sum_{a=1}^{m} \, {1\over p_a'.k-i\eps} \, p_a^{\prime\mu} p_a^{\prime \nu}
 \right]
 \, .
\een
In \refb{efingrabs} we can replace $k$ by $\omega \ n$ with $n=(1,\hat n)$.
Comparing the second and last line of \refb{efingrabs} we see that the term proportional to $\ln R$ exponentiates to
a multiplicative factor of 
\be\label{ephasecou}
\exp\left[-2\, i\, \omega\, G\, \ln R \sum_{b=1}^n p_b.n\right]\, .
\ee
Using \refb{e55gr}, \refb{e1.9pre}
and  \refb{e1.13ag} we get the late and early time behaviour of the gravitational
wave-form:
\ben\label{e3.66}
e^{\mu\nu}(t, R, \hat n) &=& {2\, G\over R} \, \left[-\sum_{a=1}^{n}  
p_a^\mu \, p_a^\nu\, 
{1\over n.p_a} + \sum_{a=1}^{m}  
p_a^{\prime\mu} \, p_a^{\prime\nu}\, 
{1\over n.p_a'} \right]\non\\
&& \hskip -1in -\, {4\, G^2\over R\, u} \left[
\sum_{a=1}^n \sum_{b=1\atop b\ne a}^n 
{ p_a.p_b\over 
\{(p_a.p_b)^2 
-p_a^2 p_b^2\}^{3/2}} \, 
\left\{{3\over 2} p_a^2 p_b^2 - (p_a.p_b)^2\right\} \, {n_\rho p_a^\mu \over n.p_a}\,
 (p_b^\rho p_a^\nu - p_b^\nu p_a^\rho) \right.
\non\\ && \left. \hskip -.5in - 
 \sum_{b=1}^n\, p_b.n  \left\{\sum_{a=1}^{n} \, {1\over p_a.n} \, p_a^\mu p_a^\nu
 - \sum_{a=1}^{m} \, {1\over p_a'.n} \, p_a^{\prime\mu} p_a^{\prime \nu}\right\}
 \right]\, , \quad \hbox{as $u\to \infty$}\, ,\non\\
 e^{\mu\nu}(t, R, \hat n) &=&  {4\, G^2\over R\, u} \Bigg[ \sum_{a=1}^m \sum_{b=1\atop b\ne a}^m 
{ p_a'.p_b'\over 
\{(p_a'.p_b')^2 
-p_a^{\prime 2} p_b^{\prime 2}\}^{3/2}} \, 
\left\{{3\over 2} p_a^{\prime 2} p_b^{\prime 2} - (p_a'.p_b')^2\right\} \non\\
&& \hskip 1in \times\ {n_\rho p_a^{\prime \mu} \over n.p'_a}\,
 (p_b^{\prime\rho} p_a^{\prime\nu} - p_b^{\prime\nu} p_a^{\prime\rho})
 \Bigg], \qquad 
 \hbox{as $u\to -\infty$}\, ,
\een
where, from \refb{e2.14} and \refb{e55gr}, \refb{ephasecou}, 
\be 
u = t - R + 2\, G\, \ln R \sum_{b=1}^n p_b.n\, .
\ee
In \refb{e3.66} we have adjusted the overall additive constant in the expression for $e^{\mu\nu}$ such
that it vanishes in the far past.

\sectiono{Generalizations} \label{sgen}

In this section we shall derive the classical soft photon theorem. We shall
also generalize the soft graviton theorem to include the effect of
electromagnetic interactions among the incoming and the outgoing particles. 
In order to simplify our formul\ae\  we shall drop the
regulator factors of $e^{i\ell.(r_a-r_b)}$, $e^{ik.r_a}$ etc., with the
understanding that momentum integrals have an upper cut-off $L^{-1}$ and a lower 
cut-off 
$R^{-1}$.

\subsection{Soft photon theorem with electromagnetic interactions} \label{sgen.1}

As in \S\ref{sgrav}, 
we consider a scattering event in asymptotically flat space-time in which $m$ particles
carrying masses $\{m'_a; 1\le a\le m\}$, four velocities $\{v'_a\}$,
four momenta $\{p'_a=m'_a \, v'_a\}$ and charges $\{q'_a\}$ come close, undergo interactions,
and disperse as 
$n$ particles carrying masses $\{m_a; 1\le a\le n\}$, four velocities $\{v_a\}$,
four momenta $\{p_a\}$ and charges $\{q_a\}$.   
Our goal will be to compute
the early and late time electromagnetic wave-form emitted during this scattering event.
In this section we shall proceed by ignoring the gravitational interaction between the
particles, but this will be included in \S\ref{sgen.2}. Since the analysis proceeds
as in \S\ref{sgrav}, we shall be brief, pointing out only the main differences. In particular,
as in \S\ref{sgrav}, we can treat the incoming particles as outgoing particles with
four velocities $\{-v'_a\}$,
four momenta $\{-p'_a\}$ and charges $\{-q'_a\}$. This allows us to
drop the sum over incoming particles by extending the
sum over $a$ from 1 to $m+n$.

In the Lorentz gauge $\eta^{\alpha\beta}\p_\alpha a_\beta=0$, the equations replacing \refb{eitergr} are:
\be\label{eiterem}
j^\mu(x) = \sum_a q_a\, \int d\sigma \, \delta^{(4)}(x - X_a(\sigma)) \, {dX_a^\mu\over d\sigma},
\qquad \square\, a_\mu = -j_\mu, \qquad m_a\, {d^2 X_a^\mu\over d\sigma^2} =q_a F^\mu_{~\nu}(X_a(\sigma))\,  {dX_a^\nu\over d\sigma}\, .
\ee
We introduce the
Fourier transforms via:
\be
a_\mu(x) =\int{d^4 k\over (2\pi)^4} \, e^{ik.x}\, \wh a_\mu(k), \quad 
j_\mu(x) =\int{d^4 k\over (2\pi)^4} \, e^{ik.x}
\, \wh j_\mu(k)\, .
\ee
This gives:
\be\label{e4.3}
\wh j^\mu(k) = \int d^4 x \, e^{-ik.x} \, j^\mu (x) 
= \sum_a q_a\, \int d\sigma \, e^{-ik.X(\sigma)} \, {dX_a^\mu\over d\sigma}\, .
\ee
The generalization of \refb{e55gr} for 
the asymptotic electromagnetic field is:
\be\label{eamuexp}
\tilde a^\mu(\omega, R, \hat n) = {1\over 4\pi R}\, e^{i\omega R} \, \wh j^\mu(k)
+\OO(R^{-2})\, .
\ee

We proceed to find iterative solutions to \refb{eiterem}
in a power series expansion in the
charges, beginning with the leading order
solution for $X^\mu$ given in \refb{e511grpre}. Substituting this into \refb{e4.3} we find
the leading order expression
for $\wh j^\mu(k)$:
\be\label{eleadem}
\wh j^\mu(k) =\sum_{a=1}^{m+n} q_a\, p_a^\mu \, 
{1\over i (k.p_a-i\eps)} \,.
\ee
This is the leading soft factor. 
Using this we can get the analogs of 
\refb{eabmugr} and \refb{eGam}:
\be\label{eabmu}
a^{(b)}_\mu(x) = -\int {d^4 \ell\over (2\pi)^4} \, e^{i\ell . x}\, G_r(\ell)  q_b\, p_{b\mu} \, 
{1\over i(\ell.p_b-i\eps)}\, ,
\ee
\be\label{efnurho}
F^{(b)}_{\nu\rho}(x) = \p_\nu \, a^{(b)}_\rho - \p_\rho \, a^{(b)}_\nu
= 
- \int{d^4 \ell\over (2\pi)^4} e^{i\ell.x}\, G_r(\ell)  \, {q_b\over (\ell.p_b-i\eps)}
(\ell_\nu p_{b\rho}-\ell_\rho p_{b\nu})\, ,
\ee
where $a^{(b)}_\mu$ and $F^{(b)}_{\mu\nu}$ denote the gauge field and field strength produced by the $b$-th
particle. The analogs of \refb{eYY1}, \refb{eYY2} take the form:
\be\label{eYY1new}
{dY_a^\mu(\sigma)\over d\sigma} = -{q_a\over m_a}\, 
\int_\sigma^\infty\, d\sigma' \, F^\mu_{~\nu}(v_a\, \sigma'+r_{a}) \, 
v_a^\nu\, ,
\ee
and
\be \label{eYY2new}
Y_a^\mu(\sigma) =-{q_a\over m_a}\, 
\int_0^\sigma\, d\sigma' \, \int_{\sigma'}^\infty\, d\sigma'' \, F^\mu_{~\nu}(v_a\, \sigma''+r_{a}) \, v_a^\nu\,  .
\ee
Using these results we can
proceed as in \S\ref{sgrav.3} to compute the next order correction 
$\Delta\wh j^\mu(k)$ to $\wh j^\mu(k)$. Since the analysis is identical to those in
\S\ref{sgrav.3} we only quote the analog of \refb{ejmfingr}:
\ben\label{ejm}
\Delta\wh j^\mu(k) &=& \sum_{a=1}^{m+n} \sum_{b=1\atop b\ne a}^{m+n}  
{q_a^2 q_b} \, \int{d^4 \ell\over (2\pi)^4} \,
{1\over \ell.p_b-i\eps} \, G_r(\ell) \, 
\Bigg[ k_\rho \,  {1\over  (k-\ell).p_a-i\eps}\, {1\over k.p_a-i\eps}\,  {1\over \ell.p_a+i\eps}
\nonumber \\
&& \hskip 2in \times\, 
(\ell_\nu p_{b\rho}-\ell_\rho p_{b\nu})\, p_a^\nu\, p_a^\mu \nonumber \\ &&
\hskip .5in -  {1\over  (k-\ell).p_a-i\eps} {1\over \ell.p_a+i\eps}
(\ell_\nu p_{b}^\mu-\ell^\mu p_{b\nu}) p_a^\nu
 \Bigg]\, .
\een
This can be evaluated exactly as in \S\ref{sgrav.3}, leading to the final result analogous
to \refb{efingratot}:
\ben\label{efinqed}
\Delta\wh j^\mu(k) &=& {1\over 4\pi} \, \ln(\omega+i\eps)\,
\,\sum_{a=1}^n \sum_{b=1\atop b\ne a}^n  {q_a^2 q_b} \, p_a^2\, 
p_b^2 \, {1\over 
\{(p_a.p_b)^2 
-p_a^2 p_b^2\}^{3/2}} {k^\rho\over k.p_a}\left\{p_{b\rho} p_{a}^\mu-p_{a\rho} p_{b}^\mu\right\}\nonumber \\ &+&
 {1\over 4\pi} \, \ln(\omega-i\eps)\,
\,\sum_{a=1}^m \sum_{b=1\atop b\ne a}^m  {q_a^{\prime 2} q'_b} \, p_a^{\prime 2}\, 
p_b^{\prime 2} \, {1\over 
\{(p_a'.p_b')^2 
-p_a^{\prime 2} p_b^{\prime 2}\}^{3/2}} {k^\rho\over k.p'_a}
\left\{p_{b\rho}' p_{a}^{\prime\mu}-p'_{a\rho} p_{b}^{\prime\mu}\right\}\, .\nonumber \\
\een

\subsection{Gravitational contribution to the soft photon theorem} \label{sgen.2}

We shall now study the effect of gravitational interaction on the soft photon theorem.
This modifies the last two equations in \refb{eiterem} as follows. First of all the equation
for $a_\mu$ get modified to:
\be \label{emodem1}
\p_\nu \left(\sqrt{-\det g} g^{\nu\rho} g^{\mu\sigma} F_{\rho\sigma}\right)
 =-j^\mu  \, .
\ee
Using Lorentz gauge condition $\eta^{\rho\sigma}\p_\rho a_\sigma=0$, this
may be written as
\be
\eta^{\mu\nu}\, \eta^{\rho\sigma}\, \p_\rho \, \p_\sigma \, a_\nu =-j^\mu -  j_h^\mu\, ,
\ee
where
\be\label{emodem3}
j_h^\mu\equiv \p_\nu \left\{\eta^{\alpha\beta} \, h_{\alpha\beta} \, \eta^{\nu\rho} 
\eta^{\mu\sigma} F_{\rho\sigma} -  2\, \left(h^{\nu\rho} 
\eta^{\mu\sigma} +  \eta^{\nu\rho} 
h^{\mu\sigma}\right) F_{\rho\sigma}
\right\} +\hbox{higher order terms}\, .
\ee
The equation for $X^\mu$ is modified to:
\be \label{emodem2}
m_a\, {d^2 X_a^\mu\over d\sigma^2} =q_a F^\mu_{~\nu}(X_a(\sigma))\,  {dX_a^\nu\over d\sigma} - m_a\, \Gamma^\mu_{\nu\rho}(X_a(\sigma))\, {dX_a^\nu\over d\sigma}\,
{dX_a^\rho\over d\sigma}\, .
\ee

We shall now expand the above equations in powers of $h_{\alpha\beta}$ and then 
raise and lower all indices by
the flat metric $\eta$. We begin with the analysis of \refb{emodem2}.
To the order that we are working, we can study the effect of the two terms on the
right hand side of \refb{emodem2} separately. The effect of the first term on $\wh j^\mu$
has already been analyzed in \S\ref{sgen.1}. The effect of the second term on
$\wh T^X_{\mu\nu}$ has been studied in \S\ref{sgrav.3}, but this can be easily extended 
to $\wh j^\mu$. The additional contribution to $\wh j^\mu$ is given by:
\begin{eqnarray} \label{ej1add}
\Delta^{(1)} \wh j^{\mu}\left(k\right)& = & -8\pi G \sum\limits_{a=1}^{m+n}\sum\limits_{b=1\atop
b\ne a}^{m+n} q_{a} \int\frac{d^{4}\ell }{\left(2\pi\right)^{4}} \, G_r(\ell)\, \frac{1}{\ell \cdot p_{b}- i\epsilon}\nonumber\\
&& \hspace{0cm}\left[-\frac{\biggl\{2 k\cdot p_{b} \; \ell \cdot p_{a} \; p_{a}\cdot p_{b} - k\cdot \ell \left(p_{a}\cdot p_{b}\right)^{2} - \frac{1}{2}\left(2k\cdot p_{a}\; \ell \cdot p_{a} - k\cdot \ell  \, 
p_{a}^{2}\right)p_{b}^{2}\biggr\}p_{a}^{\mu}}{\left[k\cdot p_{a}-i\epsilon\right]\left[\left(k-\ell \right)\cdot p_{a}-i\epsilon\right]\left[ \ell \cdot p_{a}+i\epsilon\right]}\right. \nonumber\\
&& \hspace{0cm} \left.+ \frac{2p_{b}^{\mu}\ell \cdot p_{a}\; p_{a}\cdot p_{b} - \ell ^{\mu}\left(p_{a}\cdot p_{b}\right)^{2} - \frac{1}{2}\left(2 p_{a}^{\mu} \, \ell \cdot p_{a} - \ell ^{\mu}\, p_{a}^{2}\right)p_{b}^{2}}{\left[\left(k-\ell \right)\cdot p_{a}-i\epsilon\right]\left[ \ell \cdot p_{a}+i\epsilon\right]}\right] .
\end{eqnarray}
This integral can be evaluated as in \S\ref{sgrav.3} and yields the result:
\begin{eqnarray} \label{efinqedfromgr}
&& \Delta^{(1)} \wh j^{\mu}\left(k\right) \nonumber \\
& = & -\ G\, \log(\omega+i\eps)
\sum\limits_{a=1}^n\sum\limits_{b=1\atop b\ne a}^n q_{a}\frac{k_{\rho}}{k\cdot p_{a}}
\frac{p_{a}\cdot p_{b}}{\bigl\{\left(p_{a}\cdot p_{b}\right)^{2} - p_{a}^{2}p_{b}^{2}
\bigr\}^{\frac{3}{2}}}\left(p_{a}^{\mu}p_{b}^{\rho} - p_{a}^{\rho}p_{b}^{\mu}\right)\bigl\{2\left(p_{a}\cdot p_{b}\right)^{2} - 3p_{a}^{2}p_{b}^{2}\bigr\} \nonumber \\
&-& G\, \log(\omega-i\eps)
\sum\limits_{a=1}^m\sum\limits_{b=1\atop b\ne a}^m q'_{a}\frac{k_{\rho}}{k\cdot p'_{a}}
\frac{p'_{a}\cdot p'_{b}}{\bigl\{\left(p'_{a}\cdot p'_{b}\right)^{2} - p_{a}^{\prime 2}
p_{b}^{\prime 2}
\bigr\}^{\frac{3}{2}}}\left(p_{a}^{\prime\mu}p_{b}^{\prime\rho} 
- p_{a}^{\prime\rho}p_{b}^{\prime\mu}\right)
\bigl\{2\left(p'_{a}\cdot p'_{b}\right)^{2} - 3p_{a}^{\prime 2}p_{b}^{\prime 2}\bigr\} \, .
\nonumber \\
\end{eqnarray}

We now turn to the evaluation of $j_h^\mu$ given in \refb{emodem3}.
Using the expressions for $h_{\mu\nu}$ and $F_{\mu\nu}$ given in \refb{eabmugr} 
and
\refb{efnurho}, we get:
\begin{equation}\label{j-grav-interactionpre}
\wh{j}_{h}^{\mu}\left(k\right) = 8\pi G\sum\limits_{a,b=1}^{m+n}q_{b}\int\frac{d^{4}\ell }{\left(2\pi\right)^{4}}
G_{r}\left(\ell \right)G_{r}\left(k-\ell \right)\frac{1}{\left[\ell \cdot p_{a}-i\epsilon\right]\left[\left(k-\ell \right)
\cdot p_{b}-i\epsilon\right]}\mathcal{F}^{\mu}\left(k,\ell \right),
\end{equation}
where
\ben
\FF^\mu &\equiv& p_b^\mu \left\{p_a^2\, k.(k-\ell) - 2\, p_a.(k-\ell)\, p_a.k\right\}
+(k^\mu-\ell^\mu) \, \left\{2\, k.p_a\, p_a.p_b - p_a^2\, k.p_b\right\} \non\\
&& +\ p_a^\mu \left\{2\,p_a.(k-\ell) \, k.p_b - 2\, k.(k-\ell)\, p_a.p_b\right\}\, .
\een
We shall analyze this by expressing $G_r(\ell) \, G_r(k-\ell)$ as in \refb{e345gr}. 
The term proportional to $\delta(\ell^2)$ can be analyzed as in appendix \ref{sreal}, and one can show
that in this case there is no contribution from this term. This gives
\begin{equation}\label{j-grav-interaction}
\wh{j}_{h}^{\mu}\left(k\right) = 8\pi G\sum\limits_{a,b=1}^{m+n}q_{b}\int\frac{d^{4}\ell }{\left(2\pi\right)^{4}}
G_{r}\left(\ell \right)^*
G_{r}\left(k-\ell \right)\frac{1}{\left[\ell \cdot p_{a}-i\epsilon\right]\left[\left(k-\ell \right)\cdot p_{b}-i\epsilon\right]}\mathcal{F}^{\mu}\left(k,\ell \right)\, .
\end{equation}
This integral could give logarithmic contributions from three regions: 
$R^{-1}<<|\ell^\mu|<<\omega$, $R^{-1}<<|k^\mu-\ell^\mu|<<\omega$ and
$\omega<< |\ell^\mu|<< L^{-1}$. However, since $\FF^\nu$ vanishes
as $k-\ell\to 0$, there is no logarithmic divergence from the 
$R^{-1}<<|k^\mu-\ell^\mu|<<\omega$ region. Furthermore, since $\FF^\nu$ does not have 
any
quadratic term in $\ell$,  this rules out logarithmic contribution from the region
$\omega<< |\ell^\mu|<< L^{-1}$.
Therefore the only possible source of logarithmic divergence is the 
region $R^{-1}<<|\ell^\mu|<<\omega$. In this region,
\be
\mathcal{F}^{\nu} \simeq -\, 2\, \biggl\{\left(k\cdot p_{a}\right)^{2}p_{b}^{\nu} 
- k\cdot p_{a}\; k\cdot p_{b}\; p_{a}^{\nu}\biggr\}\, .
\ee 
We have ignored the terms proportional to $k^{\nu}$ because such terms can be 
removed by gauge transformation $a_\mu\to a_\mu +\p_\mu \phi$ for 
appropriate function $\phi$. 
We can now evaluate the logarithmic contribution to the integral using the 
method described below \refb{efapproxa}, and the result is
\ben \label{ejhfin}
\wh j_h^\mu(k) &=& - \, 2\, G\, \ln(\omega+i\eps)\, 
\sum_{a=1}^n k.p_a\, \left[\sum_{b=1}^n q_b \, p_b^\mu\, {1\over k.p_b} - 
\sum_{b=1}^m q'_b  \, p_b^{\prime\mu}\, {1\over k.p'_b}\right]
\, ,
\een
after using charge conservation $\sum_{b=1}^{m+n} q_b=0$.

\subsection{Electromagnetic wave-form at early and late time}

Adding \refb{eleadem}, \refb{efinqed}, \refb{efinqedfromgr} and \refb{ejhfin}, 
and using $k=\omega\, n$ and \refb{eamuexp}, \refb{e1.9pre}, \refb{e1.13ag} we get
\ben \label{eemabs1}
4\, \pi\, R\, a^\mu(t, R, \hat n) &\simeq& -\sum_{a=1}^{n} q_a\, p_a^\mu 
{1\over n.p_a} + \sum_{a=1}^{m} q'_a\, p_a^{\prime\mu}
{1\over n.p_a'}\nonumber \\  &+&
{1\over u} \, \Bigg[-{1\over 4\pi} \,
\,\sum_{a=1}^n \sum_{b=1\atop b\ne a}^n  {q_a^2 q_b} \, p_a^2\, 
p_b^2 \, {1\over 
\{(p_a.p_b)^2 
-p_a^2 p_b^2\}^{3/2}} {n^\rho\over n.p_a}\left\{p_{b\rho} p_{a}^\mu-p_{a\rho} p_{b}^\mu\right\} \nonumber \\
&+& G\,
\sum\limits_{a=1}^n\sum\limits_{b=1\atop b\ne a}^n q_{a}\frac{n_{\rho}}{n\cdot p_{a}}
\frac{p_{a}\cdot p_{b}}{\bigl\{\left(p_{a}\cdot p_{b}\right)^{2} - p_{a}^{2}p_{b}^{2}
\bigr\}^{\frac{3}{2}}}\left(p_{a}^{\mu}p_{b}^{\rho} - p_{a}^{\rho}p_{b}^{\mu}\right)\bigl\{2\left(p_{a}\cdot p_{b}\right)^{2} - 3p_{a}^{2}p_{b}^{2}\bigr\} \nonumber \\
&+& 2\, G\,
\sum_{a=1}^n \, n.p_a\, \Bigg\{\sum_{b=1}^n q_b \, p_b^\mu \, {1\over n.p_b}\, - 
\sum_{b=1}^m q'_b  \, p_b^{\prime\mu}\, {1\over n.p'_b}\Bigg\}\Bigg]
\, , \hskip .3in \hbox{as $u\to \infty$}\, ,
\een
and
\ben \label{eemabs2}
4\, \pi\, R\, a^\mu(t, R, \hat n) &\simeq& {1\over u} \, \Bigg[
{1\over 4\pi} \,
\,\sum_{a=1}^m \sum_{b=1\atop b\ne a}^m  {q_a^{\prime 2} q'_b} \, p_a^{\prime 2}\, 
p_b^{\prime 2} \, {1\over 
\{(p_a'.p_b')^2 
-p_a^{\prime 2} p_b^{\prime 2}\}^{3/2}} {n^\rho\over n.p'_a}
\left\{p_{b\rho}' p_{a}^{\prime\mu}-p'_{a\rho} p_{b}^{\prime\mu}\right\} \nonumber \\ && \hskip -1in
- \, G\, 
\sum\limits_{a=1}^m\sum\limits_{b=1\atop b\ne a}^m q'_{a}\frac{n_{\rho}}{n\cdot p'_{a}}
\frac{p'_{a}\cdot p'_{b}}{\bigl\{\left(p'_{a}\cdot p'_{b}\right)^{2} - p_{a}^{\prime 2}
p_{b}^{\prime 2}
\bigr\}^{\frac{3}{2}}}\left(p_{a}^{\prime\mu}p_{b}^{\prime\rho} 
- p_{a}^{\prime\rho}p_{b}^{\prime\mu}\right)
\bigl\{2\left(p'_{a}\cdot p'_{b}\right)^{2} - 3p_{a}^{\prime 2}p_{b}^{\prime 2}\bigr\}
\Bigg]\, , \nonumber \\
&& \hskip 1in \hbox{as $u\to -\infty$}\, .
\een
This gives the wave-form of the electromagnetic field at early and late time.
The term on the right hand side of the first line gives the constant shift in the vector 
potential, and is
responsible for electromagnetic 
memory\cite{1307.5098,1505.00716,1507.02584}. The rest of the terms are tail terms.

\subsection{Electromagnetic contribution to the soft graviton theorem} \label{sgen.3}

We shall now analyze the effect of electromagnetic interaction on the soft graviton theorem. 
This affects our earlier analysis of soft graviton theorem in two ways. First of all,  the 
Lorentz force on the outgoing and incoming particles changes the particle 
trajectories, producing
an
additional contribution to $\wh T^{X\mu\nu}$. Analysis of this follows the same
procedure that led to \refb{ejmfingr}, \refb{ej1add}, and the final result is given by:
\ben\label{etmnemfin}
\Delta^{(1)}\wh T^{\mu\nu}(k) &=& \sum_{a=1}^{m+n} \sum_{b=1\atop b\ne a}^{m+n}  
{q_a q_b} \, \int{d^4 \ell\over (2\pi)^4} \, {1\over \ell.p_b-i\eps} \, G_r(\ell) \, \nonumber \\
&& 
\Bigg[  {1\over  (k-\ell).p_a-i\eps}\ {1\over k.p_a-i\eps}\  {1\over \ell.p_a+i\eps}
 \, 
(p_a.\ell \,  k. p_{b}-k.\ell \, p_a.p_{b})\,\, p_a^\mu p_a^\nu\nonumber \\ &&
 -  {1\over  (k-\ell).p_a-i\eps} \ {1\over \ell.p_a+i\eps}\ 
(p_a. \ell \,  p_{b}^\mu-\ell^\mu \, p_a.p_{b})\, p_a^\nu \nonumber \\ &&
 -  {1\over  (k-\ell).p_a-i\eps} \ {1\over \ell.p_a+i\eps}
(p_a. \ell \,  p_{b}^\nu-\ell^\nu \, p_a.p_{b})\, p_a^\mu
 \Bigg]\, .
\een
Evaluation of this using the method described below \refb{ejmfingr}, gives
\ben\label{eemgr1}
\Delta^{(1)}\wh T^{\mu\nu}(k) &=& {1\over 4\pi} \, \ln\{L\, (\omega+i\eps)\}\,
\,\sum_{a=1}^n \sum_{b=1\atop b\ne a}^n  {q_a q_b}  \, {1\over 
\{(p_a.p_b)^2 
-p_a^2 p_b^2\}^{3/2}} \Bigg[ {k.p_b\over k.p_a}\, p_a^2 \, p_b^2 \, p_a^\mu\, p_a^\nu \nonumber \\ &&
+\ p_b^2\, p_a.p_b \, p_a^\mu \, p_a^\nu - p_a^2\, p_b^2\, (p_a^\mu p_b^\nu + p_a^\nu p_b^\mu)
\Bigg]\nonumber \\ &&
+\ {1\over 4\pi} \, \ln\{L\, (\omega-i\eps)\}\,
\,\sum_{a=1}^m \sum_{b=1\atop b\ne a}^m  {q_a' q_b'}  \, {1\over 
\{(p_a'.p_b')^2 
-p_a^{\prime 2} p_b^{\prime 2}\}^{3/2}} \Bigg[ {k.p_b'\over k.p_a'}\, p_a^{\prime 2} \, p_b^{\prime 2} 
\, p_a^{\prime \mu}\, p_a^{\prime\nu} \nonumber \\ &&
+\ p_b^{\prime 2}\, p_a'.p_b' \, p_a^{\prime\mu} \, p_a^{\prime\nu} - p_a^{\prime 2}\, p_b^{\prime 2}\, 
(p_a^{\prime\mu} p_b^{\prime\nu} + p_a^{\prime\nu} p_b^{\prime\mu})
\Bigg]\, .
\een

Second, there is an additional contribution to the stress tensor due to the electromagnetic field. Using the form of
the electromagnetic field produced by the charged particle as given in \refb{efnurho}, this additional
contribution takes the form:
\ben\label{etmnemfina}
\Delta^{(2)}\wh T^{\mu\nu}(k) &=& \int d^4 x\, e^{-ik.x} \left[ \eta^{\mu\rho} \, \eta^{\nu\sigma}\, \eta^{\alpha\beta}\,
F_{\rho\alpha}\, F_{\sigma\beta} \, -\, {1\over 4}\, \eta^{\mu\nu} \, \eta^{\rho\sigma}\, \eta^{\alpha\beta}\,
F_{\rho\alpha}\, F_{\sigma\beta}\right] +\hbox{higher order terms}
\non\\ &\simeq&\sum_{a=1}^{m+n} \sum_{b=1}^{m+n}  {q_a q_b} \, \int{d^4 \ell\over (2\pi)^4} 
\, {1\over \ell.p_b-i\eps} \, 
{1\over (k-\ell).p_a-i\eps} \, G_r(\ell) \, G_r(k-\ell) \nonumber \\ &&
\Bigg[ \ell^\mu (k-\ell)^\nu \, p_a.p_b -\ell^\mu p_a^\nu \, \,  p_b.(k-\ell) - p_b^\mu \, (k-\ell)^\nu \, \ell.p_a
+ p_b^\mu p_a^\nu \, \, \ell.(k-\ell) \nonumber \\ &&
-{1\over 2}\eta^{\mu\nu} \left\{ \ell.(k-\ell)\, p_a.p_b - \ell.p_a \, (k-\ell).p_b
\right\}
\Bigg]\, .
\een
We shall analyze this by expressing $G_r(\ell) \, G_r(k-\ell)$ as in \refb{e345gr}. 
The term proportional to $\delta(\ell^2)$ can be analyzed as in appendix \ref{sreal}, and one can show
that the contribution from this term can be interpreted as the soft graviton emission from electromagnetic wave
produced during scattering. Since this is included in the sum over $a$ in the soft factor, we do not need
to include its contribution. This allows us to replace $G_r(\ell)$ by $G_r(\ell)^*$ in \refb{etmnemfina}. 
Since each term in the numerator of the 
integrand carries a factor of $\ell$ and a factor of $(k-\ell)$, there is no logarithmic
contribution from the $|\ell^\mu|<<\omega$ and $|k^\mu-\ell^\mu|<<\omega$ regions. Therefore we focus on the
$|\ell^\mu|>>\omega$ region, and analyze the contribution using \refb{edefiab}, \refb{eiabfin}. 
The final result is:
\ben\label{eemgr2}
&& \Delta^{(2)}\wh T^{\mu\nu}(k) = -
{1\over 4\pi} \ln(\omega+i\eps)\sum_{a=1}^{n} \sum_{b=1\atop b\ne a}^{n} {q_a q_b} \, p_a.p_b\, {1\over 
\{(p_a.p_b)^2 
-p_a^2 p_b^2\}^{3/2}} \, p_a^\mu (p_b^2 \, p_a^\nu - p_a.p_b\, p_b^\nu) \nonumber \\ 
&& -
{1\over 4\pi} \ln(\omega-i\eps)\sum_{a=1}^{m} \sum_{b=1\atop b\ne a}^{m}  {q_a' q_b'} \, p_a'.p_b'\, {1\over 
\{(p_a'.p_b')^2 
-p_a^{\prime 2} p_b^{\prime 2}\}^{3/2}} \, p_a^{\prime \mu} (p_b^{\prime 2} \, p_a^{\prime \nu} 
- p_a'.p_b'\, p_b^{\prime \nu}) \nonumber \\ &&
- {1\over 4\pi} \ln(\omega+i\eps)\sum_{a=1}^{n} \sum_{b=1\atop b\ne a}^{n} {q_a q_b} \,  {1\over 
\{(p_a.p_b)^2 
-p_a^2 p_b^2\}^{1/2}}\, p_b^\mu \, p_a^\nu\nonumber \\ &&
- {1\over 4\pi} \ln(\omega-i\eps)\sum_{a=1}^{m} \sum_{b=1\atop b\ne a}^{m} {q_a' q_b'} \, {1\over 
\{(p_a'.p_b')^2 
-p_a^{\prime 2} p_b^{\prime 2}\}^{1/2}}\, p_b^{\prime \mu} \, p_a^{\prime \nu}\, .
\een
Adding \refb{eemgr1} and \refb{eemgr2} we get the net electromagnetic contribution to the soft graviton theorem:
\ben\label{eemgr3}
&& \Delta^{(1)}\wh T^{\mu\nu}(k) + \Delta^{(2)}\wh T^{\mu\nu}(k) \nonumber \\
&=& {1\over 4\pi} \, \ln(\omega+i\eps)\,
\,\sum_{a=1}^n \sum_{b=1\atop b\ne a}^n  {q_a q_b}  \, {p_a^2 \, p_b^2 \over 
\{(p_a.p_b)^2 
-p_a^2 p_b^2\}^{3/2}} \Bigg[ {k.p_b\over k.p_a}\,  p_a^\mu\, p_a^\nu -
 p_a^\mu p_b^\nu \Bigg]\nonumber \\ &&
+{1\over 4\pi} \, \ln(\omega-i\eps)\,
\,\sum_{a=1}^m \sum_{b=1\atop b\ne a}^m  {q_a' q_b'}  \, {p_a^{\prime 2} \, p_b^{\prime 2} \over 
\{(p_a'.p_b')^2 
-p_a^{\prime 2} p_b^{\prime 2}\}^{3/2}} \Bigg[ {k.p_b'\over k.p_a'}\, p_a^{\prime \mu}\, p_a^{\prime\nu} -
p_a^{\prime \mu} p_b^{\prime \nu} 
\Bigg]\, .\nonumber \\
\een
From \refb{e55gr}, \refb{e1.13ag}, we can read out the additional contribution to the gravitational 
wave-form at early and late retarded
time due to electromagnetic interactions:
\ben\label{e4.28em}
\Delta_{\rm em} \, e^{\mu\nu} &\to& -{G \over 2\pi \, R\, u} \, \,
\sum_{a=1}^n \sum_{b=1\atop b\ne a}^n  {q_a q_b}  \, {1\over 
\{(p_a.p_b)^2 
-p_a^2 p_b^2\}^{3/2}} \Bigg[ {k.p_b\over k.p_a}\, p_a^2 \, p_b^2 \, p_a^\mu\, p_a^\nu -
p_a^2\, p_b^2\, p_a^\mu p_b^\nu \Bigg]\non\\ && \hskip 2in \hbox{as $u\to\infty$},  
\non\\
&\to& {G \over 2\pi \, R\, u} \, \,\sum_{a=1}^m \sum_{b=1\atop b\ne a}^m  {q_a' q_b'}  \, {1\over 
\{(p_a'.p_b')^2 
-p_a^{\prime 2} p_b^{\prime 2}\}^{3/2}} \Bigg[ {k.p_b'\over k.p_a'}\, p_a^{\prime 2} \, p_b^{\prime 2} 
\, p_a^{\prime \mu}\, p_a^{\prime\nu} -
p_a^{\prime 2}\, p_b^{\prime 2}\, p_a^{\prime \mu} p_b^{\prime \nu} 
\Bigg] \non\\ && \hskip 2in \hbox{as $u\to-\infty$}.
\een

\sectiono{New conjectures at the subsubleading order} \label{s2}

Emboldened by the success of soft theorem in correctly predicting the tail of the gravitational
wave-form at the subleading order, we shall now propose new conjectures at the subsubleading order. It
is known that in quantum gravity, the subsubleading soft factors are not universal. Nevertheless there are
some universal terms that we could utilize\cite{1706.00759}. These are terms that are quadratic 
in the orbital angular
momenta. Our goal will be to 
make use of these universal terms to arrive at new conjectures on the late and early time tail of
gravitational radiation.
The non-universal terms do not involve orbital angular
momenta and therefore do not have logarithmic divergences. Hence
they will not affect our analysis.

Using the relation between quantum soft factors and classical gravitational wave-forms derived in 
\cite{1801.07719}, and ignoring the non-universal terms, we can
write down the following form of the gravitational wave-form to subsubleading 
order:\footnote{A non-trivial test of this formula for the scattering of massless
particles can be found in \cite{1812.08137}.}
\ben\label{emaster}
\wt e^{\mu\nu}(\omega,\vec x) &=& {2\,G\over R} \, e^{i\omega R}
\exp\left[-2\, i\, G\, \ln \{R(\omega+i\eps)\}  \sum_{b=1}^n p_b.k\right]
\non\\ && \times\, \sum_{a=1}^{m+n} \, 
\left[-i\, {p_{a}^\mu p_{a}^\nu\over p_{a}.k} - {1 \over p_{a}. k} 
J_{a}^{\rho(\nu}p_{a}^{\mu)} k_\rho +{i\over 2} \, {1 \over p_{a}. k} \, k_\rho\, k_\sigma\, J_{a}^{\mu\rho}
J_a^{\nu\sigma}
\right]\, ,
\een
where $k$ has been defined in \refb{edefkn} and
$J_a^{\rho\sigma}$ is the sum of the orbital and spin angular momenta of the $a$-th external particle:
\be
J_a^{\rho\sigma} = X_a^\rho \, p_a^\sigma - X_a^\sigma \, p_a^\rho + \Sigma_a^{\rho\sigma}\, .
\ee
The second term in the last line of \refb{emaster} differs by a sign from the 
expressions used {\it e.g.} in \cite{1801.07719}. This can be traced to the fact that in \cite{1801.07719} we
treated the charges / momenta / angular momenta carried by ingoing particles as
positive and of the outgoing particles as negative, whereas here we are following
the opposite convention.
Following \refb{edefprime}, the spin $\Sigma_a^{\prime\mu\nu}$ for incoming
particles are given by:
\be
\Sigma_a^{\prime\mu\nu} = -\Sigma_{a+n}^{\mu\nu} \quad \hbox{for $1\le a\le m$}\, .
\ee
The phase factor $\exp\left[-2\, i\, G\, \ln \{R(\omega+i\eps)\}  \sum_{b=1}^n p_b.k\right]$ in \refb{emaster}
is not determined by soft theorem, but is
determined by independent computation\cite{peters,blanchet,0912.4254}, 
and is consistent with the term in the last line of \refb{efingrabs}. 
Due to the long range gravitational force between the outgoing
/ incoming particles, $X_a^\rho$ has logarithmic corrections at late / early time\cite{1804.09193}, 
leading to\cite{1808.03288}
\ben
X_a^\rho \, p_a^\sigma - X_a^\sigma \, p_a^\rho \hskip -.05in &=& \hskip -.05in -
{G}  \,  \sum_{b\ne a\atop \eta_a\eta_b=1} \,  \ln|\sigma_a|\,  {p_b.p_a\over 
\{ (p_b.p_a)^2 -p_a^2 p_b^2\}^{3/2}}\, (p_b^\rho p_a^\sigma - p_b^\sigma p_a^\rho) \, 
\left\{2(p_b.p_a)^2  - 3 p_a^2 p_b^2 \right\}\non\\
&& +\ (r_a^\rho p_a^\sigma - r_a^\sigma p_a^\rho)  \, ,
\een
where $\sigma_a$ denotes the proper time of the $a$-th particle, and $r_a$ is the
constant that appeared in
\refb{ethisgr}. The contribution proportional to $\ln|\sigma_a|$ arises from the
correction term $Y_a$ in \refb{ethisgr}.
The conjecture of \cite{1804.09193,1808.03288} was that
the $\ln|\sigma_a|$ factor should be replaced 
by $\ln\omega^{-1}$ in \refb{emaster}. 
After including the $i\eps$ prescription described in this paper,
this conjecture translates to the rule that in \refb{emaster}, $J_a^{\rho\sigma}$ should
be replaced by:
\ben \label{ejars}
J_a^{\rho\sigma}&=&
{G}  \,  \sum_{b\ne a\atop \eta_a\eta_b=1} \,  \ln (\omega+i\eps\eta_a) {p_b.p_a\over 
\{ (p_b.p_a)^2 -p_a^2 p_b^2\}^{3/2}}\, (p_b^\rho p_a^\sigma - p_b^\sigma p_a^\rho) \, 
\left\{2(p_b.p_a)^2  - 3 p_a^2 p_b^2 \right\} \non\\
&& +\ (r_a^\rho p_a^\sigma - r_a^\sigma p_a^\rho) +\Sigma_a^{\rho\sigma}\, .
\een

We now substitute \refb{ejars}
 into \refb{emaster} and expand the expression in powers of $\omega$, including the
$\exp\left[-2\, i\, G\, \ln \{R(\omega+i\eps)\}  \sum_{b=1}^n p_b.k\right]$ term. 
Terms proportional to $\ln(\omega\pm i\eps)$ reproduce correctly 
\refb{efingrabs}.
We shall focus on
terms proportional to $\omega(\ln\omega)^2$. In the $\omega$ space these terms are 
subdominant compared to
the order $\omega^0$ terms 
that we have left out from the 
subleading terms in the gravitational wave-form. However 
after Fourier transformation, polynomials in $\omega$ produce local terms in time, while terms involving $\ln\omega$
produce tail terms that survive at late and early retarded time. 
Therefore the corrections to \refb{e3.66} at late and early 
time will be dominated by the terms proportional to $\omega (\ln\omega)^2$ in the expression for 
$\wt e^{\mu\nu}$. The order $\omega^0$ terms may have other observational signature, {\it e.g.} the spin
memory discussed in \cite{1502.06120}.

Expanding \refb{emaster} in powers of $\omega$, with $J_a^{\rho\sigma}$ given by \refb{ejars}, we get the
corrections proportional to $\omega (\ln \omega)^2$. These take the form:
\ben\label{edeltasub}
&& \Delta_{\rm subsubleading} \, \wt e^{\mu\nu} \non\\
&=& e^{i\omega R - 2\, i\, \omega \, G\, \ln R\, \sum_{d=1}^n p_d.n} \, i\, {G^3\over R}\, \omega\, \Bigg[ 4\,
\{\ln(\omega+i\eps)\}^2 \sum_{b=1}^n p_b.n \sum_{c=1}^n p_c.n \sum_{a=1}^{m+n} {p_a^\mu p_a^\nu\over p_a.n}
\non\\ &&
+ 4\, \ln (\omega +i\eps) \, \sum_{c=1}^n p_c.n \sum_{a=1}^{m+n} \sum_{b=1\atop b\ne a, \eta_a\eta_b=1}^{m+n}
\ln(\omega+i\eps\eta_a) {1\over p_a.n} {p_a.p_b\over \{(p_a.p_b)^2 - p_a^2 p_b^2\}^{3/2}} \non\\ &&
\hskip 1in \{2 (p_a.p_b)^2 - 3 p_a^2 p_b^2\} \{ n.p_b \, p_a^\mu \, p_a^\nu - n.p_a \, p_a^\mu \, 
p_b^\nu\}\non\\ && 
+ \sum_{a=1}^{m+n} \sum_{b=1\atop b\ne a, \eta_a\eta_b=1}^{m+n}
\sum_{c=1\atop c\ne a, \eta_a\eta_c=1}^{m+n} \{\ln(\omega+i\eps\eta_a)\}^2 {1\over p_a.n} \, 
{p_a.p_b\over \{(p_a.p_b)^2 - p_a^2 p_b^2\}^{3/2}} \non\\ &&
\hskip 1in  \{2 (p_a.p_b)^2 - 3 p_a^2 p_b^2\} 
{p_a.p_c\over \{(p_a.p_c)^2 - p_a^2 p_c^2\}^{3/2}} \{2 (p_a.p_c)^2 - 3 p_a^2 p_c^2\}  \non\\ &&
\hskip 1in 
\{ n.p_b \, p_a^\mu - n.p_a \, p_b^\mu\} \, \{ n.p_c \, p_a^\nu - n.p_a \, p_c^\nu \}
\Bigg]\, .
\een
Using \refb{eGlimit}, \refb{eHlimit},  we now get,
\be\label{eaddcon3}
\Delta_{\rm subsubleading}\, e_{\mu\nu}\to \cases{ u^{-2} \, \ln|u|\, F_{\mu\nu} \quad \hbox{as} \quad u\to\infty\cr
 u^{-2} \, \ln|u| \, G_{\mu\nu} \quad \hbox{as} \quad u\to-\infty}\, ,
\ee
where
\ben \label{eaddcon4}
F^{\mu\nu} &=& 2\, {G^3\over R}\, \Bigg[ 4\,
\sum_{b=1}^n p_b.n \sum_{c=1}^n p_c.n \left\{
\sum_{a=1}^{n} {p_a^\mu p_a^\nu\over p_a.n} - \sum_{a=1}^{m} {p_a^{\prime\mu} p_a^{\prime\nu}\over 
p_a'.n}\right\}
\non\\ && \hskip -.5in
+ 4\, \sum_{c=1}^n p_c.n \sum_{a=1}^{n} \sum_{b=1\atop b\ne a}^{n}
 {1\over p_a.n} {p_a.p_b\over \{(p_a.p_b)^2 - p_a^2 p_b^2\}^{3/2}} 
 \{2 (p_a.p_b)^2 - 3 p_a^2 p_b^2\} \{ n.p_b \, p_a^\mu \, p_a^\nu - n.p_a \, p_a^\mu \, p_b^\nu\}
 \non\\ && \hskip -.5in
+2\, \sum_{c=1}^n p_c.n \sum_{a=1}^{m} \sum_{b=1\atop b\ne a}^{m}
 {1\over p'_a.n} {p'_a.p'_b\over \{(p'_a.p'_b)^2 - p_a^{\prime 2} p_b^{\prime 2}\}^{3/2}} 
 \{2 (p_a'.p_b')^2 - 3 p_a^{\prime 2} p_b^{\prime 2}\} \{ n.p'_b \, p_a^{\prime\mu} \, p_a^{\prime \nu} - 
 n.p'_a \, p_a^{\prime \mu} \, p_b^{\prime \nu}\} \non\\ && \hskip -.5in
+  \sum_{a=1}^{n} \sum_{b=1\atop b\ne a}^{n}
\sum_{c=1\atop c\ne a}^{n}  {1\over p_a.n}  
{p_a.p_b\over \{(p_a.p_b)^2 - p_a^2 p_b^2\}^{3/2}}   
\{2 (p_a.p_b)^2 - 3 p_a^2 p_b^2\} {p_a.p_c\over \{(p_a.p_c)^2 - p_a^2 p_c^2\}^{3/2}} \non\\ &&
\{2 (p_a.p_c)^2 - 3 p_a^2 p_c^2\}  \{ n.p_b \, p_a^\mu  - n.p_a \, p_b^\mu \} \, 
\{ n.p_c \, p_a^\nu - n.p_a \, p_c^\nu \}
\Bigg]\, ,
\een
and 
\ben \label{eaddcon5}
G^{\mu\nu} &=& -2\, {G^3\over R}\, \Bigg[  2\, \sum_{c=1}^n p_c.n \sum_{a=1}^{m} \sum_{b=1\atop b\ne a}^{m}
 {1\over p'_a.n} {p'_a.p'_b\over \{(p'_a.p'_b)^2 - p_a^{\prime 2} p_b^{\prime 2}\}^{3/2}} 
 \{2 (p_a'.p_b')^2 - 3 p_a^{\prime 2} p_b^{\prime 2}\}  \non\\ && \hskip 1in 
 \{ n.p'_b \, p_a^{\prime\mu} \, p_a^{\prime \nu} - 
 n.p'_a \, p_a^{\prime \mu} \, p_b^{\prime \nu}\} \non\\ && \hskip -.5in
-  \sum_{a=1}^{m} \sum_{b=1\atop b\ne a}^{m}
\sum_{c=1\atop c\ne a}^{m}  {1\over p'_a.n}  
{p'_a.p'_b\over \{(p'_a.p'_b)^2 - p_a^{\prime 2} p_b^{\prime 2}\}^{3/2}}   
\{2 (p'_a.p'_b)^2 - 3 p_a^{\prime 2} p_b^{\prime 2}\} {p'_a.p'_c\over \{(p'_a.p'_c)^2 - p_a^{\prime 2} p_c^{\prime 2}\}^{3/2}} \non\\ &&
\{2 (p'_a.p'_c)^2 - 3 p_a^{\prime 2} p_c^{\prime 2}\}  
\{ n.p'_b \, p_a^{\prime\mu} - n.p'_a \, p_b^{\prime\mu} \} \, 
\{ n.p'_c \, p_a^{\prime \nu}  - n.p'_a \, p_c^{\prime\nu} \}
\Bigg]\, .
\een
One can check that in case of binary black hole merger, regarded as a process in which a single massive object
decays into a massive object and many massless particles (gravitational waves), $F^{\mu\nu}$ and
$G^{\mu\nu}$ vanish. 

One could attempt to prove these results following the same procedure employed in this paper. 
For this we need to iteratively solve the equations of motion to one order higher
than what has been done in this paper. Terms of order $\omega\ln\omega$ would also receive contributions from the
expansion of the factors of $e^{ik.r_a}$ in various expressions in this paper, 
{\it e.g.} in \refb{ejmfingr}, to first order in $k$, and
will therefore depend on the additional data 
$\{r_a\}$. However the $\omega (\ln\omega)^2$ terms given in \refb{edeltasub} do not suffer from any such
ambiguity.

It may be possible to find higher order generalization of these
results using the exponentiated soft factor discussed in 
\cite{1812.06895,1903.12419}.

\sectiono{Numerical estimate}

Before concluding the paper, we shall give estimates of the coefficient of the $1/u$ term in \refb{e3.66int}
in some actual physical processes.
\begin{enumerate}
\item Hypervelocity stars: When a binary star system comes close to the supermassive black hole at the center of the
milky way, often one of them gets captured by the black hole and the other escapes with a high velocity, producing
a hypervelocity star\cite{hyper}. 
This can be taken as a two body decay of a single bound object.
Using the
mass of the central black hole to be of order $10^5 \, M_\odot$, the mass of the hypervelocity star to be of 
order $M_\odot$, its velocity to be of order $3\times 
10^{-3}c$, and the distance of the earth from the  galactic center to be of
order $25000$ light-years, one can estimate the coefficient of the $1/u$ term in
\refb{e3.66int} to be of order $10^{-22}$ days. The minimum value of $u$ needed for \refb{e3.66int} to
hold -- namely when the kinetic energy dominates the gravitational potential energy\cite{1806.01872} -- 
in this case is about
a day. This gives a strain of order $10^{-22}$, which is at the edge of the detection sensitivity of the
future space-based gravitational wave
detectors.
\item Core collapse supernova: 
This case was already discussed in \cite{1806.01872}. 
During this process the residual neutron
star often gets a high velocity kick which could be of the order of 1000 km/sec, balanced by ejected matter in
opposite direction at a speed up to 5000 km/sec.\cite{1206.2503}. 
Taking the neutron star to have a mass of order $M_\odot$ and
the supernova to be in our galaxy so that its distance from the earth is of the order of $10^5$ light years, the
coefficient of the $1/u$ term was computed to be of order $10^{-22}$ sec. The minimum value of $u$ for which the
asymptotic formula holds was found to be of order 
1 sec. Therefore the strain at this time will be of order $10^{-22}$, which is at the edge of the detection limit
of the current gravitational wave detectors.
\item Binary black hole merger: As already pointed out, for binary black hole merger, the coefficient of the $1/u$ term
vanishes due to cancellation between various terms. However the individual terms in \refb{e3.66int} have the same
order of magnitude as the memory effect when the asymptotic formula 
can be trusted, which is of the order of the light crossing
time of the horizon. Therefore the observation of the memory effect without observation of the $1/u$ tail 
is a prediction of general theory of relativity that can be
tested in future gravitational wave experiments. 

\item Bullet cluster: The bullet cluster\cite{harvard} consists of a pair of 
galaxy clusters, each with mass of about
$10^{14} M_\odot$\cite{1209.0384}, passing through each other at a speed of about $10^{-2} \, c$. The system is
situated at a distance of about
$4\times 10^9$ light-years from the earth. Using this data we get the 
coefficient of the $1/u$ term in \refb{e3.66int}
to be of the order of $10^{-6}$ year. The retarded time $u$ for this system, -- the time that has elapsed  since
the centers of the two clusters passed each other, --  is about $1.5\times 10^8$ years. 
This gives the
current value of the strain produced by the bullet cluster in our neighborhood 
to be about $10^{-14}$. While this is much
larger than the sensitivity of the current gravitational wave detectors, what the latter detect is not the strain but the 
change in the strain -- more precisely its second $u$ derivative that enters the expression for the Riemann tensor.
For the bullet cluster this is too small an effect to be observed by the conventional gravitational wave detectors.

\end{enumerate}

\bigskip

{\bf Acknowledgement:}
We would like to thank Alok Laddha for collaboration at the initial stages of this work,  many useful
discussions, and comments on an earlier version of the manuscript. 
We would also like to thank Alfredo Guevara, Dileep Jatkar and Justin Vines for useful discussions.
The work of A.S. was
supported in part by the 
J. C. Bose fellowship of 
the Department of Science and Technology, India and the Infosys chair professorship.

\appendix

\sectiono{Evaluation of some integrals \label{sintegral}}

In this appendix we shall review the evaluation of several integrals following  \cite{1808.03288}. 

We begin with the integral
\be\label{e3.29Japp}
J_{ab}= 
\int{d^4 \ell\over (2\pi)^4} \, {1\over \ell.p_b-i\eps} \, G_r(\ell) \, {1\over 
(\ell-k).p_a+i\eps}
\simeq
\int{d^4 \ell\over (2\pi)^4} \, {1\over \ell.p_b-i\eps} \, {1\over (\ell^0+i\eps)^2 -
\vec\ell^2} \, {1\over 
\ell.p_a+i\eps}\, ,
\ee
with the understanding that 
the integration over $\ell$ is restricted to the region $L^{-1}>>|\vec \ell_\perp|>>\omega$. Simple power
counting, together with the contour deformation arguments given in the
paragraph containing \refb{e3.34},
then shows that the logarithmic contribution can come only from the 
region $|\ell^\mu|\sim |\ell_\perp|$ for all $\mu$. However, since the $\ell^0$ and $\ell^3$ integrals converge for
fixed $\ell_\perp$, we shall take the range of these integrals to be unrestricted.

First consider the case where $a$ represents an incoming particle 
and $b$ represents an
outgoing particle. In this case $p_a^0=-p_{a-n}^{\prime 0}<0$ and $p_b^0>0$. 
In the $\ell^0$ plane the poles of $G_r(\ell)$ are in the
lower half plane, and the zeroes of $\ell.p_a+i\eps$ and $\ell.p_b-i\eps$ are also in the lower half
plane. Therefore we can close the $\ell^0$ integration
contour in the upper half plane and the integral vanishes.

If $a$ represents an outgoing particle 
and $b$ represents an
incoming particle, then the zeroes of $\ell.p_a+i\eps$ and $\ell.p_b-i\eps$ are  in the upper half
plane. Therefore the $\ell^0$ integral does not vanish automatically. 
We can evaluate this by choosing a special frame in which $p_a$ and $-p_b$ both carry spatial momenta along
the third direction, with velocities 
$\beta_a$ and $\beta_b$ respectively. Then the integral takes the form:
\ben \label{ea.2new}
&& \int{d^4 \ell\over (2\pi)^4} \, {1\over p_a^0\, p_b^0} \, {1\over \ell^0 - \beta_a\ell^3 -i\eps_1}\,
{1\over \ell^0 - \beta_b \, \ell^3 - i\eps_2} \, {1\over (\ell^0+i\eps)^2 -
\vec\ell^2} \\ &=&
i\, \int{d^3 \ell\over (2\pi)^3} \, {1\over p_a^0\, p_b^0} \, {1\over (\beta_a-\beta_b)\ell^3 +i(\eps_1-\eps_2)}\, 
\left[ -{1\over (1-\beta_a^2) (\ell^3)^2 + \vec\ell_\perp^2} + {1\over (1-\beta_b^2) (\ell^3)^2 + \vec\ell_\perp^2}
\right]\, ,\non
\een
where $\ell_\perp\equiv (\ell^1,\ell^2)$. 
In the second step we have evaluated the $\ell^0$ integral by closing its integration contour
in the upper half plane. 
Since $\beta_a,\beta_b\le 1$, the denominators of the terms inside the square bracket 
never vanish and we have dropped the $i\eps$ factors in that term.
If we now express $\{(\beta_a-\beta_b)\, \ell^3 +i(\eps_1-\eps_2)\}^{-1}$ as a sum of its principal value and 
a term proportional to $\delta(
(\beta_a-\beta_b) \, \ell^3)$, then
the contribution to the integral from 
the principle value term vanishes due to $\ell^3\to -\ell^3$ symmetry. The term proportional to
$\delta((\beta_a-\beta_b) \, \ell^3)$ forces $\ell^3$ to vanish, in which case the two terms inside the square
bracket cancel. Therefore $J_{ab}$ vanishes also in this case. 

If $a$ and $b$ both refer to outgoing particles, then the zero of $\ell.p_a+i\eps$ is in 
the upper half plane and the zero of $\ell.p_b-i\eps$ is in 
the lower half plane. Therefore if we close the contour in the upper half plane so as to avoid contribution from the
residues at the poles of $G_r(\ell)$, we only pick up the
residue at $\ell.p_a+i\eps=0$. 
If we choose a frame in which $p_a=p_a^0(1, 0,0, \beta_a)$ and $p_b=p_b^0(1,0,0,\beta_b)$,
then the pole is at $\ell^0=\beta_a\ell^3+i\eps$, and the resulting integrand takes the form
\be 
- i \,  \int{d^3 \ell\over (2\pi)^3} \, {1\over p_a^0\, p_b^0}\ 
{1\over  (\beta_a-\beta_b) \, \ell^3+i\eps}\  {1\over (1-\beta_a^2) (\ell^3)^2 + \ell_\perp^2}\, .
\ee
If we express $((\beta_a-\beta_b)\, \ell^3 +i\eps)^{-1}$ term as a sum of its principal value and $-i\pi \delta(
(\beta_a-\beta_b) \, \ell^3)$, then
the contribution from the principal value part vanishes due to $\ell^3\to -\ell^3$ symmetry. The term proportional to
$\delta((\beta_a-\beta_b)\, \ell^3)$ 
forces $\ell^3$ to vanish. Integration over $\ell_\perp\equiv (\ell^1, \ell^2)$ in the
range $\omega<<|\ell_\perp|<<L^{-1}$ now generates
a factor of $2\pi \ln(\omega\, L)^{-1}$. This gives
\be \label{ejabfin}
J_{ab} = {1\over 4\pi} \ln(\omega \, L) \, {1\over p_a^0 p_b^0} \, {1\over |\beta_a-\beta_b| } 
= {1\over 4\pi} \ln(\omega\, L) \, {1\over \sqrt{(p_a.p_b)^2 -
p_a^2 p_b^2}}\, ,
\ee
where in the last step we have reexpressed the result in the covariant form.

If $a$ and $b$ both refer to incoming 
particles, then the zero of $\ell.p_a+i\eps$ is in 
the lower half plane and the zero of $\ell.p_b-i\eps$ is in 
the upper half plane. Therefore if we close the contour in the upper half plane, we only pick up the
residue at $\ell.p_b-i\eps=0$.
The integral can be evaluated similarly using the same frame as used above  
and yields the same result \refb{ejabfin}.

We can also determine the $i\eps$ prescription for the $\ln\omega$
term by noting that in \refb{e3.29Japp}, the factor
\be\label{eieps}
\{ (k-\ell).p_a-i\eps\}^{-1} = \{- p_a^0 \omega + \vec p_a.\vec k -\ell.p_a -i\eps\}^{-1}
\ee
preserves the $i\eps$ prescription under addition of a positive (negative) imaginary part to $\omega$ for
positive (negative) $p_a^0$. Therefore the singularity
in the complex $\omega$ plane must be located in the lower (upper) half plane for positive (negative) $p_a^0$. 
This shows that
$\ln\omega$ in \refb{ejabfin} stands for
$\ln(\omega+i\eps\eta_a)$ where $\eta_a=1$ for outgoing particles and $\eta_a=-1$ for incoming particles.
The final result may be written as:
\be \label{ejabfinfin}
J_{ab} = {1\over 4\pi}\, \delta_{\eta_a,\eta_b}\, \ln\{(\omega+i\, \eps\, \eta_a)L\}
 \, {1\over \sqrt{(p_a.p_b)^2 -
p_a^2 p_b^2}}\, .
\ee

Next we consider the integral:
\be\label{eKab}
K_{b}\equiv 2\, \int {d^4\ell\over (2\pi)^4} \, G_r(k-\ell) \, 
{1\over p_b.\ell-i\eps}\, G_r(\ell) 
\simeq \int {d^4\ell\over (2\pi)^4} \, {1\over k.\ell + i\eps}\, 
{1\over p_b.\ell-i\eps}\, {1\over (\ell^0+i\eps)^2 
- \vec\ell^2} \, ,
\ee
with $\omega$ providing the upper cut-off to the integral and $R^{-1}$ providing the lower cut-off. The last expression is obtained by making the approximation 
$|\ell^\mu|<<\omega$ since the logarithmic contribution arises from this region.
This integral has the same structure as \refb{e3.29Japp} with $p_a$ replaced by $k$ and can be evaluated
similarly. There are however a few differences:
\begin{enumerate}
\item  Due to the changes in the cut-off, $\ln(\omega\, L)$ factor in \refb{ejabfin} will be
replaced by $-\ln(\omega R)$.
\item The $i\eps$ prescription for the integral can be determined by noting that in the expression for
$G_r(k-\ell)= \{(k^0-\ell^0+i\eps)^2 - (\vec k-\vec\ell)^2\}^{-1}$ in \refb{eKab}, 
if we add a positive imaginary part to $k^0=\omega$ then it does not change 
the $i\eps$ prescription for the poles, but adding a negative imaginary part will change the $i\eps$
prescription. Therefore the factors of $\ln\omega$ will correspond to $\ln(\omega+i\eps)$.
\item Since $k$ represents an outgoing momentum, it follows from the arguments given below \refb{ea.2new}
that in order for the integral in \refb{eKab} 
to be non-vanishing, $p_b$ must also represent an outgoing
momentum.
\item Since $k^2=0$, the denominator factor in \refb{ejabfinfin} simplifies to
\be
\sqrt{(k.p_b)^2 - k^2\, p_b^2} = - k.p_b\, ,
\ee
with the minus sign arising from the fact that when $k$ and $p_b$ both represent outgoing
momenta, $k.p_b$ is negative.
\end{enumerate}
With these ingredients we can express the final result for $K_b$ as:
\be\label{eKabfin}
K_b = {1\over 4\pi}\, \delta_{\eta_b, 1}\, \ln\{(\omega+i\eps)\, R\} \, {1\over k.p_b}\, .
\ee

Finally we shall analyze the integral
\be\label{eKpab}
K_{b}'\equiv 2\, \int {d^4\ell\over (2\pi)^4} \, G_r(k-\ell) \, 
{1\over p_b.\ell-i\eps}\, G_r(\ell)^* 
\simeq \int {d^4\ell\over (2\pi)^4} \, {1\over k.\ell + i\eps}\, 
{1\over p_b.\ell-i\eps}\, {1\over (\ell^0-i\eps)^2 
- \vec\ell^2} \, ,
\ee
with $\omega$ providing the upper cut-off to the integral and $R^{-1}$ providing the lower cut-off. To evaluate this
integral, note that $(K_b')^*$ is formally equal to $K_b$ with $(k,p_b)$ replaced by $(-k,-p_b)$. The latter result
can be read out for those of $J_{ab}$ with incoming momenta. This gives
\be \label{eKpabfin}
K_{b}' ={1\over 4\pi}\, \delta_{\eta_b, 1}\, \ln\{(\omega+i\eps)\, R\} \, {1\over k.p_b}\, .
\ee

\sectiono{Contribution from real gravitons} \label{sreal}

In the analysis in \S\ref{sgrsubgr}, we had left out the contribution of the second term of 
\refb{e345gr} in
\refb{etmninitial}. This is given by:
\ben \label{etmninitialb}
\widehat{T}_{\rm extra}^{\mu\nu}(k)\ &=&\ 16\, i\, \pi^2\, G\,
\sum_{a,b}\int \f{d^{4}\ell}{(2\pi)^{4}}\ G_{r}(k-\ell)
\delta(\ell^2) \left\{H(\ell^0) - H(-\ell^0)\right\} \non\\ &&
 \f{1}{p_{b}.\ell-i\epsilon}\ \f{1}{p_{a}.(k-\ell)-i\epsilon} \ \mathcal{F}^{\mu\nu, \alpha\beta , \rho\sigma}(k,\ell)
\ \Big{\lbrace} p_{b\alpha}p_{b\beta}-\f{1}{2}p_{b}^{2}\eta_{\alpha\beta}\Big{\rbrace}
\ \Big{\lbrace}p_{a\rho}p_{a\sigma}-\f{1}{2}p_{a}^{2}\eta_{\rho\sigma}\Big{\rbrace}\, ,\non\\
\een
where $\mathcal{F}^{\mu\nu , \alpha\beta , \rho\sigma }(k,\ell)$ has been defined in \refb{edefffmn}, and it is understood that the integration over the momenta $\ell^\mu$
is restricted to the range much below the cut-off $L^{-1}$, so that we can drop the
exponential factors of $e^{-ik.r_{a}}$ and $e^{i\ell.(r_{a}-r_{b})}$ that regulate the
ultraviolet divergence in \refb{etmninitial}.
We shall now analyze possible logarithmic contributions to this term from different regions of integration.

First of all, since each term in $\mathcal{F}^{\mu\nu , \alpha\beta , \rho\sigma }(k,\ell)$ 
defined in \refb{edefffmn} has a factor of $(k-\ell)$, a simple power counting shows that there are
no logarithmic contributions from the region $|k^\mu-\ell^\mu|<<\omega$. Therefore we need
to analyze contributions from the regions $R^{-1}<<|\ell^\mu|<<\omega$ and $\omega << |\ell^\mu|
<< L^{-1}$. Power counting shows that in order to analyze logarithmic contribution from the region 
$R^{-1}<<|\ell^\mu|<<\omega$, we can replace the numerator by its $\ell\to 0$ limit. Therefore we need to
 analyze an integral of the form
\be\label{ediffe0}
\EE_0 =  \int \f{d^{4}\ell}{(2\pi)^{4}}\  \f{1}{p_{b}.\ell-i\epsilon}\ \f{1}{p_{a}.(k-\ell)-i\epsilon}\ 
 G_{r}(k-\ell)
\delta(\ell^2) \left\{H(\ell^0) - H(-\ell^0)\right\} \, .
\ee
This can be reexpressed as:
\be 
\EE_0 =  {1\over 2\pi i} \, \int \f{d^{4}\ell}{(2\pi)^{4}}\  \f{1}{p_{b}.\ell-i\epsilon}\ 
\f{1}{p_{a}.(k-\ell)-i\epsilon}\  G_{r}(k-\ell)
\left\{G_r(\ell)^* - G_r(\ell)\right\}\, .
\ee
For $|\ell^\mu|<<\omega$ the contribution reduces to one of the integrals 
defined in \refb{eKab} or \refb{eKpab} and can be evaluated using \refb{eKabfin}
or \refb{eKpabfin}. The result vanishes due to the cancellation between the contributions
coming from the $G_r(\ell)$ and $G_r(\ell)^*$ terms. 
Therefore there is no logarithmic contribution from the $|\ell^\mu|<<\omega$ region.

We now focus on the region $|\ell^\mu|>>\omega$. Power counting shows that the integral has
linear divergence in this region. So we have to evaluate it carefully by keeping also the subleading terms
in this limit. First let us consider the subleading contribution arising 
from the terms in 
$\mathcal{F}^{\mu\nu , \alpha\beta , \rho\sigma }(k,\ell)$ that are linear in $\ell$. These involve integrals
of the form:
\be \label{edefe1}
\EE_1 =  \int \f{d^{4}\ell}{(2\pi)^{4}}\  \f{1}{p_{b}.\ell-i\epsilon}\ \f{1}{p_{a}.(k-\ell)-i\epsilon}\ 
G_{r}(k-\ell)
\delta(\ell^2) \left\{H(\ell^0) - H(-\ell^0)\right\} \, \ell^\kappa\, .
\ee
In the region $|\ell^\mu|>>\omega$,
we can approximate the integral as:
\be \label{edefe1a}
\EE_1 \simeq  -
\int \f{d^{4}\ell}{(2\pi)^{4}}\  \f{1}{p_{b}.\ell-i\epsilon}\ \f{1}{p_{a}.\ell+i\epsilon}\ 
{1\over 2k.\ell - i\epsilon\ell^0}\
\delta(\ell^2)\, \left\{H(\ell^0) - H(-\ell^0)\right\} \, \ell^\kappa\, .
\ee
Now, since $\delta(\ell^2)$ factor puts the momentum $\ell$ on-shell, $p_a.\ell$ and
$p_b.\ell$ never vanish in the integration region of interest and therefore we
can drop the $i\eps$ factors.\footnote{The only exception is when $p_a$ and / or $p_b$
represents a massless particle and $\ell$ becomes parallel to $p_a$ and / or $p_b$
producing a collinear divergence; but such divergences are known to cancel in
gravitational theories\cite{1109.0270}.} $k.\ell$ can vanish only when $\ell$ is parallel to $k$,
but by examining the numerator factor \refb{edefffmn} we find that there are always additional
suppression factors in this limit that kill potential singularity at $k.\ell=0$. Therefore the
$i\eps\ell^0$ factor can be dropped from this term as well. For example
the presence of a $p_a.k$ or $p_a.\ell$ factor in the numerator will mean that the ratio $p_a.k/p_a.\ell$
or $p_a.\ell/p_a.\ell$
becomes $a$ independent in the limit when $\ell$ is parallel to $k$, and the result then
vanishes after summing over $a$ using momentum conservation $\sum_a p_a=0$. 
A similar result holds for terms proportional to $p_b.k$ or $p_b.\ell$.
Also, a combination of terms of the form
$k^\mu \xi^\nu + \xi^\nu k^\mu - k.\xi \, \eta^{\mu\nu}$
will produce a term in the gravitational wave-form that is pure gauge and therefore
can be removed. Therefore we can remove such terms appearing at the level
of the integrand itself.  
Once the $i\eps$ factors are removed from all the
denominators, the integrand of \refb{edefe1a} becomes an odd function of $\ell$ and
therefore vanishes after integration over $\ell$.

We now turn to the contribution from terms in $\mathcal{F}^{\mu\nu , \alpha\beta , \rho\sigma }(k,\ell)$ 
that are quadratic in $\ell$.
The corresponding integrals take the form:
\be\label{ediffe2}
\EE_2 =  \int \f{d^{4}\ell}{(2\pi)^{4}}\   \f{1}{p_{b}.\ell-i\epsilon}\ \f{1}{p_{a}.(k-\ell)-i\epsilon}\ 
G_{r}(k-\ell)\,
\delta(\ell^2) \left\{H(\ell^0) - H(-\ell^0)\right\} \, \ell^\kappa \ell^\tau\, .
\ee
This has potential linear divergence from the region $|\ell^\mu|>>\omega$. Therefore
we need to expand the $(p_{a}.(k-\ell)-i\epsilon)^{-1}$ factor 
in powers of $p_a.k$ to the first subleading order:
\be\label{eexpanden}
\f{1}{p_{a}.(k-\ell)-i\epsilon} = - \f{1}{p_{a}.\ell+i\epsilon} 
- \f{p_a.k}{(p_{a}.\ell+i\epsilon)^2}\, .
\ee
We can argue as before that due to the presence of the $\delta(\ell^2)$ factor we can
drop all the $i\eps$ factors in the denominator. In this case the contribution from the
last term in \refb{eexpanden} to the integral \refb{ediffe2} vanishes by $\ell\to -\ell$
symmetry. On the other hand, when we substitute the first term on the right hand
side of \refb{eexpanden} into \refb{ediffe2}, the integrand is an even function of 
$\ell$. In this case  terms proportional to $H(\ell^0)$ and $-H(-\ell^0)$ give identical
contributions, and we get:
\be\label{ediffe2new}
\EE_2 \simeq  -\, \int \f{d^{4}\ell}{(2\pi)^{4}}\   \f{1}{p_{b}.\ell-i\epsilon}\ 
\f{1}{p_{a}.\ell+i\epsilon}\ 
{1\over k.\ell-i\eps}\ 
\delta(\ell^2) \ H(\ell^0)  \, \ell^\kappa \ell^\tau\, .
\ee
Note that we have kept the $i\eps$ factors even though the presence of $\delta(\ell^2)$ makes them
irrelevant.

Now from \refb{etmninitialb} and \refb{edefffmn}
we see that the indices $\kappa$ and $\tau$ must either
be free indices $\mu$, $\nu$, or be contracted with the index of $p_a$ or $p_b$, or be
contracted with each other. If they are contracted with each other then we have a
factor of $\ell^2$ and the contribution vanishes due to the $\delta(\ell^2)$ factor.
If any one of them is contracted with $p_b$, then we have a factor of $p_b.\ell$ in the
numerator that kills the denominator factor of $p_b.\ell-i\eps$. After summing over
$b$ and using momentum conservation law $\sum_b p_b=0$, this 
contribution also vanishes. A similar argument can be given for terms where
either $\ell^\kappa$ or $\ell^\tau$ is contracted with $p_a$. 
The only term that survives is where $(\kappa,\tau)$ take values $(\mu,\nu)$.
Using this we can bring the contribution to \refb{etmninitialb} to the 
form
\be\label{eb5pre}
\wh T^{\mu\nu}_{\rm extra} 
=  {G\over \pi^2} \, \int \left\{{d^4\ell} \, \delta(\ell^2) H(\ell^0)\right\} 
\left\{\sum_{a,b=1}^{m+n} {1 \over (p_a.\ell - i\eps)\, (p_b.\ell + i\eps)}\right\}
\, \left\{ (p_a.p_b)^2 -{1\over 2} p_a^2 p_b^2\right\}
{\ell^\mu \ell^\nu\over i (k.\ell-i\eps)}\, .
\ee

We shall now show that this contribution can be interpreted as the effect of soft emission from the gravitational
radiation produced during the scattering, and is therefore already accounted for when we include in the sum
over $a$ in the soft factor the contribution from the gravitational radiation 
produced during the scattering. 
For this we note that the 
flux of radiation in a phase space
volume $\delta(\ell^2)\, H(\ell^0)\, d^4\ell$ 
carrying polarization $\ve_{\mu\nu}$ is given by
\be\label{esub2}
{G\over \pi^2}\, \left\{d^4\ell \, \delta(\ell^2) H(\ell^0)\right\} 
\left\{\sum_{a=1}^{m+n} {p_a^\rho p_a^\sigma \over p_a.\ell - i\eps}\right\}
\, \left\{\sum_{b=1}^{m+n} {p_b^\kappa p_b^\tau \over p_b.\ell + i\eps}\right\}\, (\ve_{\kappa\tau})^*
\ve_{\rho\sigma} \, .
\ee
This equation can be derived by using the relation between the leading soft factor \refb{eleadgr}
and the flux of
radiation\cite{1801.07719}. In this case the first factor inside the curly bracket gives the phase space volume, 
and the rest of the factors gives the flux of
radiation produced in the scattering.  
Since we shall be interested in only the total flux,  
we can sum over polarizations using the formula
\be\label{epolsum}
\sum_\ve (\ve_{\kappa\tau})^*
\ve_{\rho\sigma} = {1\over 2} \, 
\left(\eta_{\kappa\rho}\eta_{\tau\sigma}
+ \eta_{\kappa\sigma}\eta_{\tau\rho} - \eta_{\kappa\tau} \eta_{\rho\sigma}\right)\, ,
\ee
yielding the standard result for the total 
flux of gravitational radiation given {\it e.g.} in eq.(10.4.22) of \cite{weinbergbook}:
\be\label{esub1}
{G\over \pi^2}\, \left\{d^4\ell \, \delta(\ell^2) H(\ell^0)\right\} 
\left\{\sum_{a=1}^{m+n} {p_a^\rho p_a^\sigma \over p_a.\ell - i\eps}\right\}
\, \left\{\sum_{b=1}^{m+n} {p_b^\kappa p_b^\tau \over p_b.\ell + i\eps}\right\}\,  {1\over 2} \, 
\left(\eta_{\kappa\rho}\eta_{\tau\sigma}
+ \eta_{\kappa\sigma}\eta_{\tau\rho} - \eta_{\kappa\tau} \eta_{\rho\sigma}\right) \, .
\ee
The leading soft theorem
\refb{eleadgr}, applied to this radiation flux, 
now shows that the 
contribution to the $\wh T^{\mu\nu}$ due to
the radiation is obtained by multiplying \refb{esub1} by $-i/(\ell.k-i\eps)$ and integrating over $\ell$. This
gives the net leading contribution to the soft factor due to radiation to be
\be\label{eb5}
\wh T^{R\mu\nu} 
= {G\over \pi^2} \, \int \left\{{d^4\ell} \, \delta(\ell^2) H(\ell^0)\right\} 
\left\{\sum_{a,b=1}^{m+n} {1 \over (p_a.\ell - i\eps)\, (p_b.\ell + i\eps)}\right\}
\, \left\{ (p_a.p_b)^2 -{1\over 2} p_a^2 p_b^2\right\}
{\ell^\mu \ell^\nu\over i (k.\ell-i\eps)}\, .
\ee
This agrees with \refb{eb5pre}, showing that the extra contribution 
\refb{etmninitialb} is already
accounted for by including in the sum over $a$ in the soft factor the contribution due to
radiation.\footnote{As in \cite{1801.07719}, this can also be expressed as angular 
integrals over appropriate functions of the radiative gravitational field and its derivatives, but we shall not
describe this here.}

\sectiono{Position space analysis of  $\wh T^{X\mu\nu}$}

In \S\ref{sgrav}, \S\ref{sgen} 
we have carried out our analysis in momentum space. This has the advantage that the
expressions we obtain are similar to the ones that appear in the evaluation of Feynman diagrams, and various
general techniques developed for computing amplitudes in quantum field theory may find applications here.
Nevertheless it is instructive to see how some of these computations can also be performed directly in
position space. In this appendix
we shall show how to carry out the analysis of sections \S\ref{s3.3} and \S\ref{sgrav.3} directly in position space.

Our first task will be to 
compute the gravitational fields produced by the incoming and outgoing particles during a scattering,
and study their effect on the motion of the other particles.  At the leading order, the incoming and outgoing
particle trajectories are given by \refb{e511grpre}, or equivalently \refb{e511gr}. 
Using retarded Green's function in flat space-time, we
get the following expression for the gravitational field produced by the $b$-th particle on the forward light-cone
of the trajectory of the particle\cite{1808.03288}:
\be \label{edefeamnew}
e^{(b)}_{\mu\nu}(x)=2\, G\, m_b\, {v_{b\mu} v_{b\nu}\over \sqrt{(v_b.x)^2 + x^2}}, \quad 
h^{(b)}_{\mu\nu} = e^{(b)}_{\mu\nu} - {1\over 2} \, \eta_{\mu\nu} \, e^{(b)\rho}_{~\rho}\, .
\ee
The associated Christoffel symbol is given by, in the weak field approximation,
\ben\label{echri}
\Gamma^{(b)\alpha}_{\rho\tau}(x) &=& -{2\, G\, m_b} \,  {1\over 
\{ (v_b.x)^2 + x^2\}^{3/2}}\, \eta^{\alpha\mu} \, 
\left[\left\{v_{b\mu} v_{b\tau} + {1\over 2} \eta_{\mu\tau} \right\} \left\{x_\rho + v_b.x\, v_{b\rho}\right\}
\right . \nonumber \\ && \hskip -.5in \left.
+ \left\{v_{b\mu} v_{b\rho} + {1\over 2} \eta_{\mu\rho} \right\} \left\{x_\tau + v_b.x\, v_{b\tau}\right\}
- \left\{v_{b\rho} v_{b\tau} + {1\over 2} \eta_{\rho\tau} \right\} \left\{x_\mu + v_b.x\, v_{b\mu}\right\}
\right]\,.
\een
Since the field has support on the forward light-cone of the trajectory, 
it follows that in sufficiently far future and far past
of the scattering event, the outgoing particles are affected by the gravitational field of the outgoing particles and
the incoming particles are affected by the gravitational field of the incoming particles.

Let $Y_a^\mu$ denote the correction to the particle trajectory 
\refb{e511gr} due to the gravitational field produced by the other particles: 
\be\label{ethisgrnew}
X_a^\mu(\sigma) = v_a^\mu\, \sigma +r_a^\mu + Y_a^\mu(\sigma)\, .
\ee
We shall use the compact notation described in \refb{edefprime}, 
and define $\eta_a$ to be a number that takes value $1$ for outgoing particles ($1\le a\le n$)
and $-1$ for incoming particles ($n+1\le a\le m+n$).
Then $Y_a^\mu$ satisfies the differential equation and boundary conditions:
\be\label{eYeqgrnew}
{d^2 Y_a^\mu\over d\sigma^2} =-\Gamma^\mu_{\nu\rho}(v_a\, \sigma+r_a)\,  v_a^\mu \,
v_a^\nu, \quad Y_a^\mu\to 0\, \, \hbox{as $\sigma\to  0$} , \quad  \, {d Y_a^\mu\over d\sigma}
\to 0 \, \,
\hbox{as $\sigma\to\infty$}\, ,
\ee
where 
\be \label{e4.11grnew}
\Gamma^\mu_{\nu\rho}=\sum_{b=1\atop b\ne a, \eta_a\eta_b=1}^{m+n}  \Gamma^{(b)\mu}_{\nu\rho} \, .
\ee
The constraint $\eta_a\eta_b=1$ reflects that the outgoing particles are affected by the gravitational field
of the outgoing particles and the incoming particles are affected by the gravitational field of the incoming
particles.
Using \refb{echri}, \refb{eYeqgrnew} and \refb{e4.11grnew} we get, for $\sigma>>|r_a|\sim L$:
\be
{d^2 Y_a^\alpha(\sigma)\over d\sigma^2} \simeq
{2\, G\over \sigma^2}  \, \sum_{b=1\atop b\ne a, \eta_a\eta_b=1}^{m+n}  \, m_b \, {1\over 
\{ (v_b.v_a)^2 -1\}^{3/2}}\, \left[- {1\over 2} v_a^\alpha + {1\over 2} v_b^\alpha \left\{2(v_b.v_a)^3  
- 3 v_b.v_a\right\}\right]\, . 
\ee
This gives
\be \label{edyds}
{d Y_a^\alpha(\sigma)\over d\sigma} \simeq
 -{2\, G\over \sigma}  \, \sum_{b=1\atop b\ne a, \eta_a\eta_b=1}^{m+n}  \, m_b \, {1\over 
\{ (v_b.v_a)^2 -1\}^{3/2}}\, \left[- {1\over 2} v_a^\alpha + {1\over 2} v_b^\alpha \left\{2(v_b.v_a)^3  
- 3 v_b.v_a\right\}\right]\, . 
\ee

Now in \refb{e4.10grpre} we have the expression for
$\wh T^X_{\mu\nu}$ to subleading order:
\be\label{e4.10grprepos}
\wh T^{X\mu\nu}(k)  =
 \sum_{a=1}^{m+n} m_a\, \int_0^\infty d\sigma \, e^{-ik.(v_a\, \sigma+r_a)} \, \left[v_a^\mu v_a^\nu - ik.Y_a(\sigma) \, v_a^\mu
v_a^\nu
+  {dY_a^\mu\over d\sigma}\, v_a^\nu + v_a^\mu \, {dY_a^\nu\over d\sigma}\right]\, .
\ee
As discussed below \refb{eleadgr}, the integration
over $\sigma$ is made well defined by replacing $\omega$ by $\omega+i\eps$ for outgoing particles and
by $\omega-i\eps$ for incoming particles. We now manipulate the second term by writing
\be
 e^{-ik.(v_a\, \sigma+r_a)} =  {i\over k.v_a}\, {d\over d\sigma} e^{-ik.(v_a\, \sigma+r_a)}
 \ee
 and integrating over $\sigma$ by parts. The boundary term at infinity vanishes due to the replacement of $\omega$
 by $\omega+ i\eps\eta_a$, while the boundary term at $\sigma=0$ gives a finite contribution in the $\omega\to0$
 limit and is not of interest to us. With this \refb{e4.10grprepos} can be expressed as
 \be\label{e4.10grpre1}
\wh T^{X\mu\nu}(k) =
 \sum_{a=1}^{m+n} m_a\, \int_0^\infty d\sigma \, e^{-ik.(v_a\, \sigma+r_a)} \, \left[v_a^\mu v_a^\nu - {1\over k.v_a}\, 
 k.{dY_a\over d\sigma}\, v_a^\mu
v_a^\nu
+  {dY_a^\mu\over d\sigma}\, v_a^\nu + v_a^\mu \, {dY_a^\nu\over d\sigma}\right]\, .
\ee
After integration over $\sigma$ the first term gives the leading term. In the other terms we can substitute the
expression \refb{edyds} for $dY_a/d\sigma$. Since the integrand is proportional to $1/\sigma$ in the range
$L<<\sigma<<\omega^{-1}$, we get contribution proportional to $\ln ((\omega+ i\eps\eta_a)^{-1}/L)$. 
Therefore, with the help of \refb{edyds}, the logarithmic correction to $\wh T^X$, given by the last three
terms in \refb{e4.10grpre1}, takes the form:
\ben \label{etxposi}
&& \Delta \wh T^{X\mu\nu}(k) = 2\,G\, 
 \sum_{a=1}^{m+n} m_a\, \ln \{L(\omega+ i\eps\eta_a)\}\, 
\sum_{b=1\atop b\ne a, \eta_a\eta_b=1}^{m+n}  \, m_b \, {1\over 
\{ (v_b.v_a)^2 -1\}^{3/2}}\, \non\\ && 
\Bigg[ - {v_a^\mu v_a^\nu\over k.v_a} \, k_\alpha \left\{
- {1\over 2} v_a^\alpha + {1\over 2} v_b^\alpha \left\{2(v_b.v_a)^3  
- 3 v_b.v_a\right\} \right\} + \left\{
- {1\over 2} v_a^\mu + {1\over 2} v_b^\mu \left\{2(v_b.v_a)^3  
- 3 v_b.v_a\right\} \right\} v_a^\nu  \non\\ && \hskip 1in + \left\{
- {1\over 2} v_a^\nu + {1\over 2} v_b^\nu \left\{2(v_b.v_a)^3  
- 3 v_b.v_a\right\} \right\} v_a^\mu \Bigg]\, . 
\een
After using the relations $p_a=m_a v_a$ and some simplification we get:
\ben\label{efingrapre}
\Delta\wh T^{X\mu\nu}(k) &=& 2\, G \, \sum_{a=1}^{m+n} \ln\{L(\omega+i\eps\eta_a)\}\,\sum_{b=1\atop b\ne a, \eta_a\eta_b=1}^{m+n}
{1\over 
\{(p_a.p_b)^2 
-p_a^2 p_b^2\}^{3/2}} \non\\ && \times \,  \Bigg[{k.p_b\over k.p_a}\, p_a^\mu p_a^\nu \, p_a.p_b
\left\{{3\over 2} p_a^2 p_b^2 - (p_a.p_b)^2\right\}  \nonumber \\ &&
\hskip .1in
+ {1\over 2} p_a^\mu p_a^\nu \, p_a^2 \, (p_b^2)^2 - \{ p_a^\mu p_b^\nu + p_a^\nu p_b^\mu\} \,  p_a.p_b\,
\left\{{3\over 2} p_a^2 p_b^2 - (p_a.p_b)^2\right\} 
\Bigg]\, .
\een
This is in perfect agreement with \refb{efingratot}.

\end{document}